\documentclass[a4paper,fleqn,usenatbib]{mnras}

\usepackage{newtxtext,newtxmath}

\usepackage[T1]{fontenc}
\usepackage{ae,aecompl}

\usepackage{graphicx}	
\usepackage{amsmath}	
\usepackage{amssymb}	
\usepackage{epstopdf}
\usepackage{hyperref}
\usepackage{color}
\usepackage{ulem}

\newcommand{\authorfix}{\textcolor{black}}
\newcommand{\finalfix}{\textcolor{black}}

\title[Structural models of spiral galaxies]{Photometric and kinematic \finalfix{DiskFit} models of four nearby spiral galaxies}

\author[W.\ Peters \& R.\ Kuzio de Naray]{
Wesley Peters,$^{1}$\thanks{E-mail: peters@astro.gsu.edu (WP); kuzio@astro.gsu.edu (RKD)}
Rachel Kuzio de Naray,$^{1}$\footnotemark[1]
\\
$^{1}$Department of Physics \& Astronomy, Georgia State University, PO Box 5060, Atlanta, GA 30302-5060, USA\\
}

\date{Accepted 2017 March 29. Received 2017 March 28; in original form 2016 December 15}

\pubyear{2016}

\begin{document}
\label{firstpage}
\pagerange{\pageref{firstpage}--\pageref{lastpage}}
\maketitle

\begin{abstract}
We present optical \textit{BVRI} photometry, H$\alpha$ IFU velocity fields, and H$\alpha$ long-slit rotation curves for a sample of four nearby spiral galaxies having a range of morphologies \authorfix{and inclinations}. \finalfix{We show that the \texttt{DiskFit} code can be used to model the photometric and kinematic data of these four galaxies and explore how well} the photometric data can be decomposed into structures like bars and bulges and to look for non-circular motions in the kinematic data. In general, we find good agreement between our photometric and kinematic models for most parameters. We find the best consistency between our photometric and kinematic models for NGC~6674, a relatively face-on spiral with clear and distinct bulge and bar components. We also find excellent consistency for NGC~2841, and find a bar $\sim$10$\degr$ south of the disc major axis in the inner 20$\arcsec$. Due to geometric effects caused by its high inclination, we find the kinematic model for NGC~2654 to be less accurate than its photometry. We find the bar in NGC~2654 to be roughly parallel to the major axis of the galaxy. \authorfix{We are unable to photometrically model our most highly inclined galaxy, NGC 5746, with \texttt{DiskFit} and instead use the galaxy isophotes to determine that the system contains a bar} $\sim$5$\degr$ to $\sim$10$\degr$ east of the disc major axis. The high inclination and extinction in this galaxy \authorfix{also} prevent our kinematic model from accurately determining parameters about the bar, though the data are better modeled when a bar is included.
\end{abstract}

\begin{keywords}
galaxies: kinematics and dynamics --- galaxies: structure
\end{keywords}



\section{Introduction}
\label{sec:intro}

Morphology provides an immense amount of information about a galaxy.  For instance, the strength and pattern speed of bars can be used to determine the degree of disc maximality \citep{athanassoula2003, block2004, byrd2006}. The evolutionary history of a galaxy may also be imprinted on the bulge and reflected in the shape of its light profile (i.e. classical bulges vs pseudo bulges) \citep{fisher2008, gadotti2009, weinzirl2009, barentine2012, perez2013, menendez2014}. \authorfix{It is therefore crucial to have a reliable method for accurately measuring and quantifying the structural properties of a galaxy.}

When galaxies are moderately-inclined to face-on, their structural components are easily visible.  Measurements of inclinations, position angles, lengths, ellipticities, etc, are relatively straight-forward. Using ellipse isophote-fitting techniques or photometric decomposition of galaxy images \citep{peng2002, gadotti2008}, surface brightness profiles can be obtained along with the disc, bulge, and bar parameters.  Kinematic data can be modeled with tilted-ring codes \citep[e.g. ROTCUR or TiRiFiC:][]{begeman1989, joza2007} to quantify non-circular motions caused by bars.

When galaxies are viewed close to edge-on, however, their morphologies are obscured and it becomes much more complicated to characterize bars and bulges.  Oftentimes, indirect measurements must be made by relying on the \authorfix{combined} interpretation of visible features in \authorfix{both} the photometry and kinematics.

For example, edge-on galaxies with boxy/peanut-shaped bulges are interpreted as being barred \citep{combes1981, combes1990, bundy2015, lutticke2000b, bureau2006}. Peanut or X-shaped bulges are thought to be bars that are viewed side-on and that have buckled.  Boxy bulges are seen when a bar is oriented at an intermediate angle relative to the observer, and round bulges will be seen when the bar is pointed directly at the observer.  The kinematics of barred edge-on galaxies may display a figure-of-eight pattern or a parallelogram shape in a long-slit spectrum, or twisted isovelocity contours in a two-dimensional velocity field \citep{kuijken1995, bureau1999b, athanassoula1999, merrifield1999, bureau2005}.

In this paper, we model optical photometry and H$\alpha$ kinematics of four nearby spiral galaxies selected to have very different morphological properties in order to explore how well various techniques can return measurements of galaxy components as a function of inclination.

\authorfix{In particular, w}e model our data with the publicly available code \texttt{DiskFit}. We choose \texttt{DiskFit} because it is able to work with \textit{both} photometric and kinematic data.

The photometric side of \texttt{DiskFit} can decompose galaxy photometry into any combination of disc, bulge, and bar components and assumes only a S\'{e}rsic profile for the bulge.  In addition to returning details of the structure of each of these components, \texttt{DiskFit} calculates the amount of light coming from each component as a function of radius as well as the total percentage of overall galaxy light coming from each.

The kinematic side of \texttt{DiskFit} can fit for axisymmetric and nonaxiysmmetric motions in two-dimensional velocity fields, as well as perform fits for radial flows and outer-disc warps. Similar to its photometric side, \texttt{DiskFit} returns measurements of the galaxy morphology as well as the velocity as a function of radius.

\authorfix{We are interested in determining what limits \texttt{DiskFit} has when modeling observed data. Is there an inclination limit? How much dust or obscuration can there be? For photometry, does the chosen filter matter? For kinematics, how much does the radial or spatial resolution affect the results? Are both photometry and kinematic data required, or do they provide a consistent picture of the galaxy such that having only one type of data is sufficient?}

Our paper is organized as follows. In Section 2 we describe our sample and provide basic information about the galaxies. We discuss our observations and data reduction procedures in Section 3. An overview of \texttt{DiskFit}, details about its capabilities, and its application to our photometric and kinematic data are presented in Section 4. We discuss the results of our modeling for each galaxy in Section 5 and finally summarize our findings in Section 6.

\section{Sample}
\label{sec:sample}

Our sample consists of four spiral galaxies \authorfix{specifically} selected to show very different properties (see Table~\ref{gal_info} and Fig.~\ref{gal_pics}) \authorfix{in order to test the limits of \texttt{DiskFit}}. Some of our galaxies have obvious features, whereas others have hidden components. 

NGC~2841 and NGC~6674, have inclinations between 50\degr and 65\degr, making their structure easy to see. NGC~2841 has a rather undefined, flocculent spiral structure and dust throughout.  NGC~6674 displays a prominent bar and ring in its photometry, as well as diffuse spiral arms. \authorfix{Given the moderate inclinations of these two galaxies, we anticipate that their morphology will be easy to quantify.}

\authorfix{The other two galaxies, NGC 2654 and NGC 5746, are much more edge-on systems with inclinations of at least 80\degr.} NGC~2654 showcases a few star forming regions in its disc, as well as a faint dust lane. NGC~5746 extends roughly 6$\arcmin$ on the sky and possesses a massive dust lane that cuts through the central bulge. Due to their high inclinations, it is difficult to see distinct components in these galaxies. There are, however, indirect indications that both of these systems contain bars; both galaxies have boxy/peanut-shaped bulges, which, as described before, are interpreted as being bars viewed close to edge-on.

\authorfix{All four of these galaxies are well-studied in the literature and these past results will provide important independent checks on the results of the \texttt{DiskFit} modeling.}

\begin{table}
	\centering
	\caption{Galaxy Sample.  Hubble types and inclinations are taken from SIMBAD \citep{wenger2000} and references therein.}
  	\label{gal_info}
	\begin{tabular}{ccccc}
 	\hline
   	Galaxy 	& R.A. 	& Dec.  	& Hubble Type 	& Inc. \\
    			& (J2000)	& (J2000)	&                       & ($\degr$)\\
  	\hline
  	NGC~2654 & 08 49 11.9 & +60 13 16 & SBab & 79 \\
  	NGC~2841 & 09 22 02.6 & +50 58 35 & Sab & 65 \\ 
  	NGC~5746 & 14 44 55.9 & +01 57 18 & SABb & 80 \\  
  	NGC~6674 & 18 38 33.9 & +25 22 31 & SBb & 57\\
	\hline
	\end{tabular}
\end{table}

\section{Data}
\label{sec:data}

In order to study the photometric and kinematic structure of these galaxies in detail, we have obtained optical broadband photometry, as well as H$\alpha$ velocity fields and H$\alpha$ long-slit rotation curves. \authorfix{The \textit{BVRI} images are used to test how morphology and the quality of the \texttt{DiskFit} models change with wavelength.} The velocity field data provide a two-dimensional map of the gas velocity throughout each galaxy while the long-slit observations provide a complimentary measure of the gas velocity out to much larger radii. Basic information about the details of the observations is given in Table~\ref{data_info}. 

\begin{table*}
	\centering
	\caption{Observation details.  Slit position angles are on-sky values, rotating East from North.}
	\label{data_info}
	\begin{tabular}{ccccc}
		\hline
					&				&Photometry	&		&\\
		\hline
		Galaxy		&Date			&Instrument	&Filter	&Seeing\\
					&				&			&		&($\arcsec$)\\
		\hline
		NGC~2654	&2016 Feb 03 		&SPICAM  	&B 		&0.98\\
	                			&                      		&                	& V 		&0.81\\ 
	                 		&                      		&                	& R 		&0.87\\ 
	                 		&                      		&                	& I  		&0.90\\  
	         			&				&			&		&\\
		NGC~2841 	&2016 Mar 24  		&ARCTIC   	&B 		&1.23\\
	                 		&                      		&                	&V 		&1.12\\ 
	                 		&                      		&                	&R 		&1.10\\ 
	                 		&                      		&                	&I  		&0.99\\  
	         			&				&			&		&\\
		NGC~5746 	&2016 Mar 11  		&ARCTIC   	&B 		&1.03\\
	                 		&                      		&                	&V 		&1.27\\ 
	                 		&                      		&                	&R 		&1.19\\ 
	                 		&                      		&                	&I  		&1.16\\ 
	         			&				&			&		&\\
		NGC~6674 	&2016 Mar 11  		&ARCTIC   	&B 		&1.80\\
	                 		&                      		&                	&V 		&2.16\\ 
	                 		&                      		&                	&R 		&2.15\\ 
	                 		&                      		&                	&I  		&1.68\\       
		\hline
					&				&Spectroscopy	&		&\\
		\hline
		Galaxy 		&Date 			&Instrument 	&Grating 	&Slit P.A.\ \& Width \\
		                         &                        	&                       &            	&($\degr$, $\arcsec$)\\
		\hline
		NGC~2654 	&2009 Feb 19 		&SparsePak 	&316@63.4 	&... , ... \\
			 		&2014 Jan 24  		&DIS 		&B400/R300 	& 66, 1.5 \\
					&				&			&			&\\
		NGC~2841 	&2009 Feb 19 		&SparsePak 	&316@63.4 	&... , ... \\
					&				&			&			&\\
		NGC~5746 	&2009 May 15 		&SparsePak 	&316@63.4 	&... , ... \\
			 		&2016 May 12 		&DIS 		&B1200/R1200 &170, 1.5 \\
					&				&			&			&\\
		NGC~6674 	&2009 May 15 		&SparsePak 	&316@63.4 	&... , ... \\
			 		&2016 May 12 		&DIS 		&B1200/R1200 &143, 1.5\\
		\hline
	\end{tabular}
\end{table*}

\subsection{Photometry}
\label{sec:photometry}

We have obtained broadband \textit{BVRI} photometry of our four galaxies using the SPICAM and ARCTIC imagers on the 3.5-m telescope at Apache Point Observatory (APO)\footnote{Based on observations obtained with the Apache Point Observatory 3.5-meter telescope, which is owned and operated by the Astrophysical Research Consortium.}. SPICAM has a field of view of 4.78$\arcmin$~$\times$~4.78$\arcmin$, and ARCTIC has a field of view of 7.5$\arcmin$~$\times$~7.5$\arcmin$. Used in 2~$\times$~2 binning mode, the plate scales are 0.28$\arcsec$\ pix$^{-1}$ and 0.228$\arcsec$\ pix$^{-1}$, respectively. 

SPICAM observations for NGC~2654 were obtained on 2016 February 03. ARCTIC observations for NGC~2841, NGC~5746, and NGC~6674 were obtained on 2016 March 23, 2016 May 11, and 2016 May 11, respectively. We observed NGC~2654 with 3~$\times$~300s exposures in each filter with 15$\arcsec$ dithering between each exposure to correct for bad pixels and cosmic rays.  Similarly, two 300s exposures dithered by 15$\arcsec$ were taken in each filter for NGC~2841, NGC~5746, and NGC~6674.  Landolt standard stars \citep{landolt1992} were also observed in each filter throughout the night. Unfortunately, the data were not photometric on 2016 May 11.

\begin{figure*}
	\center
	\includegraphics[width=0.45\textwidth]{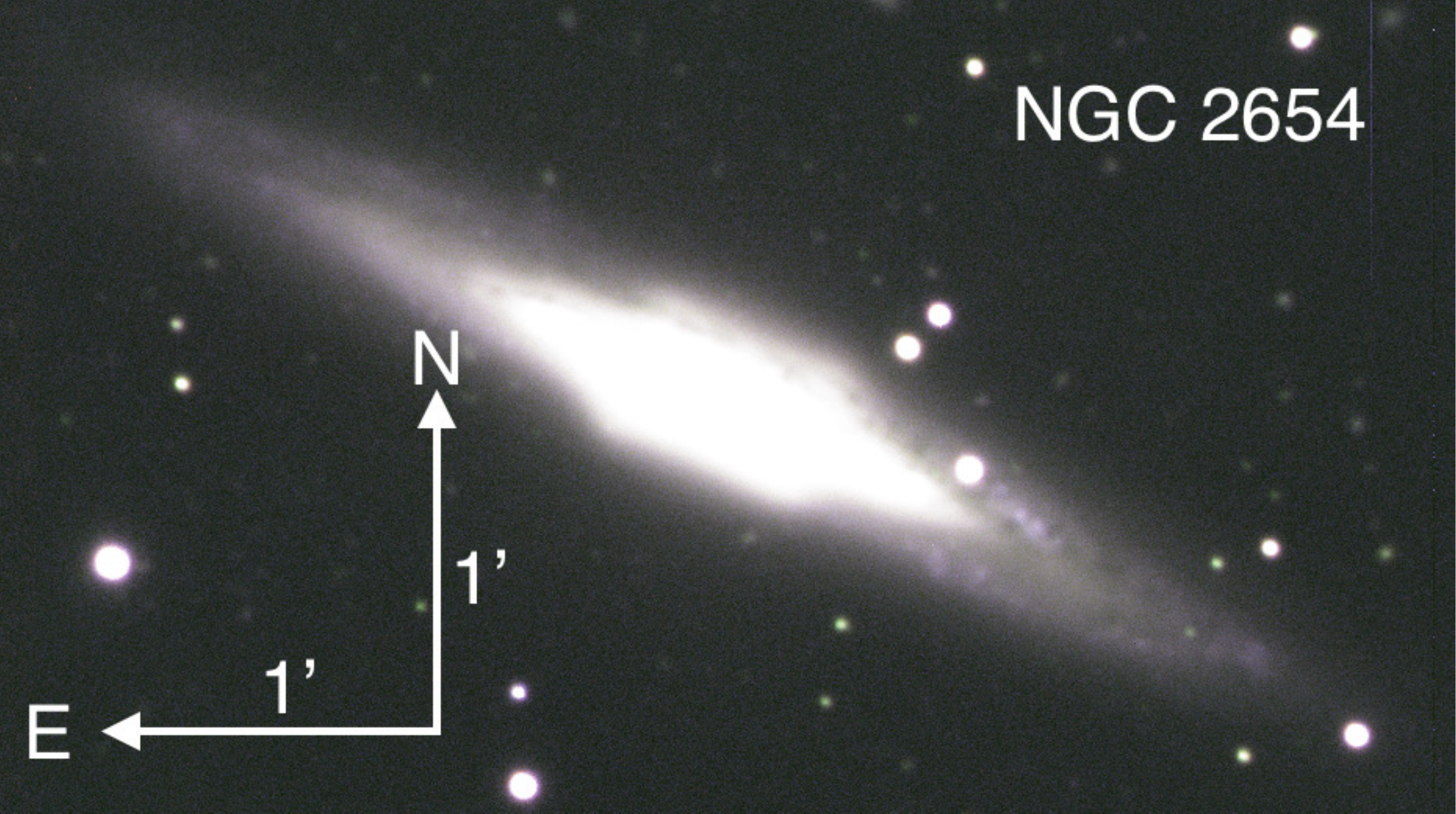} \hskip 5mm \includegraphics[width=0.45\textwidth]{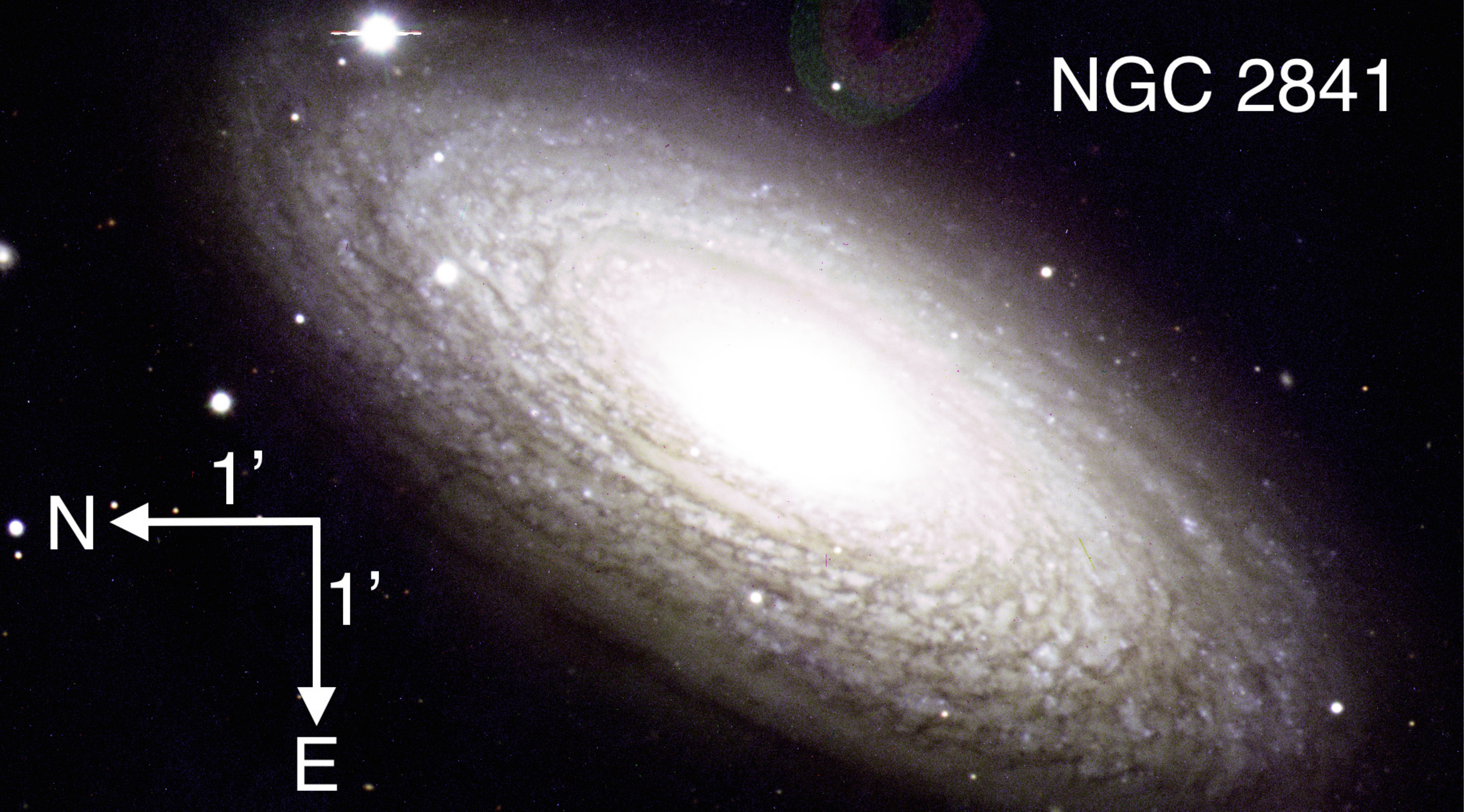} \\
	\vskip 5mm
	\includegraphics[width=0.45\textwidth]{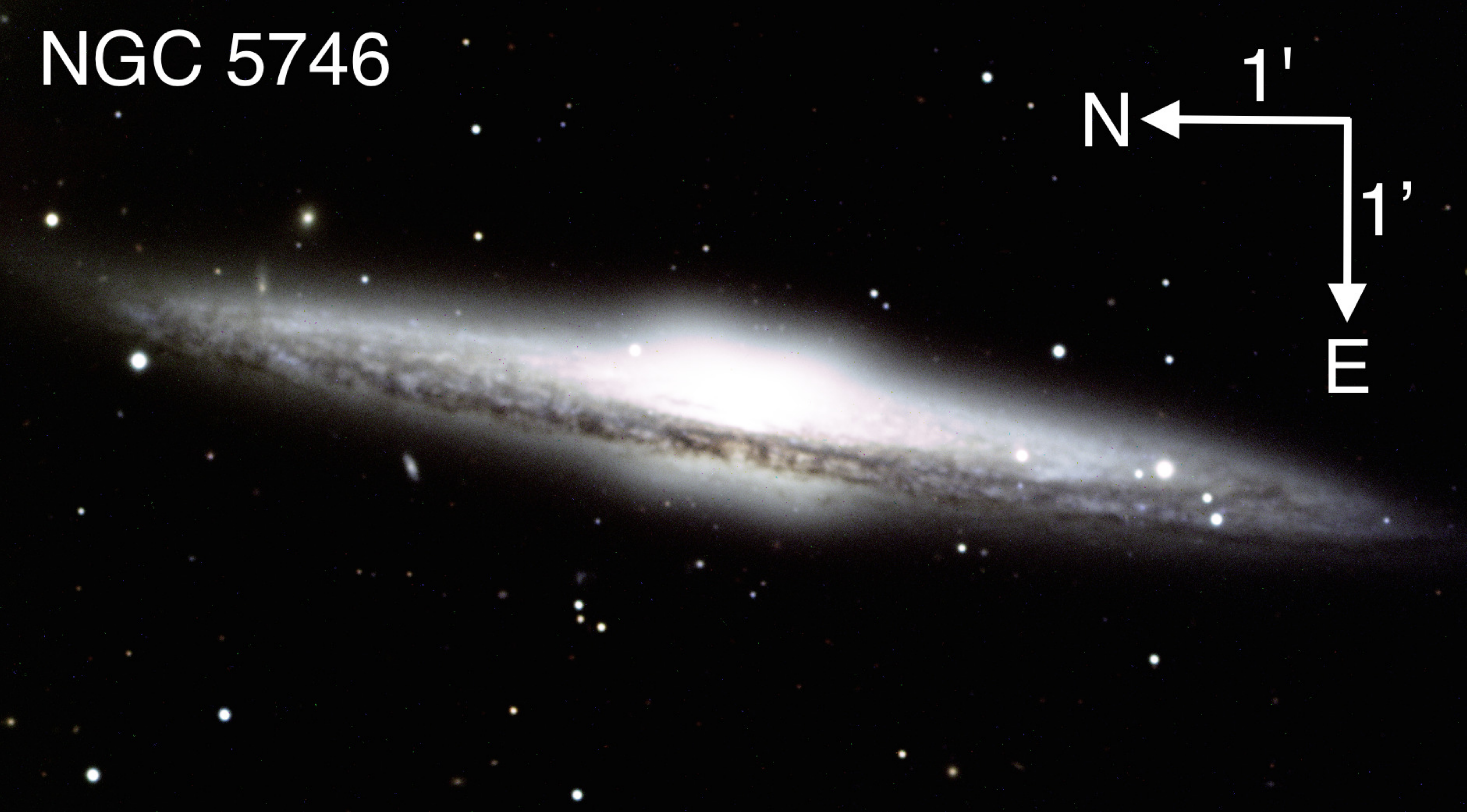} \hskip 5mm \includegraphics[width=0.45\textwidth]{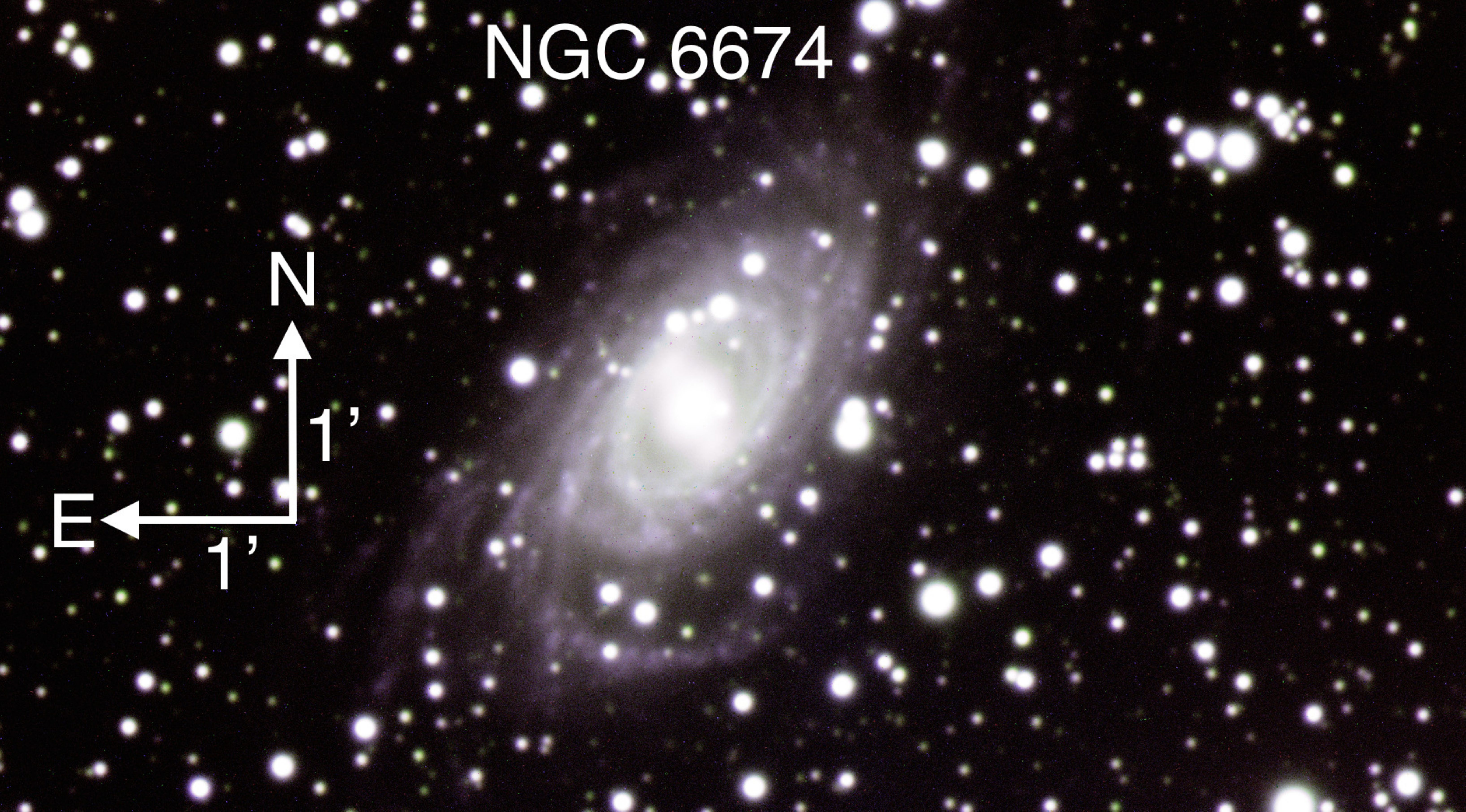}
	\caption{Combined \textit{BVR} images of our sample. Arrows indicate North and East directions. Each arrow is roughly 1$\arcmin$ in length. \textit{Top Left}: NGC~2654; \textit{Top Right}: NGC~2841; \textit{Bottom Left}: NGC~5746; \textit{Bottom Right}: NGC~6674. }
	\label{gal_pics}
\end{figure*}

The data were reduced in \texttt{IRAF}\footnote{IRAF is distributed by the National Optical Astronomy Observatory, which is operated by the Association of Universities for Research in Astronomy (AURA) under a cooperative agreement with the National Science Foundation.} using standard packages and routines. Images were bias and dark subtracted and flat fielded. We removed a constant sky background value from each galaxy image by determining the mean sky value from small star- and galaxy-free patches across each frame. Our reduced images were  photometrically calibrated using the Landolt standard stars observed during the night. This calibration was then checked against stars in the field and found to be accurate for NGC 2654 and NGC 2841 to within $\pm$0.2 mags using the PPMXL Catalog \citep{roeser2010}. Combined \textit{BVR} images of our galaxies are shown in Fig.~\ref{gal_pics}. 

\subsection{Spectroscopy}
\label{sec:spectroscopy}

We have also obtained H$\alpha$ velocity fields for our full sample, as well as long-slit rotation curves for all galaxies except NGC~2841. 

\subsubsection{H$\alpha$\ Velocity Fields}
\label{sec:havfs}

We have obtained H$\alpha$\ velocity fields of the four galaxies using the SparsePak integrated field unit (IFU) \citep{bershady2004} on the 3.5-m WIYN\footnote{The WIYN Observatory is a joint facility of the University of Wisconsin-Madison, Indiana University, the National Optical Astronomy Observatory and the University of Missouri.} telescope at Kitt Peak National Observatory (KPNO).   SparsePak is composed of 82 fibres, each 5$\arcsec$ in diameter, arranged in a fixed, main array of 70$\arcsec$~$\times$~70$\arcsec$. There is a central diamond of closely packed fibres and there are seven sky fibres placed roughly 20$\arcsec$ away from the main array.    

Observations for NGC~2654 and NGC~2841 were obtained on 2009 February 19 and observations for NGC~5746 and NGC~6674 were obtained on 2009 May 15. We used SparsePak with the 316@63.4 grating in eighth order. The central wavelengths were 6726\AA\ on 2009 February 19 and 6700\AA\ on 2009 May 15. The spectral resolution was 0.9\AA\ pix$^{-1}$ for both nights. 

\begin{figure*}
	\center
   	\includegraphics[width=0.45\textwidth]{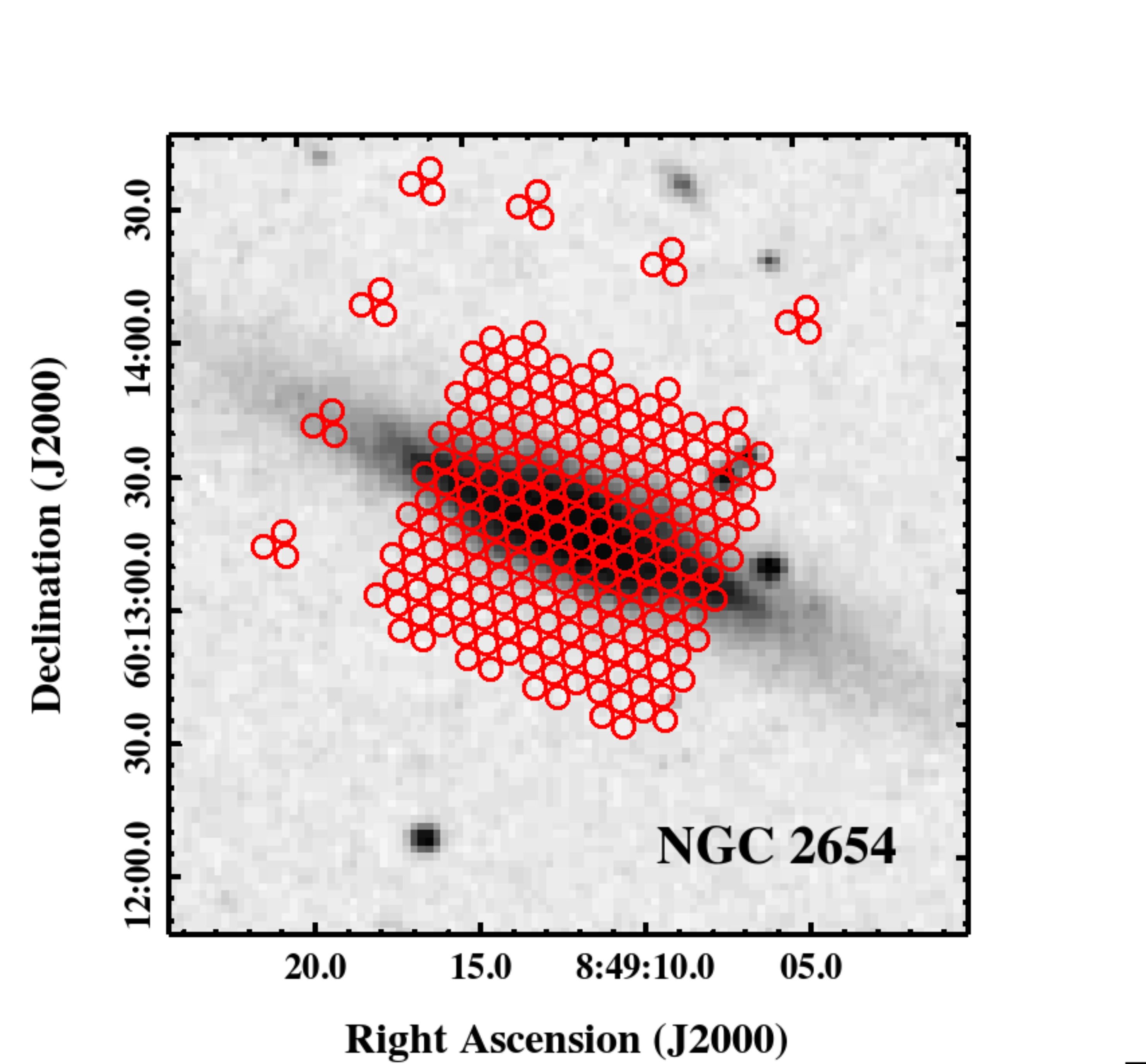} \hskip 2mm \includegraphics[width=0.45\textwidth]{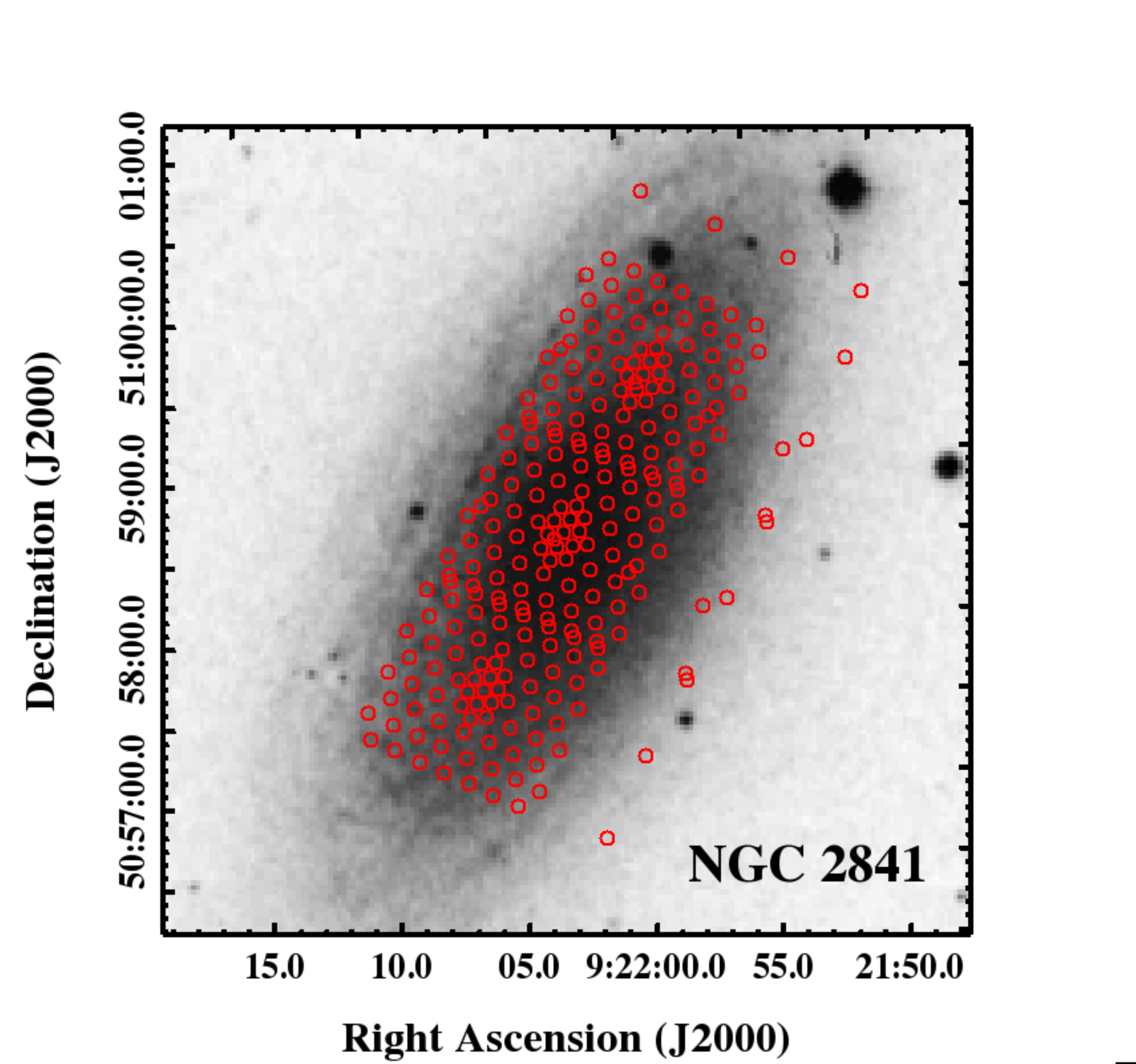}\\
	\vskip 2mm
    	\includegraphics[width=0.45\textwidth]{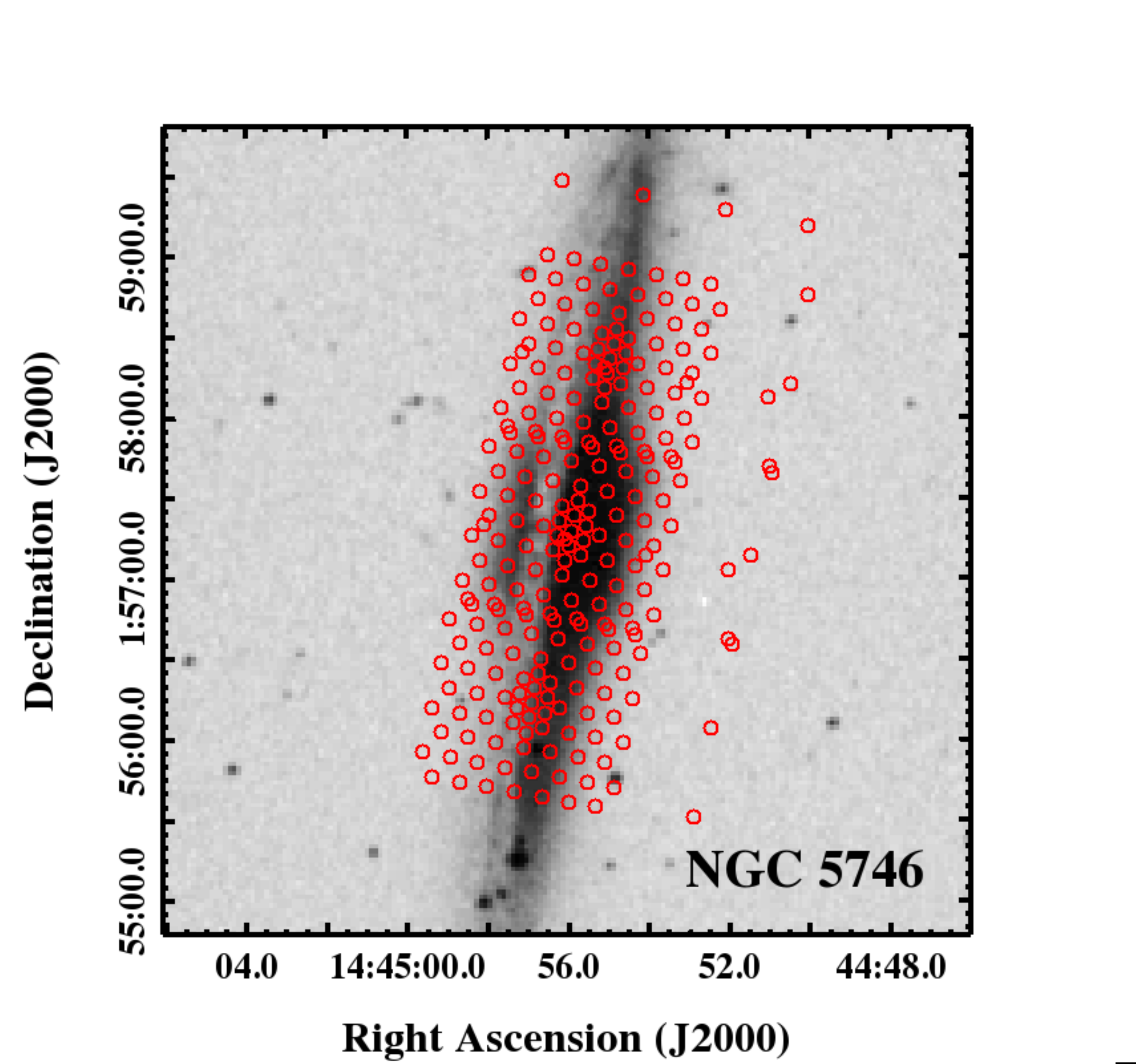} \hskip 2mm \includegraphics[width=0.45\textwidth]{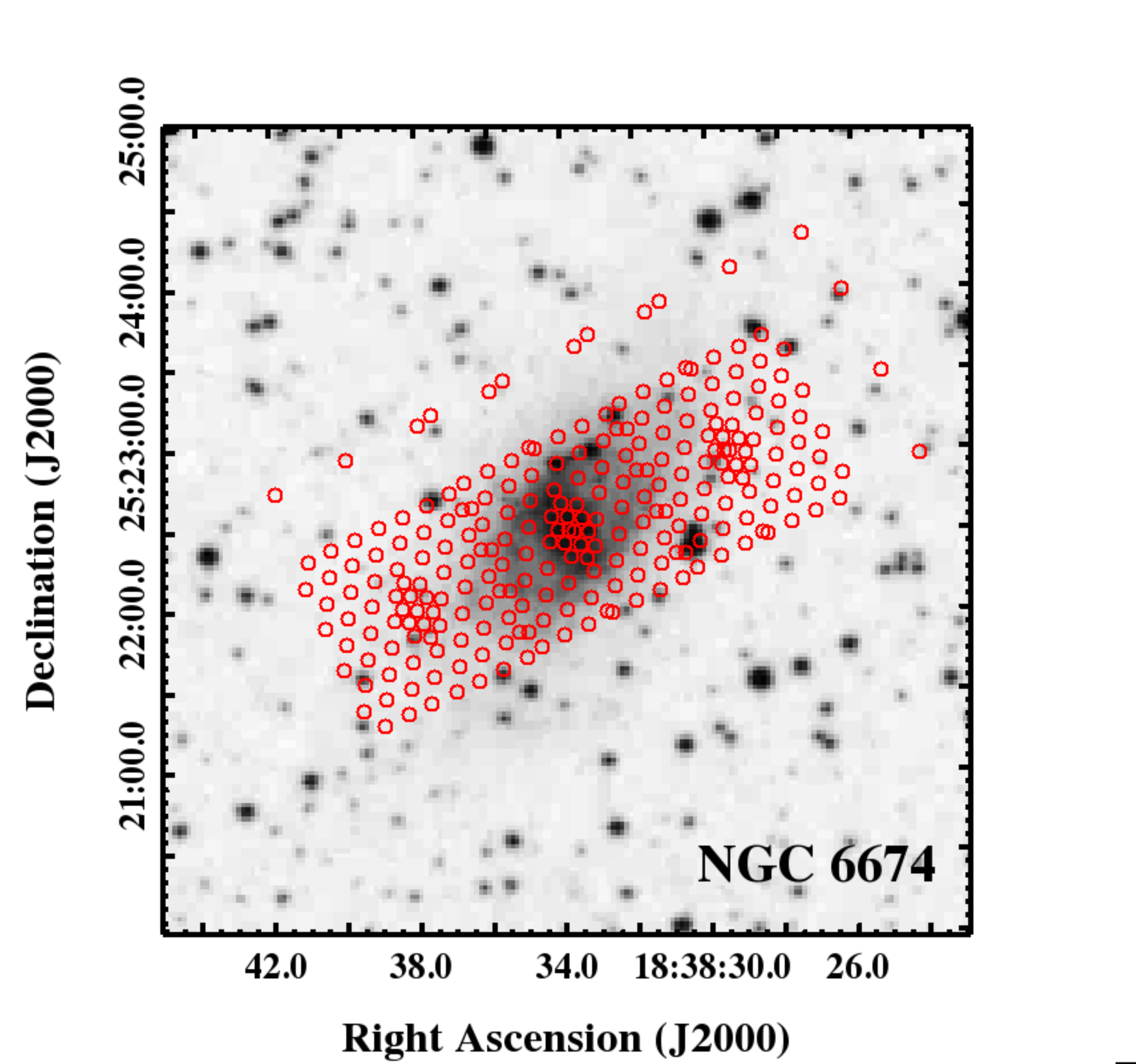}
  	\caption{Position of SparsePak array over our four galaxies. Each fibre is 5$\arcsec$ in diameter. Images are taken from SAO Digital Sky Survey. \textit{Top Left}: NGC~2654.  There were three pointings of the array, dithered such that gaps between fibres would be filled. Image is 3$\arcmin$~$\times$~3$\arcmin$. \textit{Top Right}: NGC~2841. There were three pointings of the array, offset to cover the length of the galaxy. Image is 5$\arcmin$~$\times$~5$\arcmin$. \textit{Bottom Left}: NGC~5746. The three pointings were offset from each other to cover the length of the galaxy. Image is 5$\arcmin$~$\times$~5$\arcmin$. \textit{Bottom Right}: NGC~6674. The three pointings were offset from each other to cover the length of the galaxy and the array is rotated to align the central diamond of fibres with the bar. Image is 5$\arcmin$~$\times$~5$\arcmin$.}
	\label{sparse_pak}
\end{figure*}

Each galaxy was observed with 3 pointings of the SparsePak array (see Fig.~\ref{sparse_pak}).  With the exception of NGC~6674, the array was rotated such that the central fibre grouping was aligned with the major axis of each galaxy; in the case of NGC~6674, the array was aligned with its bar. SparsePak was shifted between pointings such that each group of pointings was positioned to cover as much of each galaxy as possible, or to improve coverage of the centre of the galaxy. For NGC~2654, for example, the array was dithered so that the spaces between fibres was minimized and the centre of the galaxy received better coverage. For NGC~2841, NGC~5746 and NGC~6674 the array was offset such that the three pointings of the main array were end-to-end, as these galaxies span more than 2$\arcmin$\ on the sky. Two 1200s exposures were taken at each pointing to improve the signal to noise and a ThAr lamp was observed before and after each galaxy exposure to provide wavelength calibration.

The data were bias-subtracted and flat-fielded in \texttt{IRAF}. The spectrum in each fibre was extracted using the \texttt{DOHYDRA} package and calibrated using a wavelength solution created from the ThAr lamp exposures. Because the SparsePak sky fibres often overlap with the galaxy and do not provide a clean measurement of the sky (see, for example, NGC~2841 in Fig.~\ref{sparse_pak})  standard sky subtraction was not performed. As described below, we instead use these sky lines as an additional wavelength calibration step.

Velocities were measured in each fibre by fitting Gaussians to four optical emission lines: H$\alpha$, [NII]$\lambda$6583, [SII]$\lambda$6717 and [SII]$\lambda$6731. In addition, two reference night sky lines \citep{osterbrock1996}, one close to the H$\alpha$ and [NII] lines and the other close to the [SII] lines, were also measured. These sky lines served as a final wavelength calibration step and reduced the scatter in measured galaxy emission line velocities. The final velocity assigned to each fibre was determined by averaging  the individual emission line velocities. The largest difference between an emission line velocity and the average was assigned as the error on the velocity for that fibre. Typical errors were $\sim$6 km s$^{-1}$; very few fibres had errors greater than 10 km s$^{-1}$. For fibres where only H$\alpha$\ was observed, the velocity assigned to the fibre was the velocity of the H$\alpha$ line and the error was set to 10 km s$^{-1}$.

\begin{figure*}
	\center
	\includegraphics[width=0.45\textwidth]{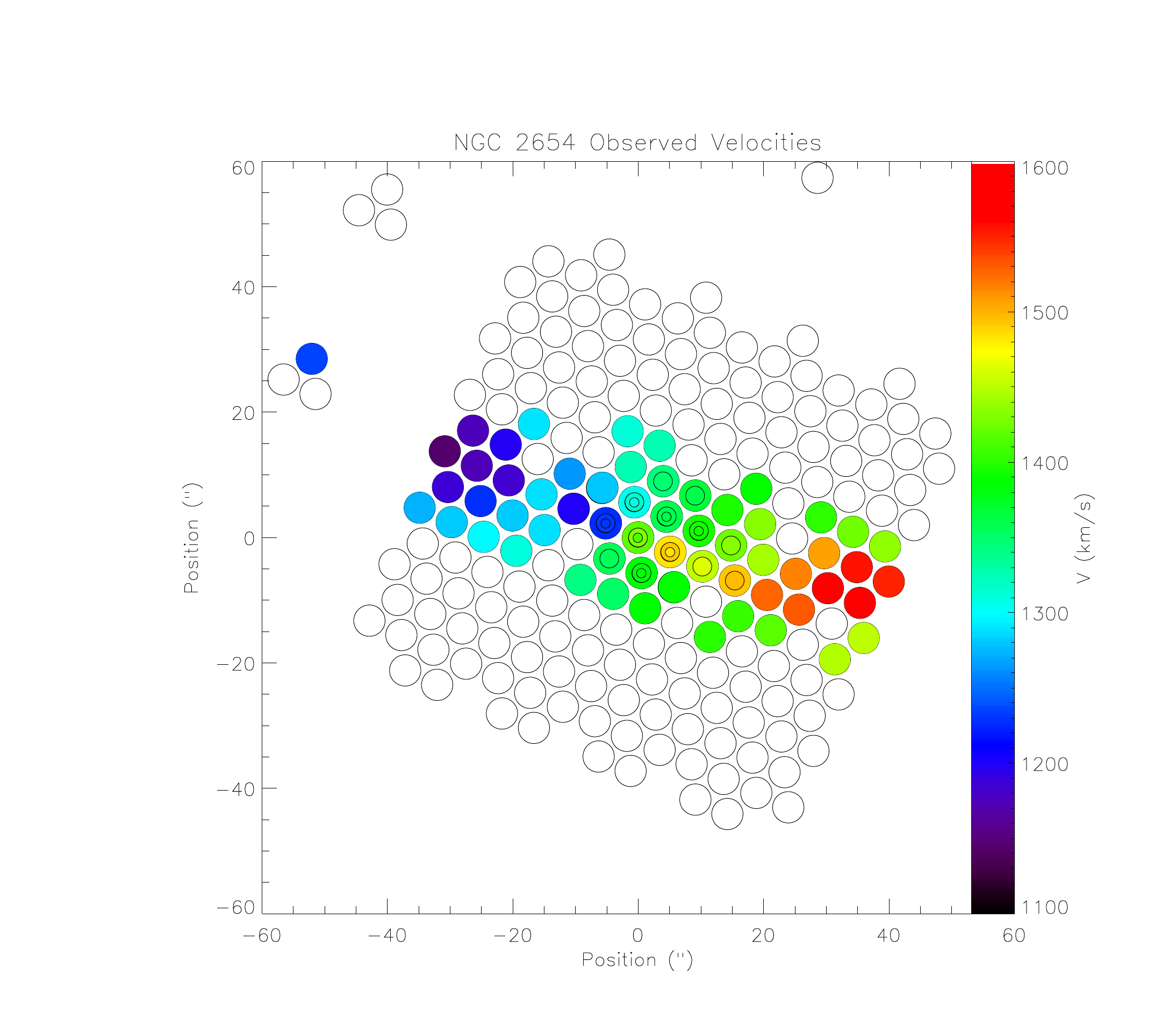} \hskip 2mm \includegraphics[width=0.45\textwidth]{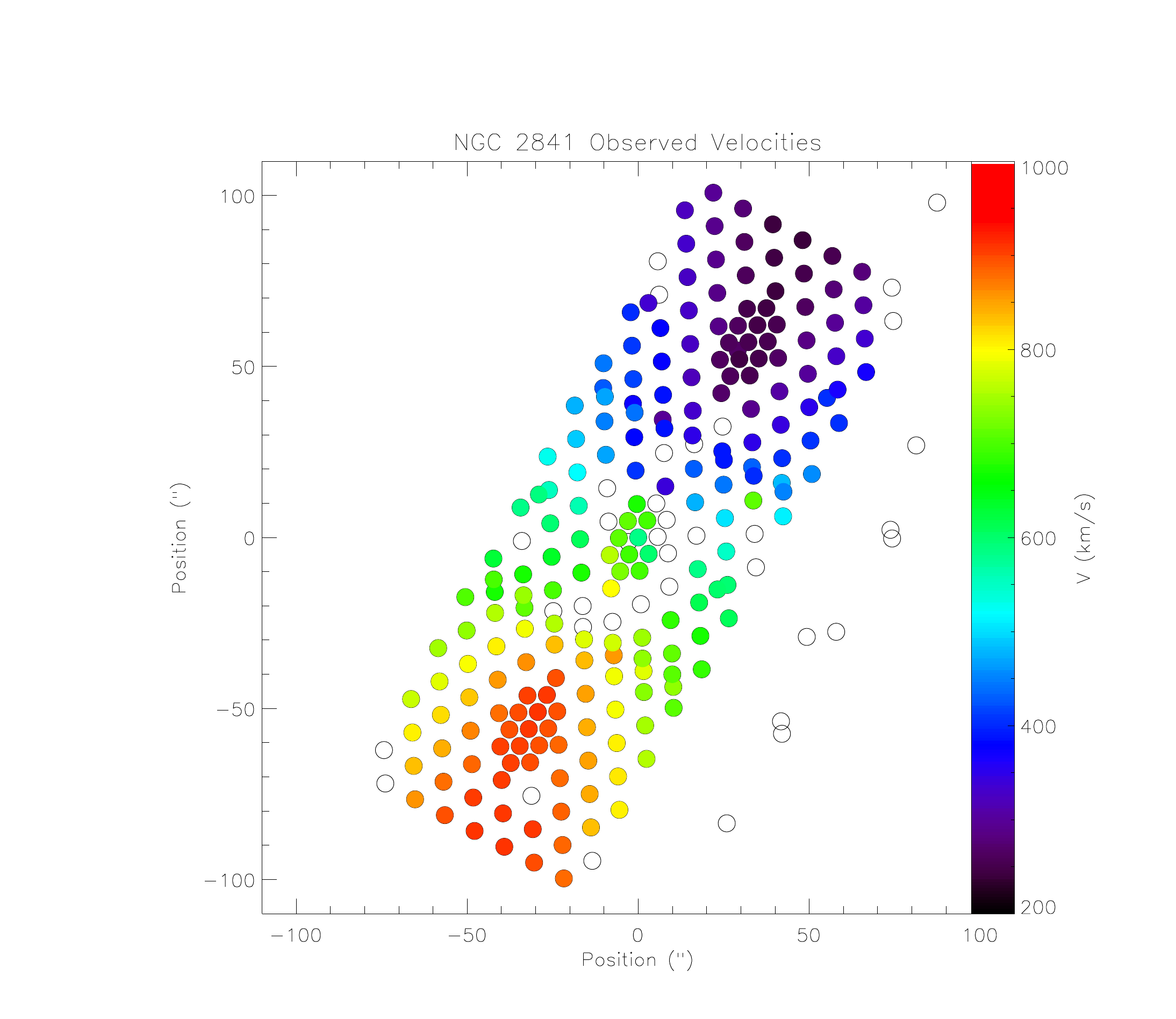}\\
	\vskip 2mm
	\includegraphics[width=0.45\textwidth]{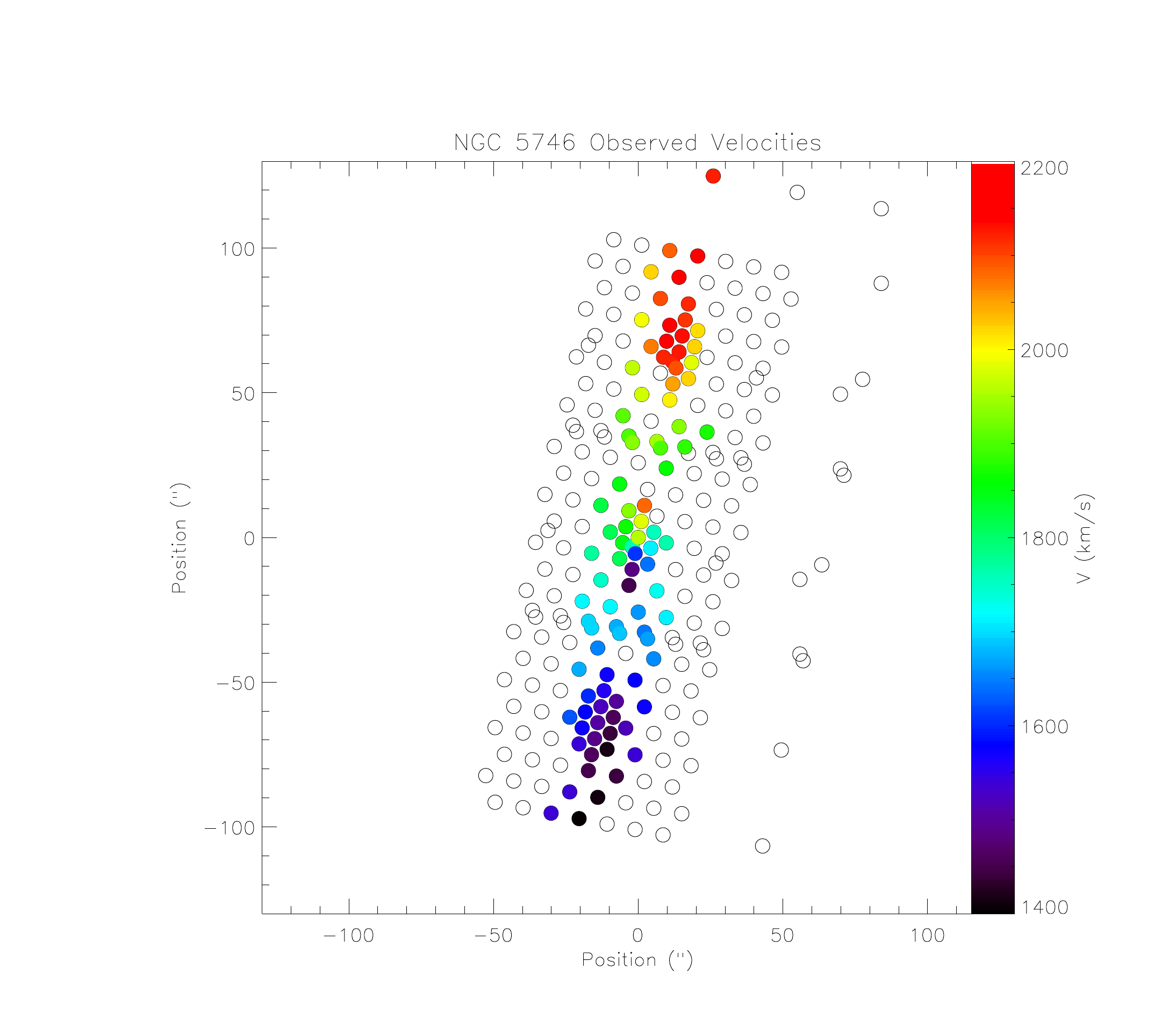} \hskip 2mm \includegraphics[width=0.45\textwidth]{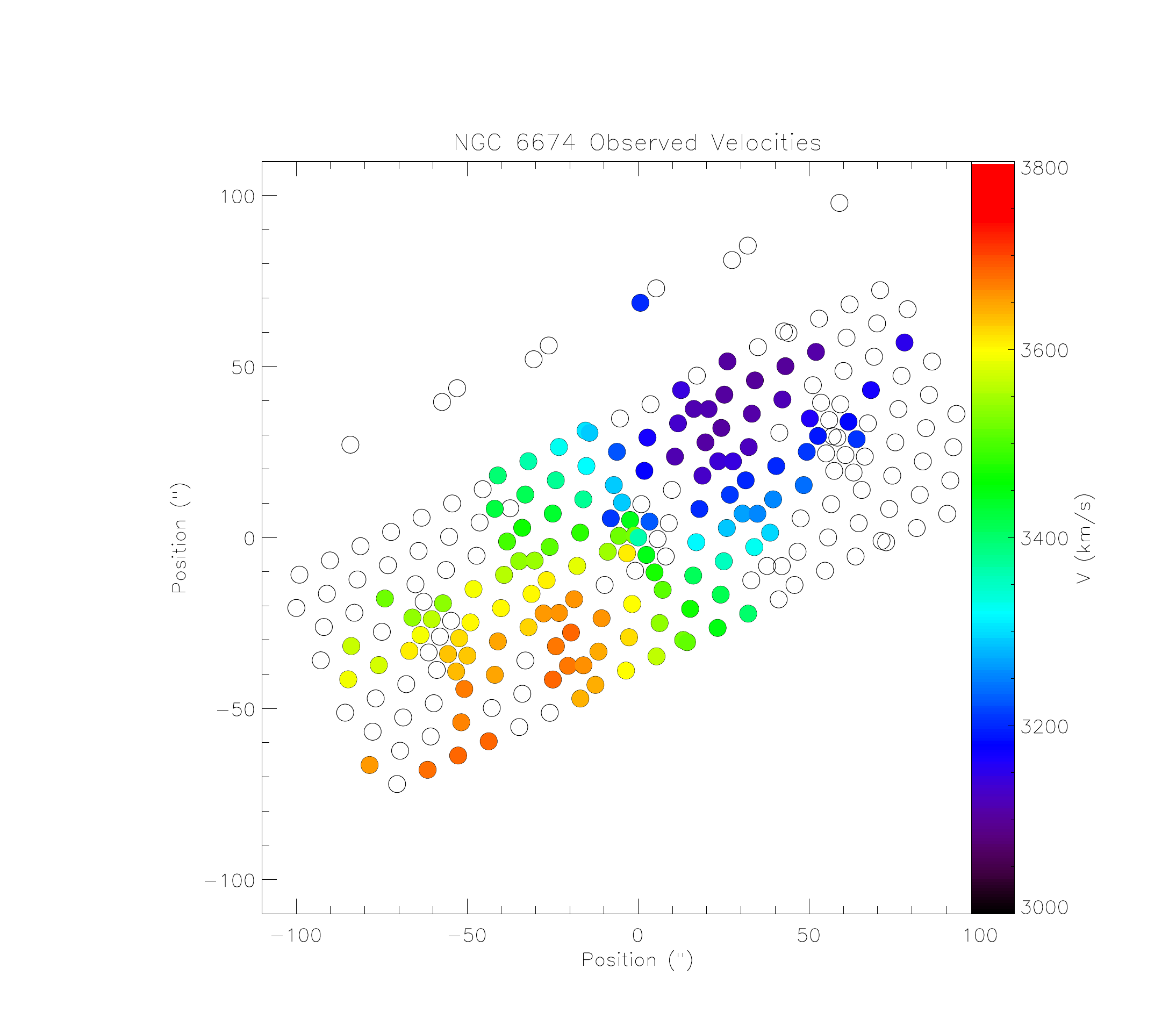}
	\caption{Observed average SparsePak velocity fields. Empty fibres indicate a lack of emission. \textit{Top Left}: NGC~2654. fibres with concentric circles were a result of the dithering pattern causing fibres to overlap; \textit{Top Right}: NGC~2841; \textit{Bottom Left}: NGC~5746; \textit{Bottom Right}: NGC~6674. \textbf{N.B.}: The velocity fields have been zoomed in with respect to Fig.~\ref{sparse_pak}.}
	\label{velocity_fields}
\end{figure*}

Images of our galaxies with the SparsePak fibres overlaid are shown in Fig.~\ref{sparse_pak}, and the  observed average velocity fields are shown in Fig.~\ref{velocity_fields}.

\subsubsection{Long-Slit Spectroscopy}
\label{sec:lsspec}

We have obtained long-slit spectroscopy for three of our galaxies using the Dual Imaging Spectrograph (DIS) on the 3.5-m telescope at APO. Observations for NGC~2654 were obtained on 2014 January 24. We used a slit width of 1.5$\arcsec$, the B400/R300 gratings and central wavelengths of 4800\AA\ for the blue arm and 7500\AA\ for the red arm. Our spectral resolution in the blue was 1.83\AA\ pix$^{-1}$\ and 2.31\AA\ pix$^{-1}$\ in the red. Observations for NGC~5746 and NGC~6674 were obtained on 2016 May 12. We used a slit width of 1.5$\arcsec$, the B1200/R1200 gratings and central wavelengths of 5050\AA\ and 6200\AA\ for the blue and red arms, respectively. Spectral resolutions were 0.62\AA\ pix$^{-1}$\ in the blue and 0.58\AA\ pix$^{-1}$\ in the red. 

The slit was aligned with the major axis of each galaxy and two 1200s exposures were taken for each target. For wavelength calibration, HeNeAr lamps were observed before and after each galaxy exposure at the same slit position angle. 

The data were bias-subtracted, flat-fielded, and wavelength calibrated in \texttt{IRAF}. As with the SparsePak data, we did not remove the sky lines from the spectra and instead used them as a secondary wavelength calibration to reduce the scatter in our measured velocities.

Velocities were measured in a method similar to that described for the H$\alpha$ velocity fields. We extracted a spectrum at each pixel along the slit and fit Gaussians to four optical emission lines in the red arm (H$\alpha$, [NII]$\lambda$6583, [SII]$\lambda$6717 and [SII]$\lambda$6731), as well as reference sky lines. We were unable to measure [OIII]$\lambda$5007 or H$\beta$\ in the blue arm as the signal-to-noise of these lines was too low and the emission was not extended enough along the slit. The final velocity assigned to each pixel was the average of the individual galaxy emission line velocities;  the assigned error was the largest difference between an emission line velocity and the average velocity. Typical errors were $\sim$10 km s$^{-1}$.  If only H$\alpha$ was measured in a pixel, the velocity was set to the H$\alpha$ velocity and an error of 10 km s$^{-1}$ was assumed. 

\section{Modeling}
\label{sec:modeling}

We use \texttt{DiskFit} to model our \textit{BVRI}\ images and IFU velocity fields in order to identify bulges and/or bars that may be present. \texttt{DiskFit} allows for non-axisymmetric modeling of both photometry and kinematic data, albeit not at the same time. \texttt{DiskFit} has been described extensively in the literature and applied to a wide range of galaxy data, as seen in \citet{reese2007}, \citet{spekkens2007}, \citet{sellwood2010}, \citet{kuzio2012}, and \citet{holmes2015}. In Sections~\ref{sec:dfphot} and \ref{sec:dfkine} below, we briefly cover the fitting mechanisms behind \texttt{DiskFit} and discuss how we select our starting values of the input parameters for each model. In Section~\ref{sec:lsrcs}, we discuss how we derive rotation curves from our long-slit DIS spectroscopy.

\subsection{\texttt{DiskFit} Photometry}
\label{sec:dfphot}

The details of the photometric modeling done through \texttt{DiskFit} can be found in \citet{reese2007}; we highlight here the relevant aspects. 

\texttt{DiskFit} can fit up to three components simultaneously in photometry: a disc, bar, and bulge.  The only component that has an assumed light profile is the bulge, which is characterized by the S\'{e}rsic function:
\begin{equation}
I(r) = I_{0}\exp{ \left\{ -B_{n} \left[ \left( \frac{r}{r_{e}} \right) ^{1/n} - 1 \right] \right\} }
\end{equation}
where $r_{e}$\ is the effective radius, \textit{n} is the S\'{e}rsic index and $I_{0}$\ is the central intensity. 

The galaxy is divided into rings with radii set by the user and then \texttt{DiskFit} determines the best fit values for the position angle of the disc (P.A.$_{\mathrm{disc}}$), inclination of the disc, P.A.$_{\mathrm{bar}}$, bar ellipticity ($\epsilon_{\mathrm{bar}}$), bulge ellipticity ($\epsilon_{\mathrm{bulge}}$), bulge effective radius ($r_{e}$), I$_{0}$ and \textit{n}. In contrast to ellipse-fitting methods, \texttt{DiskFit} returns a single value for each parameter rather than as a function of radius.

\texttt{DiskFit} does return \textit{the amount of light} in each component as a function of radius, as well as the percentage of the total light coming from each component.   We turn these intensities into surface brightnesses via:
\begin{equation}
	\mu = -2.5 \log{ \left( \frac{I}{t\ p^{2}} \right) } + Z
\end{equation}
where \textit{I} is the sky subtracted value output by \texttt{DiskFit}, \textit{t}\ is the exposure time in seconds, \textit{p} is the plate scale in arcsec pix$^{-1}$, and \textit{Z} is the zero point determined from our photometric calibration described in Section~\ref{sec:photometry}.

Finally, \texttt{DiskFit} uses a bootstrap method to determine errors on each of the parameters, the intensities as a function of radius, and the total amount of light in each component.

We consider four different models for our galaxies: disc-only, disc+bulge, disc+bar, and disc+bulge+bar. Our ring spacing was set to be greater than twice the largest seeing for each galaxy (see Table~\ref{data_info}). To begin our fits, we estimate values of the disc inclination and position angle based on a visual inspection of the galaxy images. We determine the disc centre by running a disc-only model and then hold the centre fixed to this position in the other three models. We let all other parameters vary freely.

Only for NGC~6674 was there an obvious bar in the photometry of the galaxy from which to estimate the bar position angle. For NGC~2654, NGC~2841, and NGC~5746 we start the bar position angle at the \texttt{DiskFit} manual recommended value of 45\degr. \authorfix{The bar length was started at a value estimated from the images but typically needed to be increased so that \texttt{DiskFit} would return a model with a smoothly declining light profile for the bar. The bar lengths that we report in Section 5 are not precise measurements with calculated errors, but rather an indication of where the bar contribution to the total light profile declines to a negligible amount.}

We use our galaxy images to estimate the $r_{e}$\ of the bulge and use an initial value of 0.2 for the bulge ellipticity. If these parameters began to runaway to unphysically large (or small) values, we stopped the model and re-ran the code with slightly different initial values.

Finally, we determine errors on the parameters and light profiles by generating 1000 bootstrap realizations of each model. \authorfix{Large amounts of dust, knots of star formation, and other non-axisymmetric features in the galaxies typically drive the $\chi^{2}$\ of the model-fits to very large values, so rather than relying on $\chi^{2}$ to evaluate how well the models match the data, we select as our best-fitting model the one that produces the most consistent results across all four photometric bands.}

\subsection{\texttt{DiskFit} Kinematics}
\label{sec:dfkine}

The kinematic side of \texttt{DiskFit} is detailed in \citet{spekkens2007} and \citet{sellwood2010} and is only briefly summarized here. 

In contrast to a tilted-ring model, \texttt{DiskFit} fits an entire velocity field with a physically motivated model. The underlying assumption is that the circular orbit of a star or region of gas is affected by higher order perturbations (i.e.\ ``harmonics'' ). Physically, \textit{m}=1 harmonics correspond to ``lopsided'' perturbations and \textit{m}=2 harmonics correspond to bisymmetric (i.e.\ bar) perturbations. The \texttt{DiskFit} code concerns itself with only these first two harmonics as the higher order terms are overshadowed by the \textit{m}=2 harmonic \citep{spekkens2007}. The model is given by:
\begin{multline}
V_{\mathrm{model}} = V_{\mathrm{sys}} + \sin{i} ( \bar{V_{\mathrm{t}}}\cos{(\theta)} - V_{m,\mathrm{t}}\cos{(2\theta_{b})}\cos{(\theta)} \\ 
- V_{m,\mathrm{r}}\sin{(2\theta)}\sin{(\theta)} )
\end{multline}
where V$_{\mathrm{sys}}$\ is the systemic velocity, $\bar{V_{\mathrm{t}}}$ is the mean orbital speed, $\theta$\ is the angle between a point in the disc relative to the major axis, $\theta_{b}$\ is the angle between a point in the disc relative to the bar, V$_{m,\mathrm{t}}$ and V$_{m, \mathrm{r}}$ are the tangential and radial components of the non-circular motions, and \textit{m}\ is the harmonic order. 

\texttt{DiskFit} can also fit for radial flows and symmetric warps in the outer disc. Finally, \texttt{DiskFit} also allows for turbulence in the disc, $\Delta_{\rm ISM}$, to be taken into account and does so by adding it in quadrature to the errors on the observed emission line velocities.

For this work, we focus on the axisymmetric, disc-only (\textit{m}=0), and bisymmetric, disc+bar (\textit{m}=2) models and use \texttt{DiskFit} to determine the disc inclination \textit{i}, P.A.$_{\mathrm{disc}}$, V$_{\mathrm{sys}}$, disc centre, and P.A.$_{\mathrm{bar}}$. \texttt{DiskFit} also returns the rotation velocity as a function of radius as well as the tangential and radial components of the non-circular motions in the disc+bar model. \authorfix{For completeness, we ran the models with the warp and radial flow features of \texttt{DiskFit} enabled, but because we did not find any significant improvement in the fits when doing so, we focused our efforts on flat discs having only circular motions and bar-like flows.}

For input, \texttt{DiskFit} requires a two-dimensional velocity field with errors. The velocity field is broken into rings at radii specified by the user. Ideally, the ring radii should be the effective spatial resolution of the data. For SparsePak this would be the diameter of the fibres, 5$\arcsec$. However, due to the limited number of data points in our velocity fields (on the order of 100-150) we typically had to increase the ring spacing in order to get roughly the same number of data points per ring and to ensure that each ring contained a minimum of 9 measured velocities.

We use the results of our photometric modeling as the starting points for our disc inclination, disc position angle, bar position angle, and bar length. The starting value of the systemic velocity are estimated directly from the velocity field data. Once the disc centre is found, we re-run the code with the centre held fixed for all subsequent modeling, while letting the rest of the parameters vary freely. We set $\Delta_{\rm ISM}$ to 10 km s$^{-1}$ for each galaxy and check that the results of the modeling are unaffected by the specific choice of this value. \authorfix{Similar to the photometric modeling, the bar length was identified by determining the radius at which the non-circular motions caused by the bar became negligible.}

We generate 1000 bootstrap realizations of each velocity field model in order to determine uncertainties on the parameter values and derived rotation curve points. Our DIS long-slit rotation curves are used as a reference for the model velocity field rotation curves, and we select as the best model the one that is most consistent with our DIS and literature rotation curves at intermediate and large radii. 

\subsection{Long-slit Rotation Curves}
\label{sec:lsrcs}

In this section we briefly cover how we obtain long-slit rotation curves for our galaxies. In order to derive rotation curves from our long-slit observations, we assume purely circular motions throughout the disc. This essentially simplifies Equation 3 into:
\begin{equation}
V_{\mathrm{rot}} = V_{\mathrm{sys}} + V_{\mathrm{obs}}\sin{i}
\end{equation}
where V$_{\mathrm{rot}}$ is the circular rotational velocity, V$_{\mathrm{sys}}$ is the systemic velocity, V$_{\mathrm{obs}}$ is the observed velocity, and \textit{i} is the inclination of the galaxy.

We first convert our measured line-of-sight velocities at each pixel from Section~\ref{sec:lsspec} to rotational velocities by assuming an inclination based on the inclinations determined from the DiskFit photometric and kinematic modeling.

We determine the systemic velocity of each galaxy by locating the galaxy centre in the long-slit data. This centre matches the location of the stellar continuum in the spectra. This is done by finding the midpoint between the V$_{\mathrm{flat}}$ sides of the rotation curve of the galaxy. As non-lopsided spiral galaxies are rotating at a relatively constant velocity outside their inner regions, the rotational velocities are relatively symmetric about a central pixel. The velocity associated with this pixel is assigned as the systemic velocity of the galaxy.

The data are flipped about this systemic velocity so that the red and blue-shifted velocities are now rotational velocities as a function of radius from the galaxy centre. We further iterate the systemic velocity until the flat portion of both sides of the rotation curve match. \authorfix{So as to not lose any information about possible asymmetries in the galaxies, we do not average the two sides together.} We also check the systemic velocities derived are consistent with those found from the H$\alpha$ velocity fields in Section~\ref{sec:results}.

We describe and present the final DIS rotation curves in more detail in Section~\ref{sec:results}.

\section{Results}
\label{sec:results}

Here we present the best-fitting parameters for each galaxy. The format is the same for each: first the photometry is discussed, then the kinematics, and then a comparison between the two. Each galaxy has a master table (Tables~\ref{6674_table} - \ref{5746_table}) showing the best-fitting photometric and kinematic parameters. Each galaxy also has a series of figures presenting the photometric and kinematic results. The galaxies are presented in an order to best discuss the increasing complexity of modeling as inclination increases: NGC~6674, NGC~2841, NGC~2654, and finally NGC~5746.

\subsection{NGC~6674}
\label{sec:n6674}

The best-fitting parameters for the photometry and kinematics for NGC~6674 are shown in Table~\ref{6674_table}. The best-fitting \textit{B}-band \texttt{DiskFit} photometric model of NGC~6674 is shown in Fig.~\ref{n6674_b}. The \textit{B}-band residuals for all four models are shown in Fig.~\ref{n6674_res}. The \textit{I}-band surface brightness profile is shown in Fig.~\ref{n6674_colour}. The best-fitting kinematic \texttt{DiskFit} model and both SparsePak and DIS rotation curves are shown in Fig.~\ref{n6674_kinematics}.

\begin{table*}
	\centering
	\caption{Best-fitting photometric and kinematic parameters for NGC~6674. Position angles (P.A.) and inclinations are in degrees. The bulge effective radius, $r_{e}$, and bar radius, $R_{\mathrm{bar}}$, are in arcsec and velocities are in km s$^{-1}$. \authorfix{The bar lengths are not determined explicitly by \texttt{DiskFit}, but rather are estimated by determining from the surface brightness profiles and rotation curves the radius at which the light or non-circular motions from the bar become negligible.} \textit{n} is the bulge S\'{e}rsic index. The DIS P.A.\ is the angle of the slit; the other DIS parameters are determined from the long-slit data. The bottom three parameters are the percentages of light coming from the three different galaxy components.}
	\label{6674_table}
	\begin{tabular}{rcccccccc}
		\hline
		Parameter		&		&\multicolumn{4}{c}{Photometry}		&		&\multicolumn{2}{c}{Kinematics}\\
							\cline{3-6}									\cline{8-9}
					&		&B		&V		&R		&I		&		& SparsePak		&DIS\\
		\hline
	P.A.$_{\mathrm{disc}}$ (\degr) & & 140.30 $\pm$ 1.95 & 139.90 $\pm$ 1.67 & 139.72 $\pm$ 1.49 & 139.36 $\pm$ 1.75 & & 149.25 $\pm$ 0.91 & 143  \\
	\textit{i}$_{\mathrm{disc}}$ (\degr)& & 53.43 $\pm$ 1.66 & 52.71 $\pm$ 1.47 & 54.42 $\pm$ 1.39 & 53.93 $\pm$ 1.46 & & 61.61 $\pm$ 0.76 & 60\\
	$V_{\mathrm{sys}}$ (km s$^{-1}$) & & ... & ... & ... & ... & & 3393.94 $\pm$ 1.71 & 3392 $\pm$ 15 \\
	\\
	P.A.$_{\mathrm{bar}}$ (\degr) & & 26.41 $\pm$ 3.66 & 27.10 $\pm$ 2.85 & 26.34 $\pm$ 2.85 & 27.08 $\pm$ 3.04 & & 30.36 $\pm$ 4.10 & ... \\
	$R_{\mathrm{bar}}$ ($\arcsec$) & & 28 & 28 & 28 & 28 & & 45 & ... \\
	$\epsilon_{\mathrm{bar}}$ & & 0.34 $\pm$ 0.04 & 0.35 $\pm$ 0.04 & 0.35 $\pm$ 0.04 & 0.41 $\pm$ 0.02 & & ... & ... \\
	\\
	$r_{\mathrm{e}}$ ($\arcsec$) & & 2.61 $\pm$ 1.37 & 2.07 $\pm$ 1.28 & 2.05 $\pm$ 1.13 & 1.96 $\pm$ 0.21 & & ... & ... \\
	$\epsilon_{\mathrm{bulge}}$ & & 0.07 $\pm$ 0.03 & 0.1 $\pm$ 0.12 & 0.04 $\pm$ 0.05 & 0.08 $\pm$ 0.23 & & ... & ... \\
	\textit{n} & & 0.64 $\pm$ 0.17 & 0.64 $\pm$ 0.87 & 0.63 $\pm$ 0.36 & 0.67 $\pm$ 1.37 & & ... & ... \\
	\\
	\% Disc & & 90.16 $\pm$ 1.04 & 81.79 $\pm$ 1.78 & 80.07 $\pm$ 1.48 & 79.80 $\pm$ 1.96 & & ... & ... \\
	\% Bar & & 6.40 $\pm$ 0.80 & 12.36 $\pm$ 1.41 & 12.82 $\pm$ 1.37 & 13.26 $\pm$ 2.28 & &... & ... \\
	\% Bulge & & 3.44 $\pm$ 0.99 & 5.84 $\pm$ 1.48 & 7.11 $\pm$ 0.67 & 6.93 $\pm$ 2.03 & & ... & ... \\
	\hline
	\end{tabular}
\end{table*}

\subsubsection{Photometry}
\label{sec:n6674phot}

NGC~6674 displays a prominent bar and bulge in its photometry, as well as diffuse spiral arms, as seen in the lower right panel of Fig.~\ref{gal_pics}. Not surprisingly, our best-fitting model for this galaxy is the three-component model. 

\begin{figure*}
	\center
	\includegraphics[width=0.9\textwidth]{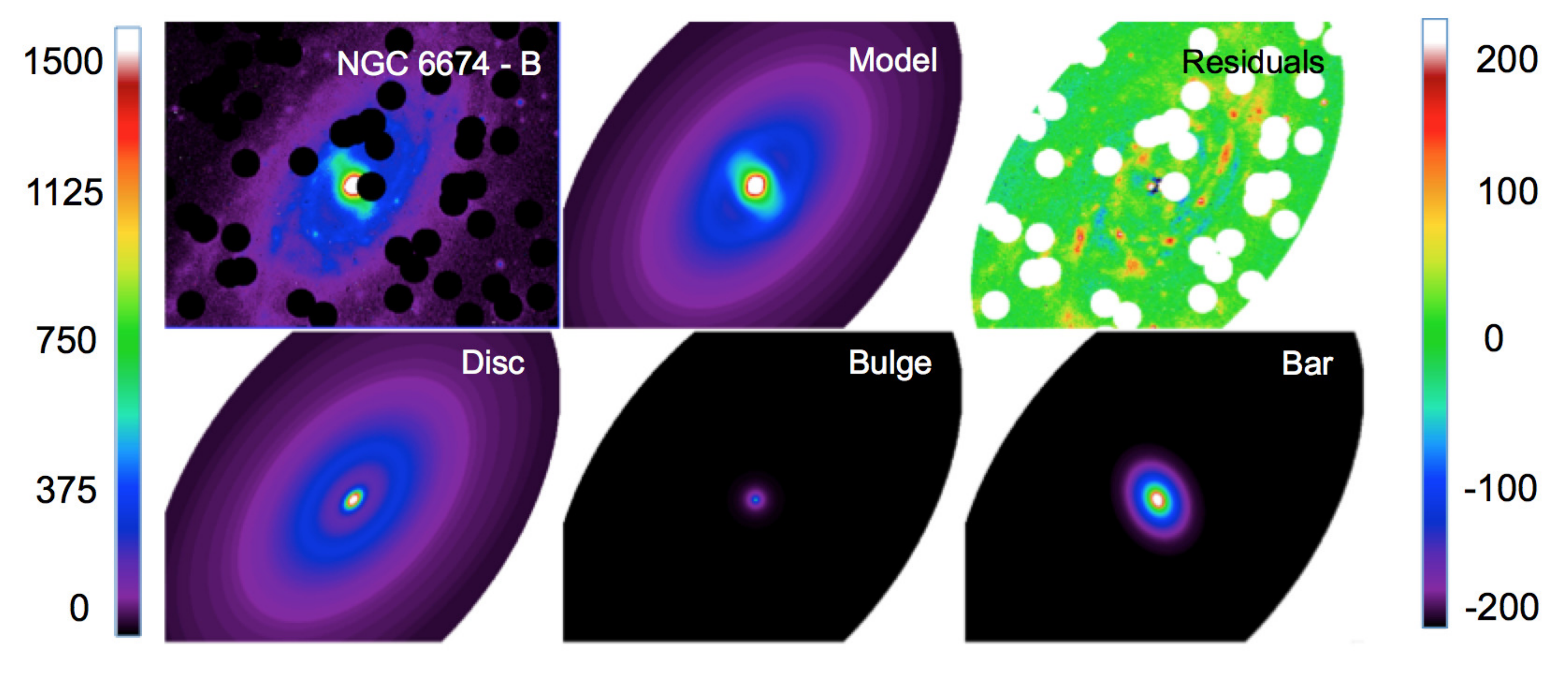}
  	\caption{Best-fitting \textit{B}-band \texttt{DiskFit} model for NGC~6674. Each frame is 3.1$\arcmin$~$\times$~2.36$\arcmin$. The units on the colour bars are ADU; the first two panels in the top row and all bottom panels are scaled with the colour bar on the left while the residual plot in the top right is scaled with the colour bar on the right. \textit{Top Left}: Sky-subtracted, star-masked \textit{B}-band ARCTIC image of NGC~6674; \textit{Top Middle}: 3-Component Disc+Bulge+Bar \texttt{DiskFit} model; \textit{Top Right}: (Data - Model) residuals; \textit{Bottom Left}: Disc component of the model; \textit{Bottom Middle}: Bulge component of the model; \textit{Bottom Right}: Bar component of the model.}
  	\label{n6674_b}
\end{figure*}

Averaged over the four optical bands, we find a P.A.$_{\mathrm{disc}}$ of $\sim$139\degr, an inclination of $\sim$53\degr, a P.A.$_{\mathrm{bar}}$ of $\sim$27\degr, a bar radius of $\sim$28$\arcsec$, an $\epsilon_{\mathrm{bar}}$ of $\sim$0.36, a bulge $r_{e}$ of $\sim$2$\arcsec$, an \textit{n} of 0.65, and an $\epsilon_{\mathrm{bulge}}$ of $\sim$0.07. In the \textit{B}-band we find the disc to contribute $\sim$90\% of the total galaxy light, the bar to contribute $\sim$6\%, and the bulge to contribute $\sim$4\%. Averaged across the other three bands, we find the disc to contribute $\sim$80\%, the bar to contribute $\sim$13\%, and the bulge to contribute $\sim$7\%. The galaxy image, three-component model, (galaxy-model) residuals, and each of the individual model components can be seen in Fig.~\ref{n6674_b} for the \textit{B}-band observations.

\begin{figure}
	\center
	\includegraphics[width=\columnwidth]{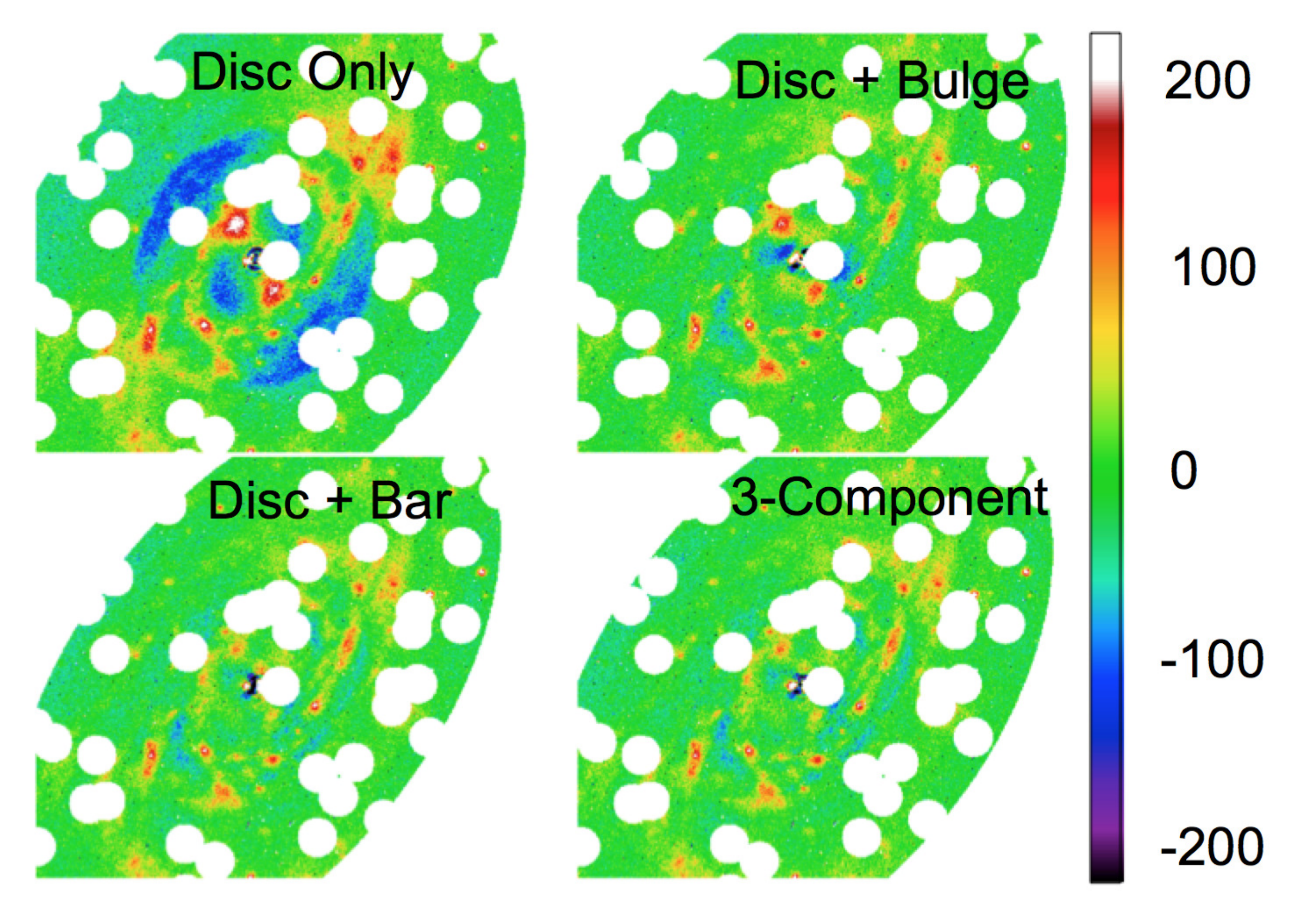}
	\caption{Residuals for the four different \texttt{DiskFit} models for the \textit{B}-band image of NGC~6674. The colour bar is in ADU. Each frame is 3.1$\arcmin$~$\times$~2.36$\arcmin$. \textit{Top Left}: Disc-only model; \textit{Top Right}: Disc+Bulge model; \textit{Bottom Left}: Disc+Bar; \textit{Bottom Right}: 3-Component Model.}
	\label{n6674_res}
\end{figure}

Looking at the residuals in Fig.~\ref{n6674_res}, it is clear that the disc-only model failed in numerous areas. The blue residuals are a result of the model being far too face-on, $\sim$45\degr, with respect to the best-fit value of $\sim$53\degr. The ring is visible in the residuals as well, showing up as the red values encircling the galaxy centre. 

When a second component is added to the model (i.e., the disc+bulge and disc+bar models), the residuals improve. However, when looking at the disc+bulge residuals (top right panel of Fig.~\ref{n6674_res}), there is a noticeable area where the model fails: the location of the bar. This shows up as the the diagonal blue area that cuts through the centre. This is all but removed when looking at the disc+bar model (bottom left panel of Fig.~\ref{n6674_res}). 

From the disc+bar model to the three-component model, there are incremental improvements to the residuals everywhere. The large red and white spots that remain even in the residuals of the three-component model are simply knots in the ring and arms of the galaxy. The ring itself is no longer visible in the residuals, but one can still trace it and the spiral arms by following these spots. The ring is visible in the full model and disc component, shown in the top middle and bottom left panels of Figure \ref{n6674_b}.

We find the bar position angle, ellipticity, and radius to be consistent between all four bands. This is also the case for the size of the bulge. The bulge is very circular ($\epsilon_{\mathrm{bulge}}$$\sim$0), although this could be caused by the \authorfix{poor} seeing \authorfix{at the time the galaxy was observed} (see Table~\ref{data_info}). The average S\'{e}rsic index, $\sim$0.65, is consistent with that of a pseudo-bulge, \authorfix{though} the errors are large. \authorfix{It is likely that the poor seeing is a contributing factor here, as well.}

\begin{figure}
	\center
    	\includegraphics[width=\columnwidth]{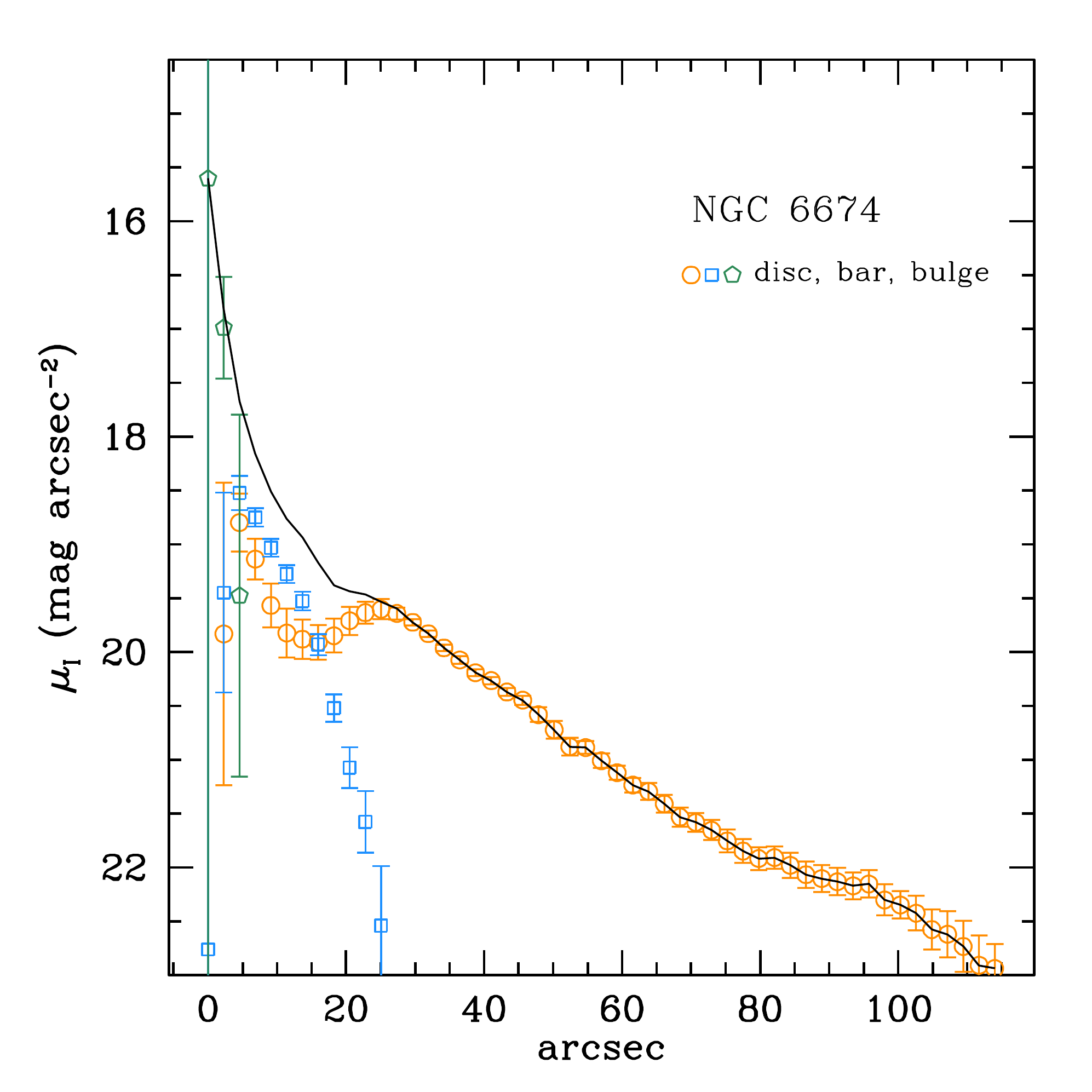}
  	\caption{\textit{I}-Band surface brightness profile for NGC~6674. The solid line shows the total surface brightness profile for NGC~6674. The orange circles, blue squares, and green pentagons show the disc, bar, and bulge components respectively. As stated in Section \ref{sec:photometry}, the data are not photometric, so the \textit{I}-band magnitudes are only approximate.}
  	\label{n6674_colour}
\end{figure}

\authorfix{As NGC 6674 was observed under non-photometric conditions, we show in Fig.~\ref{n6674_colour} the approximate \textit{I}-band surface brightness profile and do not plot \textit{(B-I)}, \textit{(V-I)}, or \textit{(R-I)}\ colour profiles.} Outside of the bulge/bar region, we find no evidence for any truncation in the disc, consistent with a Freeman Type I profile \citep{freeman1970}, or a single exponential disc. From this plot, we can see that the bulge (open green pentagons) contributes very little light to the total amount. The dip and rise in the light profile of the disc (open orange circles) near 15$\arcsec$ is the region between the bar and ring. \authorfix{As can be seen from the combined \textit{BVR}\ image in Fig.~\ref{gal_pics}, there is a decrease in the galaxy's light in the inner regions of the disc in the direction away from the bar axis (roughly NW-SE). This is visible in all four bands. In addition, the light profile of the bar (blue squares) in Fig.~\ref{n6674_colour} effectively ends at 15$\arcsec$, coinciding with the location of the ring.}

\citet{boreils1991} have also studied the photometric structure of NGC~6674.  They modeled the galaxy via elliptical isophote fitting, and give the P.A. and inclination as a function of radius, as well as plot the \textit{R}-band surface brightness profile. We find that our disc and bar components in our Fig.~\ref{n6674_colour} match the disc and bulge components from \citet{boreils1991} very well. However, our total surface brightness profile (the solid black line in Fig.~\ref{n6674_colour}) is quite different than theirs in the bar region \authorfix{(roughly between 10$\arcsec$-30$\arcsec$)}. The difference between the \textit{R}-band surface brightness profile from their work and ours is likely explained by their lack of any modeling of a bar for this galaxy. In fact, the authors state that the data within the radius of the bar, the inner 30$\arcsec$, can only be taken as indicative. 

\subsubsection{Kinematics}
\label{sec:n6674kine}

\begin{figure*}
	\center
    	\hskip 10mm \includegraphics[width=0.46\textwidth]{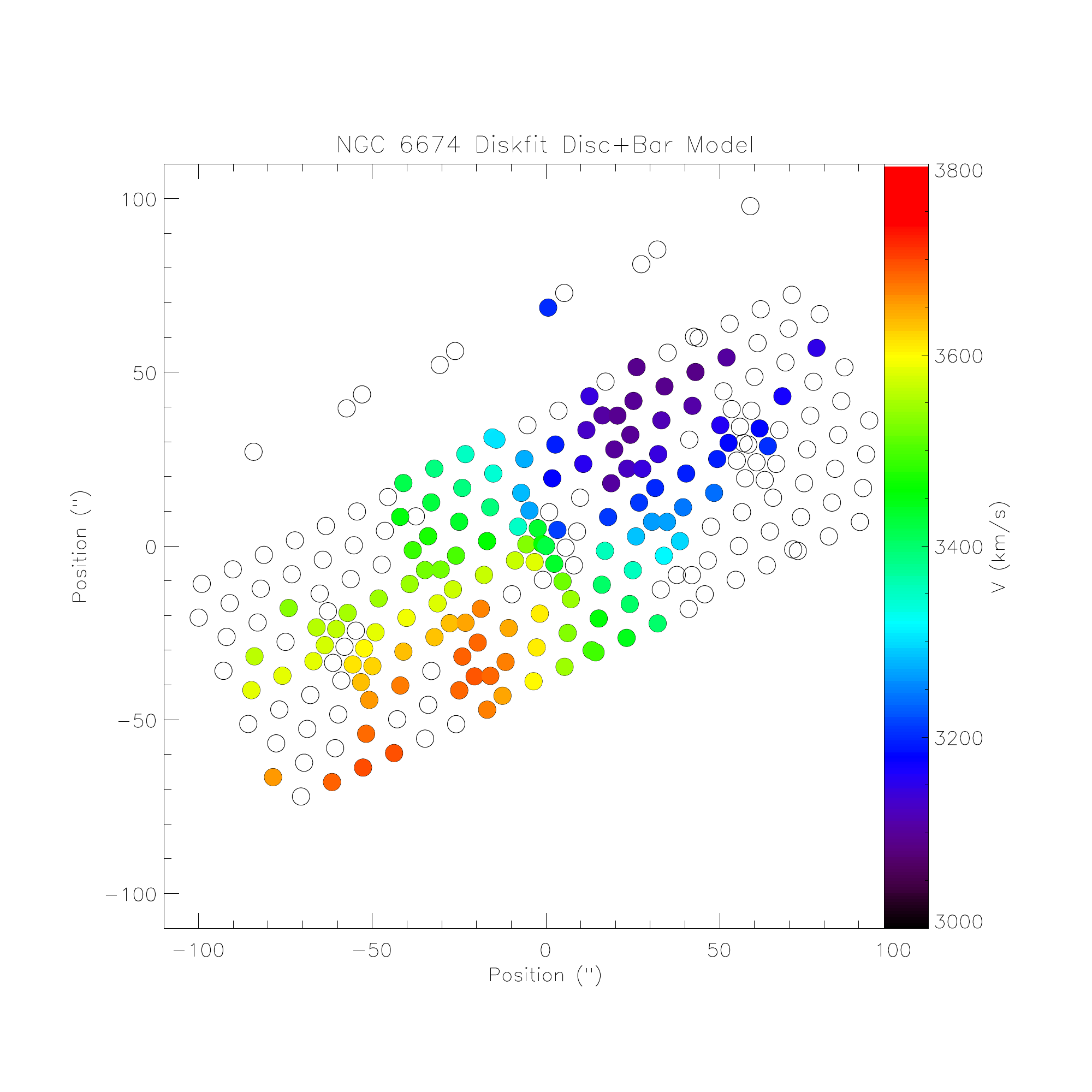} \hskip 2mm \includegraphics[width=0.46\textwidth]{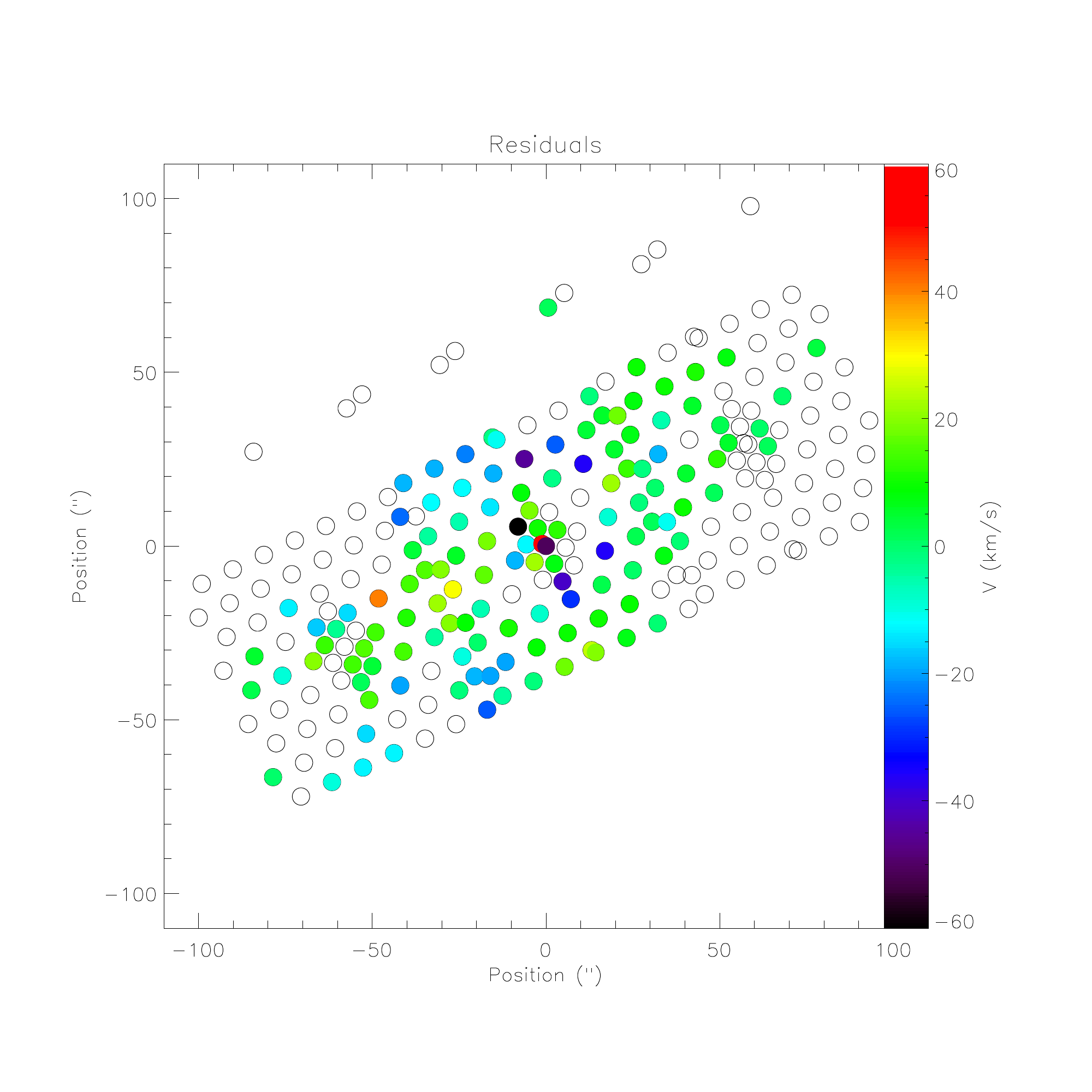}\\
	\hskip 2mm
    	\includegraphics[width=0.45\textwidth]{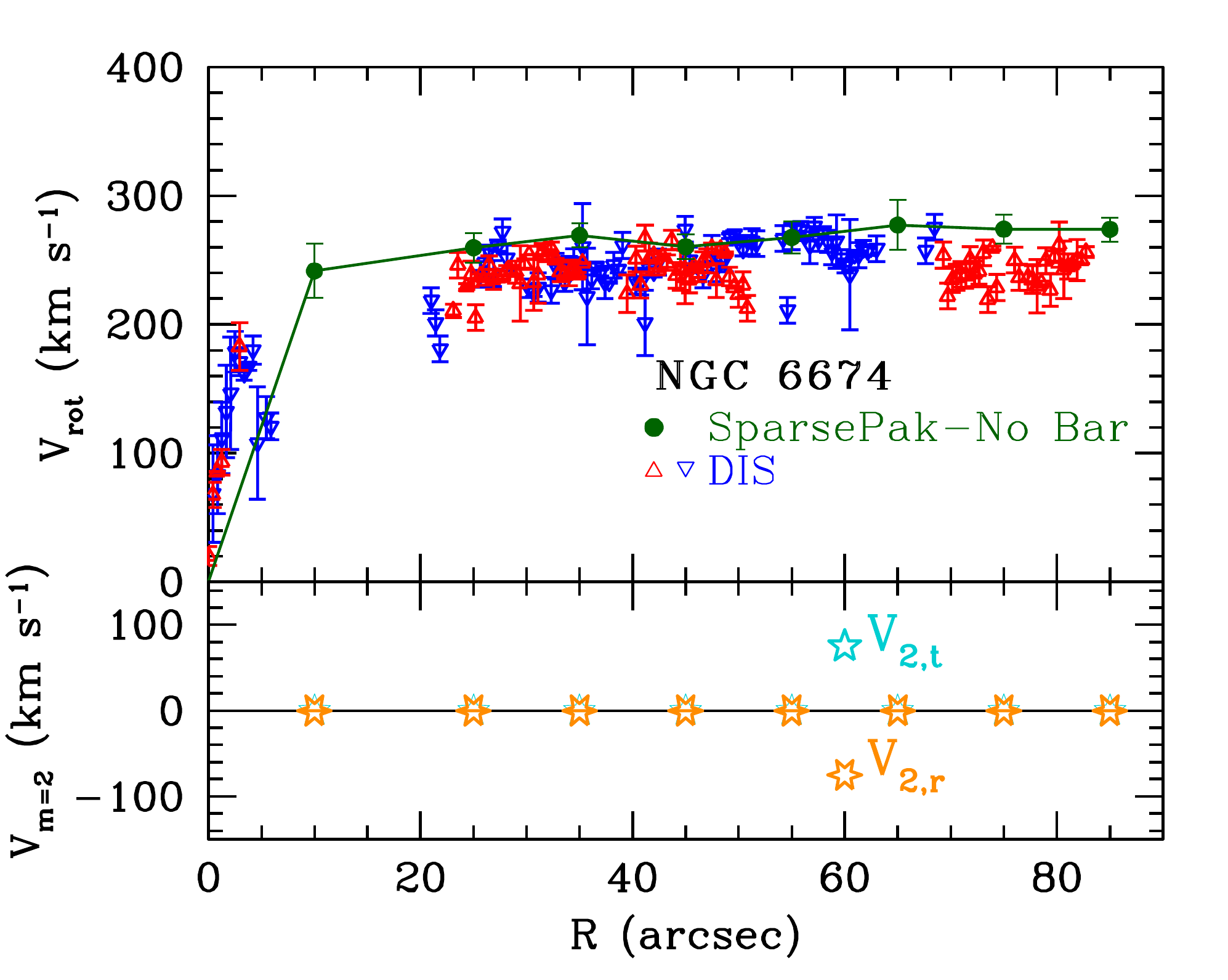} \hskip 2mm \includegraphics[width=0.45\textwidth]{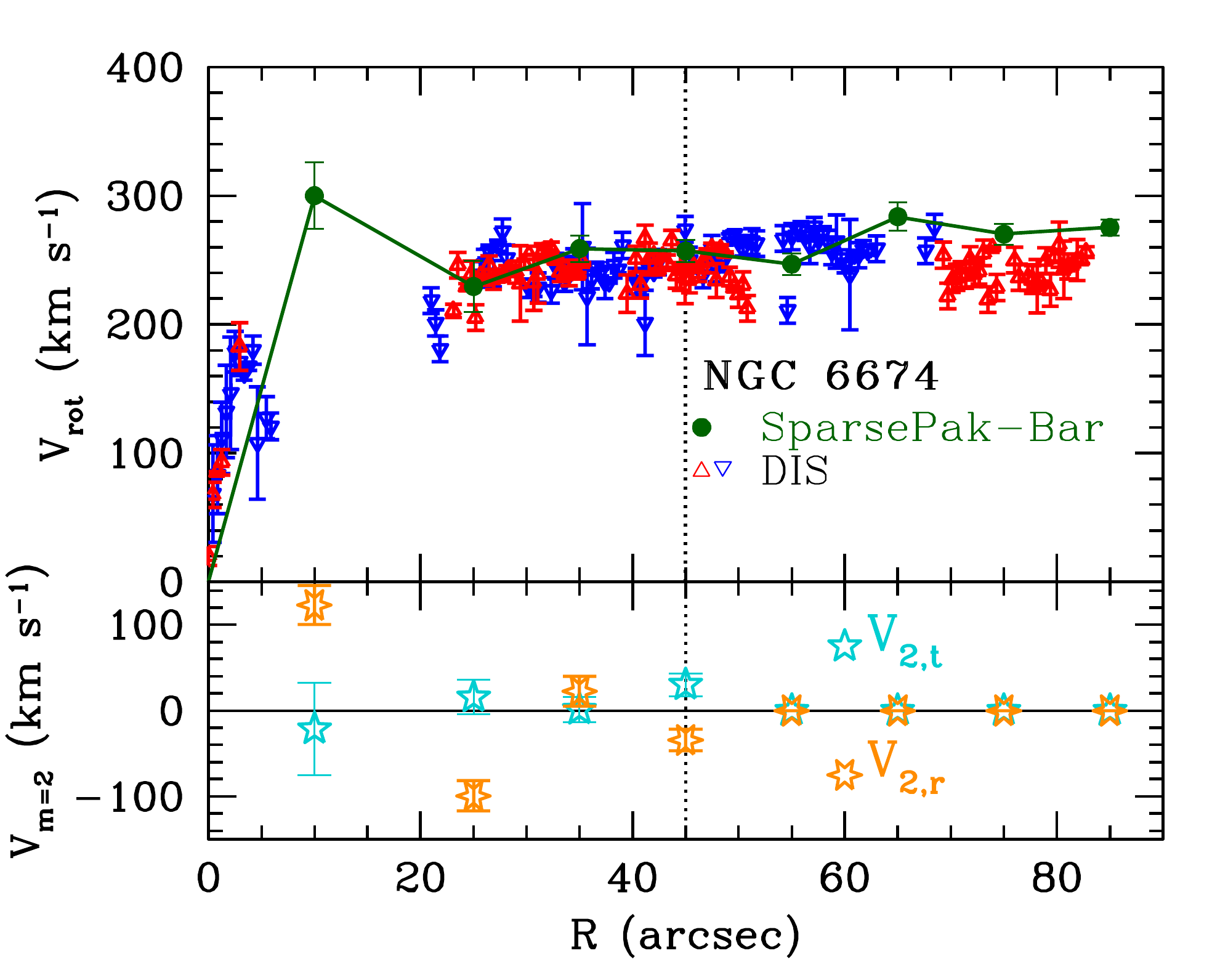}
  	\caption{Kinematic model and rotation curves for NGC~6674. \textit{Top Left}: Disc+Bar model velocities from \texttt{DiskFit}. Empty fibres in velocity field indicate no observed emission; \textit{Top Right}: Residual velocity field showing the difference between observed SparsePak velocity field and the \texttt{DiskFit} model; \textit{Bottom Left}: Derived SparsePak rotation curve without a bar (dark green circles) and observed DIS rotation curve (red and blue triangles); \textit{Bottom Right}: Derived SparsePak rotation curve including a bar and observed DIS rotation curve, the vertical line indicates the radius of the bar. The bottom portions of the bottom panels show the amplitudes of the tangential (turquoise five pointed stars) and radial (orange 6 pointed stars) components of the non-circular motions.}
  	\label{n6674_kinematics}
\end{figure*}

Our best-fitting \texttt{DiskFit} kinematic model of the SparsePak velocity field for NGC~6674 contains a disc and a bar and is shown in the top left panel of Fig.~\ref{n6674_kinematics}. We find a P.A.$_{\mathrm{disc}}$ of 149.25\degr $\pm$ 0.91\degr, an inclination of 61.61\degr $\pm$ 0.76\degr, a P.A.$_{\mathrm{bar}}$ of 30.36\degr $\pm$ 4.10\degr, a bar radius of 45$\arcsec$, and a systemic velocity of 3393.94 km s$^{-1}$ $\pm$ 1.71. Our P.A.$_{\mathrm{disc}}$ is roughly consistent with the fixed P.A. of our DIS observations as is the systemic velocity. Differences between our photometry and kinematic models will be discussed in Section~\ref{sec:n6674compare}.

The residuals for our best-fitting model are shown in the top right panel of Fig.~\ref{n6674_kinematics}. Our model does a fairly good job matching the disc of NGC~6674, with the majority of residuals around $\pm$15 km s$^{-1}$. However, there are a few fibres with residuals near -60 km s$^{-1}$ near the centre of the galaxy, within the bar. A possible explanation for this is that the large 5$\arcsec$ SparsePak fibres blur out the inner bar/ring of the galaxy. Given that NGC~6674 is roughly 56 Mpc away \citep{sorce2014}, the 5$\arcsec$ fibres correspond physically to $\sim$1.5 kpc in the galaxy. \authorfix{Thus, in the central regions of the galaxy, where gas can be moving at quite different velocities in a $\sim$1.5 kpc region, the velocities are blurred together in one fibre. To confirm this, we examined the lines present in these central fibres and found them to have large widths, indicating that there is significant blending occurring.} The fibre size, however, did not prevent the bar \textit{angle} from being found. 

Our SparsePak and DIS rotation curves are shown in the bottom panels of Fig.~\ref{n6674_kinematics}. The SparsePak rotation curve (filled dark green circles) of the disc-only model is shown in the left panel and the rotation curve from the disc+bar model is shown in the right panel; in both panels the DIS rotation curve (open red and blue triangles) is for a disc only. 

The DIS rotation curve has a rapid initial rise followed by a gap \authorfix{between $\sim$5$\arcsec$\ and 20$\arcsec$\ where there is no emission measured} and a flat portion around 250 km s$^{-1}$, consistent with HI data from \citet{combes2015}. \authorfix{The observed gap is the same one seen in the photometry. We do not find any significant asymmetry in the rotation curve for this galaxy; both the red and blue points match each other quite well. We do note, however, that the two sides have somewhat different radial distributions, with the red triangle side having a gap in data from $\sim$45$\arcsec$to $\sim$65$\arcsec$\ and the blue triangle side stopping at $\sim$65$\arcsec$.}

The SparsePak rotation curve of our disc-only model, shown in the left panel, shows a smooth rise before flattening out. At radii larger than about 25$\arcsec$, the \texttt{DiskFit} model over-predicts the rotation velocities compared to the DIS rotation curve by $\sim$15 km s$^{-1}$. 

When a bar is included in the \texttt{DiskFit} model, the shape and amplitude of the SparsePak rotation curve change significantly (see lower right panel of Fig.~\ref{n6674_kinematics}). In particular, the innermost rotation curve point shoots up to 300 km s$^{-1}$. This rapid rise and hump is consistent with bulged (e.g.\ Sa or Sb) galaxies \citep{sofue1999} and is also observed by \citet{brownstein2006} for NGC~6674. We note that while including a bar in the SparsePak model improves the agreement between the SparsePak and DIS rotation curves at intermediate radii, the SparsePak rotation curve remains higher than the DIS rotation curve at radii 65$\arcsec$ and larger. 

\authorfix{To investigate whether this was a result of the difference between the P.A.$_{\mathrm{disc}}$\ of DIS and the \texttt{DiskFit} model, we re-ran the disc+bar \texttt{DiskFit} model with the disc position angle held fixed at the DIS value from Table \ref{6674_table}. While we found this to lower the velocities of the outer SparsePak points to the level of the DIS rotation curve, the inner rotation curve was also significantly lowered, and the overall quality of the fit as indicated by $\chi^{2}$\ was much poorer ($\chi^{2} \sim 2.5$) than when the P.A. was allowed to vary freely ($\chi^{2} \sim 1.1$). The magnitude of the non-circular motions was not effected, however.}

In the lower half of the bottom right panel of Fig.~\ref{n6674_kinematics} we show the radial (six pointed stars) and tangential (five pointed stars) components of the $m=2$ non-circular motions calculated by \texttt{DiskFit}. We find significant non-circular motions (motions close to 100 km s$^{-1}$) in the inner regions of NGC~6674 with a steep fall past 25$\arcsec$. This implies that the bar length is most likely around $\sim$30$\arcsec$, which is in agreement with our photometric results.

\subsubsection{Comparison of Photometric and Kinematic Models}
\label{sec:n6674compare}

Our photometric fits give a disc position angle of $\sim$139\degr\ and an inclination of $\sim$53\degr, while the kinematic fit gives a position angle of 149\degr\ and an inclination of 61\degr. At first glance, the inconsistencies between these position angles and inclinations are somewhat concerning. However, they are in agreement with values from the literature and are simply a consequence of how \texttt{DiskFit} models a galaxy, combined with the specific structures in NGC~6674, as explained below. 

The inner region of NGC~6674 possesses a ring that is at a position angle of $\sim$137\degr. This is shown in Table 2 of \citet{boreils1991}, which gives P.A. and inclination as a function of radius. Since \texttt{DiskFit} fits a single P.A. for the entire galaxy, and not as a function of radius, our photometric fit is most heavily influenced by the inner ring present in the galaxy. This ring is blurred out in the kinematics by the SparsePak fibres, and so the kinematic side of \texttt{DiskFit} finds a P.A. that is more consistent with the majority of the outer disc of the galaxy, $\sim$147\degr. 

The difference in inclinations between our photometric and kinematic models can also be explained by looking at the inclinations given in \citet{boreils1991}: the inclinations at small radii are closer to our photometric results ($\sim$50\degr), and closer to our kinematics at large radii ($\sim$60\degr). 

\authorfix{As a test of this hypothesis, we masked out the central region of NGC 6674 and re-ran the disc-only photometric model. When the ring, bar, and bulge are no longer part of the fit, the position angle determined by \texttt{DiskFit} was $\sim$150\degr\ across all four bands; this is much closer to the value from the kinematic model. Unfortunately, constraints on the inclination are lost when this large central region is masked, and \texttt{DiskFit} settles on a value that is far too face-on (\textit{i}$\sim$30\degr).}

As is expected from this moderately inclined galaxy, the photometric and kinematic models easily find the structural components. They are both consistent with each other once certain facets of \texttt{DiskFit}'s fitting are taken into account. Because NGC~6674 has such obvious and distinct components, it serves as a prime example of how \texttt{DiskFit} can be used for photometric and kinematic modeling. Our next galaxy, NGC~2841, with a central bulge and slightly less well-defined spiral structure, also seems to fall into this category.

\subsection{NGC~2841}
\label{sec:n2841}

The best-fitting parameters for the photometry and kinematics for NGC~2841 are shown in Table~\ref{2841_table}. The best-fitting \textit{B}-band \texttt{DiskFit} photometric model is shown in Fig.~\ref{n2841_b}. \authorfix{Model components and residuals are shown in are shown in Figs.~\ref{n2841_res} and ~\ref{n2841_components}}. The \textit{I}-band surface brightness profile and (\textit{B-I}), (\textit{V-I}), (\textit{R-I}) colour profiles are shown in  Fig.~\ref{n2841_colour}. The best-fitting kinematic \texttt{DiskFit} model and rotation curve are shown in Fig.~\ref{n2841_kinematics}.

\begin{table*}
	\centering
	\caption{Best-fitting photometric and kinematic parameters for NGC~2841. Same format as Table~\ref{6674_table} though there are no long-slit DIS observations.}
	\label{2841_table}
	\begin{tabular}{rcccccccc}
		\hline
		Parameter		&		&\multicolumn{4}{c}{Photometry}		&		&\multicolumn{2}{c}{Kinematics}\\
							\cline{3-6}									\cline{8-9}
					&		&B		&V		&R		&I		&		&SparsePak		&DIS\\
		\hline
	P.A.$_{\mathrm{disc}}$ (\degr) & & 150.08 $\pm$ 1.38 & 150.41 $\pm$ 1.05 & 150.82 $\pm$ 1.81 & 153.18 $\pm$ 1.56 & & 152.22 $\pm$ 0.45 & ... \\
	\textit{i}$_{\mathrm{disc}}$ (\degr) & & 66.07 $\pm$ 1.08 & 64.44 $\pm$ 1.64 & 60.67 $\pm$ 2.28 & 63.29 $\pm$ 2.14 & & 62.72 $\pm$ 0.82 & ... \\
	$V_{\mathrm{sys}}$ (km s$^{-1}$) & & ... & ... & ... & ... & & 628.56 $\pm$ 2.04 & ... \\
	\\
	P.A.$_{\mathrm{bar}}$ (\degr) & & 159.62 $\pm$ 19.06 & 162.60 $\pm$ 9.28 & 160.49 $\pm$ 7.07 & 147.61 $\pm$ 9.95 & & 155.69 $\pm$ 4.94 & ... \\
	$R_{\mathrm{bar}}$ ($\arcsec$) & & 46 & 46 & 46 & 46 & & 65 & ... \\
	$\epsilon_{\mathrm{bar}}$ & & 0.62 $\pm$ 0.13 & 0.33 $\pm$ 0.05 & 0.62 $\pm$ 0.09 & 0.26 $\pm$ 0.12 & & ... & ... \\
	\\
	$r_{e}$ ($\arcsec$) & & 11.04 $\pm$ 1.17 & 5.02 $\pm$ 0.96 & 9.82 $\pm$ 1.85 & 8.97 $\pm$ 2.6 & & ... & ... \\
	$\epsilon_{\mathrm{bulge}}$ & & 0.27 $\pm$ 0.03 & 0.29 $\pm$ 0.07 & 0.3 $\pm$ 0.07 & 0.3 $\pm$ 0.11 & & ... & ... \\
	\textit{n} & & 1.86 $\pm$ 0.08 & 0.85 $\pm$ 0.14 & 0.93 $\pm$ 0.41 & 0.7 $\pm$ 0.74 & & ... & ... \\
	\\
	\% Disc & & 80.06 $\pm$ 2.49 & 83.10 $\pm$ 2.07 & 81.71 $\pm$ 3.69 & 85.6 $\pm$ 3.05 & & ... & ... \\
	\% Bar & & 3.82 $\pm$ 2.27 & 10.79 $\pm$ 2.63 & 14.30 $\pm$ 2.75 & 3.9 $\pm$ 3.4 & & ... & ... \\
	\% Bulge & & 16.12 $\pm$ 2.25 & 6.11 $\pm$ 1.57 & 3.98 $\pm$ 4.28 & 11.5 $\pm$ 3.74 & & ... & ... \\
	\hline
	\end{tabular}
\end{table*}

\subsubsection{Photometry}
\label{sec:n2841phot}

As seen in the top right panel of Fig.~\ref{gal_pics}, NGC~2841 is a moderately inclined, flocculent spiral galaxy with a large central bulge. Our best-fitting photometric \texttt{DiskFit} model for NGC~2841 is the three component model containing a disc, bar, and bulge.

We find good agreement across all four photometric bands on the value of the disc position angle ($\sim$151\degr), disc inclination ($\sim$63\degr), bar position angle ($\sim$160\degr), bar length ($\sim$46$\arcsec$) and the percentage of the total galaxy light coming from the disc ($\sim$83\%).  As can be seen in Table~\ref{2841_table} however, there is a moderate amount of variation in the bulge parameters and the division of light between the bulge and bar components. 

We also note that all four best-fitting bulges have relatively large ellipticities ($\geq$~0.27), making them appear a bit more bar-like than bulge-like. None of our effective radii are consistent with the results from \citet{varela1996}, who find the $r_{e}$\ in \textit{V}, \textit{R}, and \textit{I} to all be greater than 20$\arcsec$. This is likely because \citet{varela1996} did not take a bar into account, only a bulge. In our disc+bulge model, the bulge had a similar effective radius ($\sim$20$\arcsec$) in all four bands.

\begin{figure*}
	\center
	\includegraphics[width=0.9\textwidth]{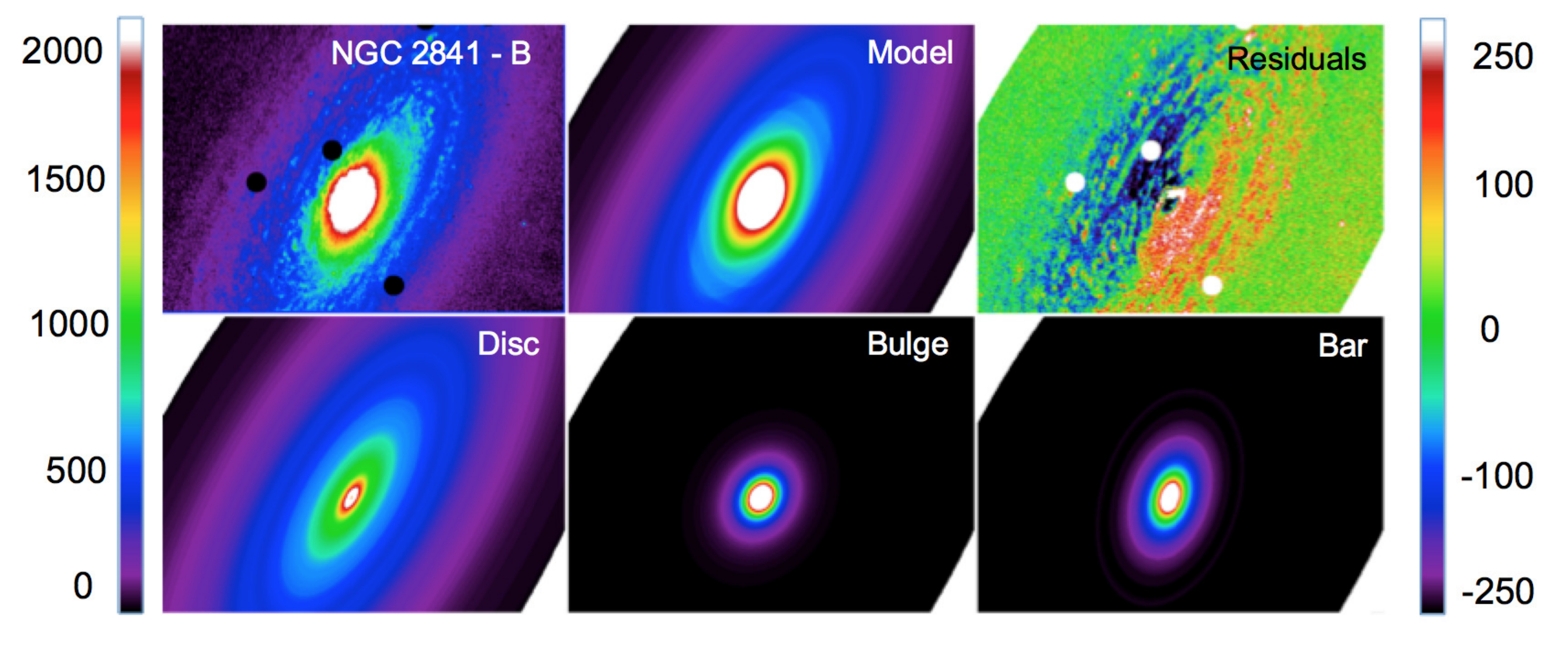}
  	\caption{Same as Fig.~\ref{n6674_b} but for NGC~2841. Each frame is 3.8$\arcmin$~$\times$~2.66$\arcmin$.}
  	\label{n2841_b}
\end{figure*}

We also find relative consistency between the \textit{V,R} and \textit{I}-bands for the S\'{e}rsic index ($\sim$0.83), although the errors are quite high. We find the \textit{B}-band value to be significantly larger than the other three bands (1.86). In Fig.~\ref{n2841_b} we show the galaxy image, three-component model, (galaxy-model) residuals, and each of the individual model components for the  \textit{B}-band.

\begin{figure}
	\center
	\includegraphics[width=\columnwidth]{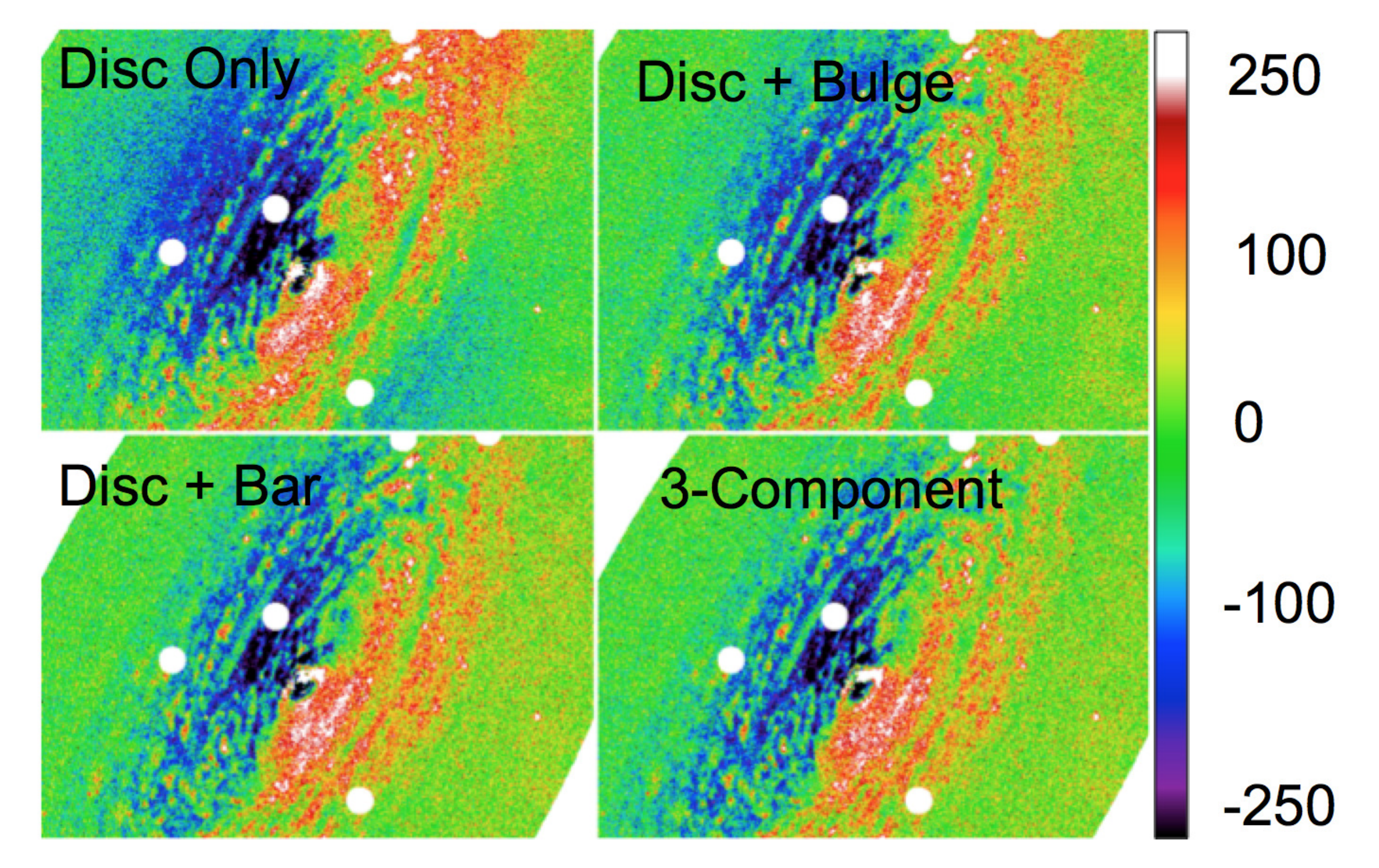}
	\caption{Same as Fig.~\ref{n6674_res} but for NGC~2841. Panels are 3.8$\arcmin$~$\times$~2.66$\arcmin$.}
	\label{n2841_res}
\end{figure}

As can be seen in Fig.~\ref{n2841_res}, all four \texttt{DiskFit} models have difficulty matching the dust lanes/spiral arms in NGC~2841 and try to compensate by making the galaxy too face-on. This leads to the large-scale blue and red residuals seen in each panel. The disc-only model (top left panel of Fig.~\ref{n2841_res}) is the most extreme example, with the inclination being driven to $\sim$50\degr. \authorfix{The rather symmetric blue/red pattern is a direct consequence of the dust in this galaxy. In forcing the galaxy to be less inclined than it really is, \texttt{DiskFit} is adding too much light to the model on the far side of the galaxy (producing the blue residuals) and takes away too much light on the near side (producing the red residuals).}

Given the undefined spiral structure and copious amounts of dust in NGC~2841, it is difficult to distinguish between the residuals for the disc+bulge, disc+bar and three-component models in Fig.~\ref{n2841_res}. While by-eye these models provide seemingly similar quality fits, the details of the models are physically very different. 

\authorfix{In addition, these large residuals demonstrate the benefit of selecting the best-fitting model as the one that is the most consistent across all bands, rather than selecting the model with the lowest $\chi^{2}$. The magnitude of the $\chi^{2}$\ for the models can be extremely high, even if the model is a good fit to the data. For example, the disc-only model for NGC 2841 has a lower $\chi^{2}$\ than any of the other models in all of the bands (ranging from $\sim$30 in the \textit{B}-band to $\sim$100 in the \textit{I}-band). However, this is obviously not physical, and the model clearly does not fit the data well.}

Due to the large, non-axisymmetric central component in this galaxy, including a component in addition to the disc improves the fit of the model. However, we find that the disc+bulge and disc+bar results are not unique: the disc+bulge model forces the bulge into a very elongated and squashed shape ($\epsilon_{\mathrm{bulge}}$\ $\sim$0.5 and $r_{e}$ $\sim$50$\arcsec$) that is nearly identical to the bar from the disc+bar model ($\epsilon_{\mathrm{bar}}$ $\sim$0.4 and $R_{\mathrm{bar}}$ $\sim$45$\arcsec$). This suggests that neither of these models are exclusively accurate. 

\begin{figure}
	\center
	\includegraphics[width=\columnwidth]{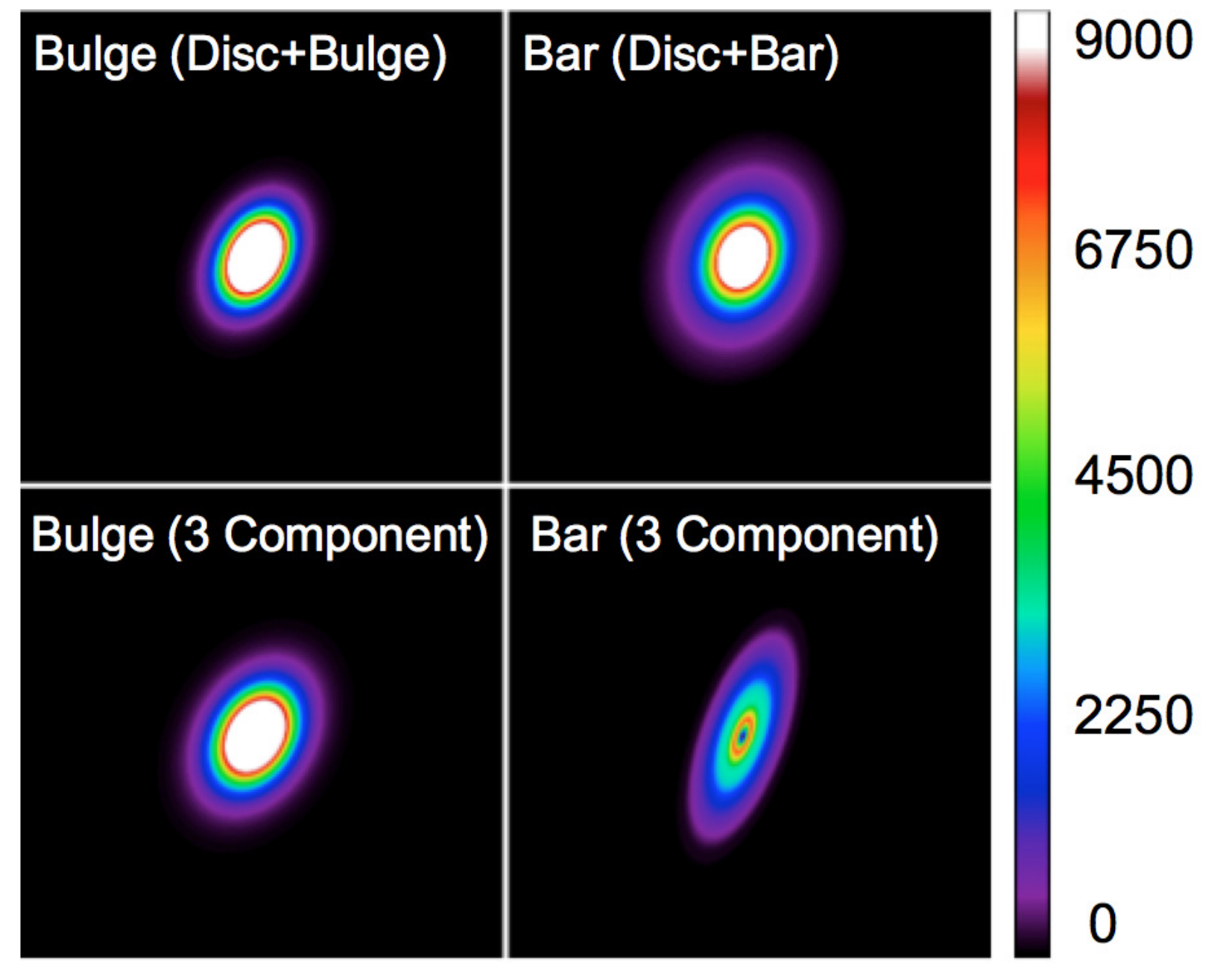}
	\caption{Individual bulge and bar components of NGC 2841 for various models in the \textit{R}-band.}
	\label{n2841_components}	
\end{figure}

\authorfix{This is best seen in Fig.~\ref{n2841_components}, where we show the individual bulge and bar components from the various models in the \textit{R}-band. We show the \textit{R}-band here as we have already shown the \textit{B}-band components of the three component model in Fig.~\ref{n2841_b}, and we want to illustrate that this degeneracy is present in all bands. The components have all been scaled to the same values for easy comparison. The top row shows the bulge from the disc+bulge model and the bar from the disc+bar model, clearly indicating the degeneracy between these two models. The bottom row shows both components from the disc+bulge+bar model, similar to what is seen in Fig.~\ref{n2841_b} but for the \textit{R}-band. These two components now look very different from each other.} Only by including all three components (disc, bulge, and bar) do we find the best agreement between all the parameters, as well as physically meaningful components of the galaxy. 

\authorfix{We note, however, that this is a complex galaxy, and while including three components does produce the best consistency between all four bands, a large amount of scatter between parameter values still remains (note the bar ellipticity and individual light percentages in Table \ref{2841_table}, for example). This could be an indication that an additional component, or perhaps a different assumed light profile for the bulge, is needed to more accurately model this galaxy.}

That including a bar in the photometric model for this galaxy provides the most consistency between all four bands is of note for several reasons. NGC~2841 does not prominently display a stellar bar, nor is the galaxy classified in catalogs and surveys as being barred. For example, it is classified as SA(r)b in the Third Reference Catalogue of Bright Galaxies \citep{deVaucouleurs1991}, as Sab on SIMBAD (see Table~\ref{gal_info}), and as Sb in the THINGS survey \citep{walter2008}. However, a possible bar in the inner 10$\arcsec$ was reported from nuclear H$\alpha$ and [NII] emission  as far back as \citet{keel1983}. This was later confirmed via triaxial bulge decomposition by \citet{varela1996} and \citet{afanasiev1999}. These authors find the bar to extend to roughly 15-33$\arcsec$ and to be offset from the P.A.$_{\mathrm{disc}}$ by $\sim$7\degr. These results are generally consistent with what we find with \texttt{DiskFit}. Curiously, there has been no further mention of NGC~2841 having a bar.

\begin{figure}
	\center
    	\includegraphics[width=\columnwidth]{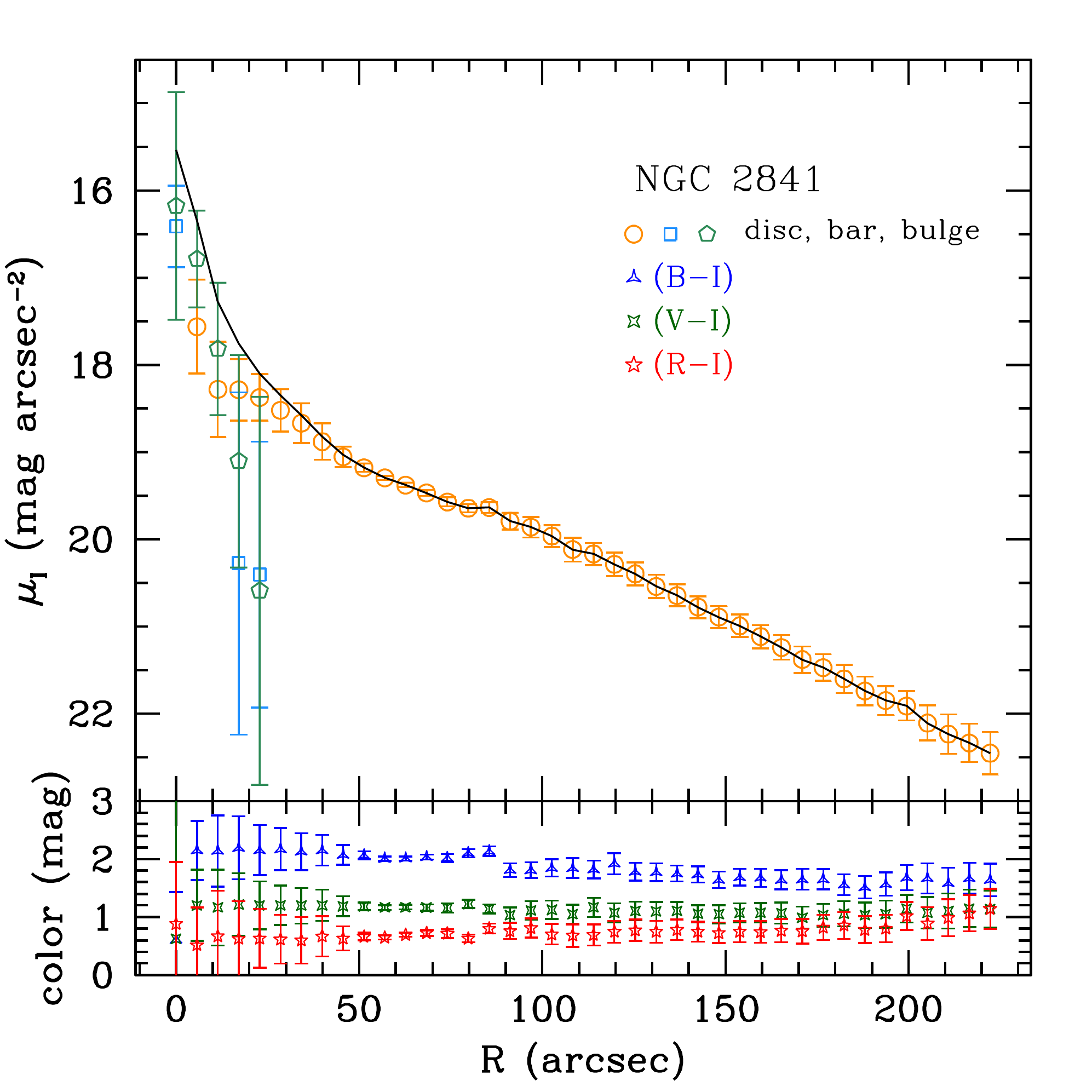}
  	\caption{Point types are the same as in Fig.~\ref{n6674_colour} but for NGC~2841. Bottom panel shows the (\textit{B-I}), (\textit{V-I}), and (\textit{R-I}) colour profiles.}
  	\label{n2841_colour}
\end{figure}

In Fig.~\ref{n2841_colour} we show the \textit{I}-band surface brightness profile of NGC~2841. There is a slight truncation around 80-90$\arcsec$, consistent with the \textit{BVRI} surface brightness profiles found by \citet{marci2001}. We find a slight negative gradient in the (\textit{B-I}) colour, while the (\textit{V-I}) and (\textit{R-I}) colours are fairly constant.

\subsubsection{Kinematics}
\label{sec:n2841kine}

\begin{figure*}
	\center
    	\hskip 10mm \includegraphics[width=0.46\textwidth]{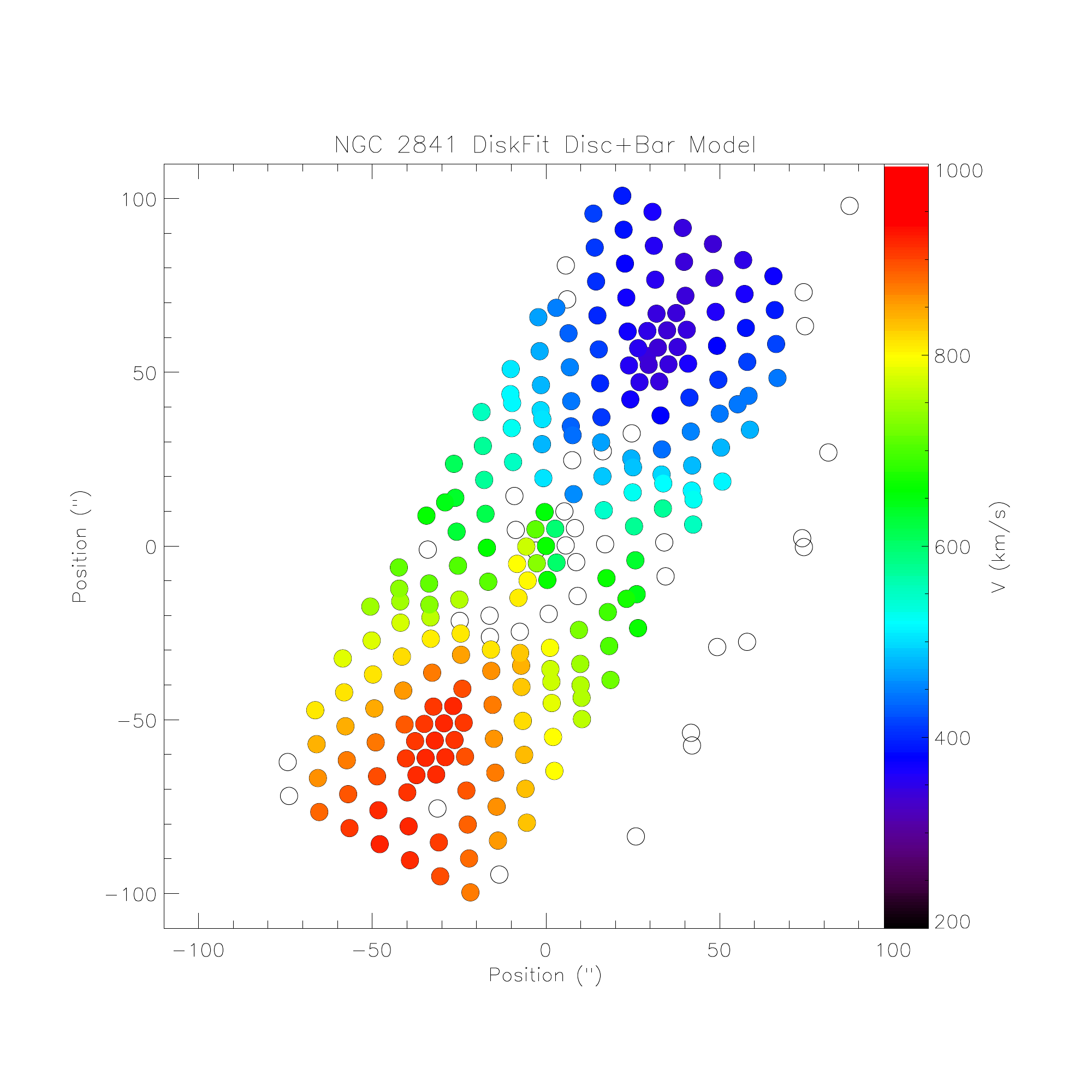} \hskip 2mm \includegraphics[width=0.46\textwidth]{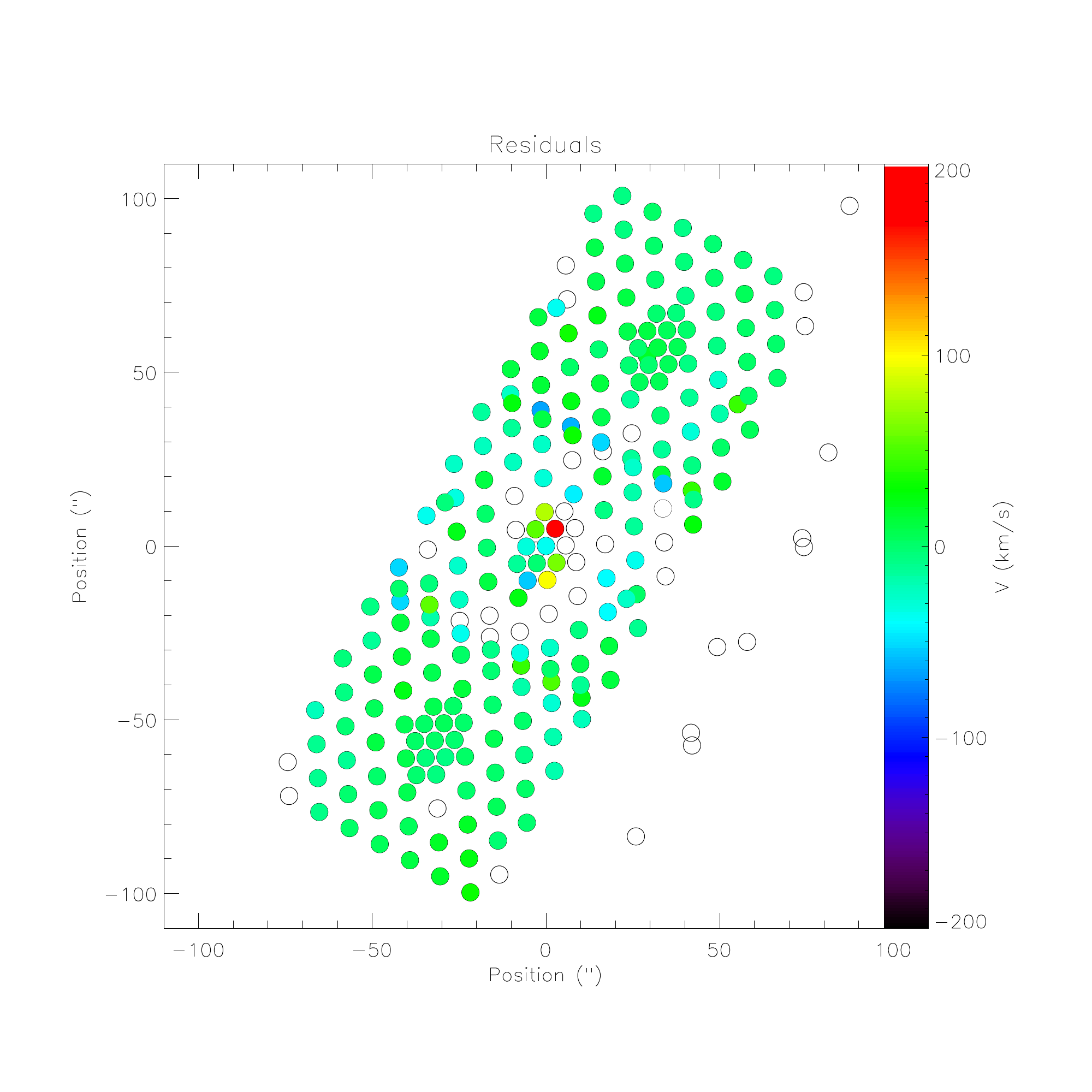}\\
	\hskip 2mm
    	\includegraphics[width=0.45\textwidth]{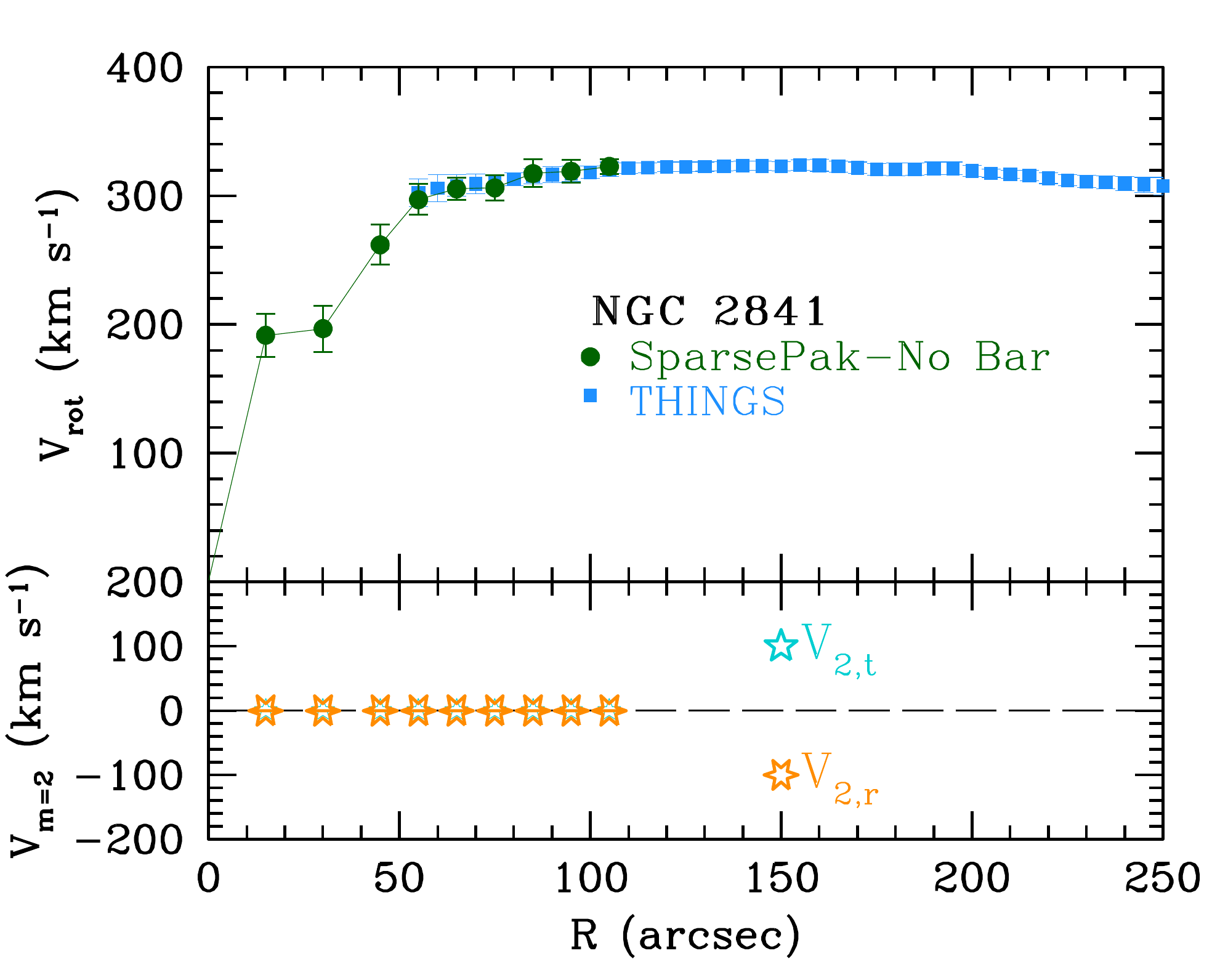} \hskip 2mm \includegraphics[width=0.45\textwidth]{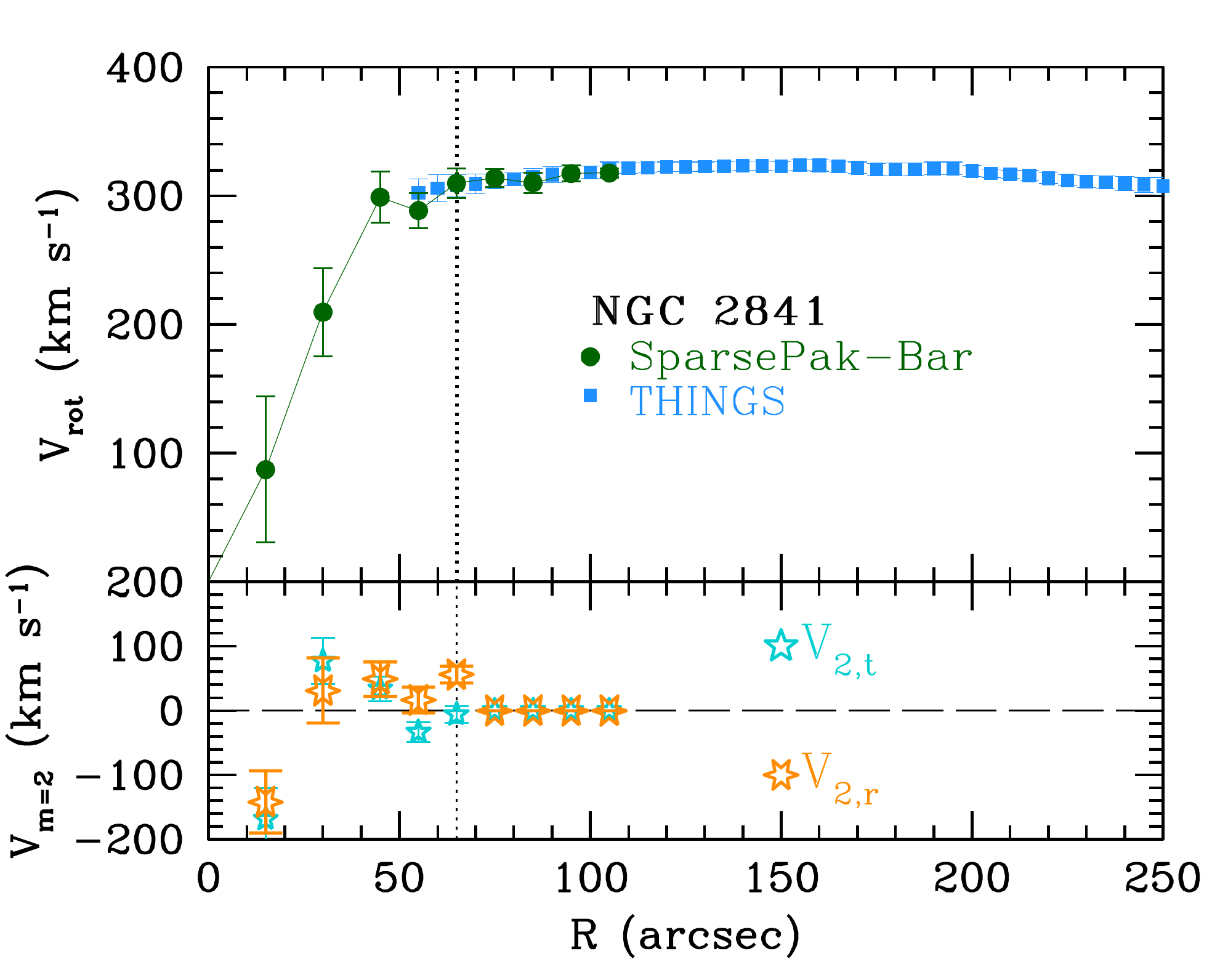}
  	\caption{Same as Fig.~\ref{n6674_kinematics} but for NGC~2841.}
  	\label{n2841_kinematics}
\end{figure*}

Our best-fitting \texttt{DiskFit} kinematic model of the SparsePak velocity field for NGC~2841 contains a disc and a bar, and is shown in the top left panel of Fig.~\ref{n2841_kinematics}. We find a P.A.$_{\mathrm{disc}}$ of 152.22\degr $\pm$ 0.45\degr, an inclination of 62.72\degr $\pm$ 0.82\degr, a P.A.$_{\mathrm{bar}}$ of 155.69\degr $\pm$ 4.94\degr, a bar radius of 65$\arcsec$, and a systemic velocity of 628.56 $\pm$ 2.04 km s$^{-1}$. 

The residuals for this model are shown in the top right panel of Fig.~\ref{n2841_kinematics}. As can be seen, the disc+bar model is a good fit nearly everywhere in the galaxy (residuals on the order of $\pm$20 km s$^{-1}$), only failing in a few central fibres (residuals $>$ 100 km s$^{-1}$). 

In the lower right and left panels, respectively, of Fig.~\ref{n2841_kinematics} we show the SparsePak (filled dark green  circles) rotation curves with and without a bar.  As we did not observe this galaxy with DIS, we supplemented our SparsePak rotation curves with its THINGS HI rotation curve \citep{deblok2008}, shown as the light blue, filled squares in the bottom two panels. 

Both our disc-only and disc+bar SparsePak rotation curves agree well with the THINGS rotation curve at radii 50$\arcsec$ and larger, and our data fill in the inner hole present in the HI. The shape of the inner SparsePak rotation curve, however, is quite different depending on whether or not a bar is included in the \texttt{DiskFit} model.

Without a bar, the rotation curve begins at nearly 180 km s$^{-1}$ and features a ``plateau'' in velocities around 200 km s$^{-1}$ before continuing to again rise until flattening at $\sim$300 km s$^{-1}$.  This kind of rotation curve structure, when combined with any twisting in a velocity field, is usually an indication that a bar is present \citep[see, for example, UGC~1551 in][]{kuzio2008}.  The rotation curve of the disc+bar model produces a more realistic rotation curve in the inner 30$\arcsec$ that shows the steep rise in velocities that would be expected for a massive galaxy like NGC~2841.  

As seen in the bottom half of the lower right panel of Fig.~\ref{n2841_kinematics}, we find significant non-circular motions in the inner region of NGC~2841.   We also find that these non-circular motions rapidly decrease in magnitude away from the inner two rings (from $\sim$150 km s$^{-1}$\ to $<$\ 75 km s$^{-1}$), likely indicating that the bar length is less than 65$\arcsec$ and closer to 30$\arcsec$. This is consistent with the bar length found by \citet{afanasiev1999}.

\subsubsection{Comparison of Photometric and Kinematic Models}
\label{sec:n2841compare}

We find very good agreement between our photometric and kinematic constraints on the disc and bar parameters, specifically on the values of the disc and bar position angles and the disc inclination (see Table~\ref{2841_table}).

Though there is a difference in bar length determined by the photometry and kinematics, we consider the values listed in Table~\ref{2841_table} to be upper limits and for the more likely true lengths to be in agreement with each other. For example, when looking at the light profile of the bar in Fig.~\ref{n2841_colour} (blue squares), we see that it only makes a significant contribution to the total galaxy light out to $\sim$30$\arcsec$, consistent with the bar length found by \citet{afanasiev1999}. And while the bar determined by the kinematic model formally extends out to 65$\arcsec$, for reasons already discussed in Section~\ref{sec:n2841kine}, the bar is likely closer to 30$\arcsec$ in length.  

Given the excellent agreement between the bar parameters determined from both the photometric and kinematic data for NGC~2841, we consider it quite likely that a bar is indeed present in NGC~2841.  

NGC~2841 shows the benefit of using \textit{both} photometry and kinematics to determine the structure and properties of a galaxy. NGC~2841 does not display an obvious stellar bar in its photometry; if only photometric modeling were performed, it would therefore be hard to claim that a bar was found in the three-component model. It is only with the addition of finding the same bar parameters in our kinematic model that we can fairly confidently claim to have found a bar. The next galaxy in our sample, NGC~2654, is a more complicated system, with its structure hidden by its high inclination.

\subsection{NGC~2654}
\label{sec:n2654}

The best-fitting parameters for the photometry and kinematics for NGC~2654 are shown in Table~\ref{2654_table}. The best-fitting \textit{B}-band \texttt{DiskFit} photometric model of NGC~2654 is shown in Fig.~\ref{n2654_b}. The \textit{B}-band residuals for all four models are shown in Fig.~\ref{n2654_res}. The \textit{I}-band surface brightness profile and (\textit{B-I}), (\textit{V-I}), (\textit{R-I}) colour profiles are shown in Fig.~\ref{n2654_colour}. The best-fitting kinematic \texttt{DiskFit} model and both SparsePak and DIS rotation curves are shown in Fig.~\ref{n2654_kinematics}.

\begin{table*}
	\centering
	\caption{Best-fitting photometric and kinematic parameters for NGC~2654. Same format as Table~\ref{6674_table}.}
	\label{2654_table}
	\begin{tabular}{rcccccccc}
		\hline
		Parameter		&		&\multicolumn{4}{c}{Photometry}		&		&\multicolumn{2}{c}{Kinematics}\\
							\cline{3-6}									\cline{8-9}
					&		&B		&V		&R		&I		&		&SparsePak		&DIS\\
		\hline
	P.A.$_{\mathrm{disc}}$ (\degr) & & 65.27 $\pm$ 0.76 & 65.57 $\pm$ 0.5 & 65.87 $\pm$ 0.91 & 66.13 $\pm$ 0.74 & & 250 $\pm$ 3.26 & 66 \\
	\textit{i}$_{\mathrm{disc}}$ (\degr) & & 80.01 $\pm$ 2.35 & 78.95 $\pm$ 1.53 & 78.90 $\pm$ 2.00 & 79.53 $\pm$ 2.51 & & 82.97 $\pm$ 1.69 & 81  \\
	$V_{\mathrm{sys}}$ (km s$^{-1}$) & & ... & ... & ... & ... & & 1356.38 $\pm$ 6.08 & 1348 $\pm$ 15 \\
	\\
	P.A.$_{\mathrm{bar}}$ (\degr) & & 65.01 $\pm$ 3.95 & 62.95 $\pm$ 2.04 & 62.93 $\pm$ 4.35 & 62.29 $\pm$ 2.73 & & 156.84 $\pm$ 21.50 & ... \\
	$R_{\mathrm{bar}}$ ($\arcsec$) & & 11.2 & 22.4 & 16.8 & 16.8 & & 10 & ... \\
	$\epsilon_{\mathrm{bar}}$ & & 0.59 $\pm$ 0.09 & 0.55 $\pm$ 0.06 & 0.57 $\pm$ 0.07 & 0.57 $\pm$ 0.06 & & ... & ... \\
	\\
	$r_{e, \mathrm{bulge}}$ ($\arcsec$) & & 12.42 $\pm$ 5.13 & 11.85 $\pm$ 4.6 & 10.24 $\pm$ 3.30 & 13.14 $\pm$ 3.84 & & ... & ... \\
	$\epsilon_{\mathrm{bulge}}$ & & 0.23 $\pm$ 0.15 & 0.23 $\pm$ 0.12 & 0.2 $\pm$ 0.13 & 0.27 $\pm$ 0.12 & & ... & ... \\
	\textit{n} & & 1.99 $\pm$ 0.55 & 2.11 $\pm$ 0.42 & 2.07 $\pm$ 0.48 & 2.00 $\pm$ 0.39 & & ... & ... \\
	\\
	\% Disc & & 68.05 $\pm$ 5.27 & 62.06 $\pm$ 5.66 & 62.58 $\pm$ 6.28 & 53.37 $\pm$ 7.87 & & ... & ... \\
	\% Bar & & 14.48 $\pm$ 2.89 & 15.61 $\pm$ 4.84 & 14.85 $\pm$ 4.21 & 12.36 $\pm$ 5.12 & & ... & ... \\
	\% Bulge & & 17.47 $\pm$ 6.79 & 22.33 $\pm$ 6.75 & 22.56 $\pm$ 8.31 & 34.28 $\pm$ 9.11 & & ... & ... \\
	\hline
	\end{tabular}
\end{table*}

\subsubsection{Photometry}
\label{sec:n2654phot}

As described in Section~\ref{sec:sample} and seen in the upper left panel of Fig.~\ref{gal_pics}, NGC~2654 is a highly-inclined galaxy with a boxy/peanut-shaped bulge. We find the three-component model to be the best description of the galaxy. 

\begin{figure*}
	\center
	\includegraphics[width=0.9\textwidth]{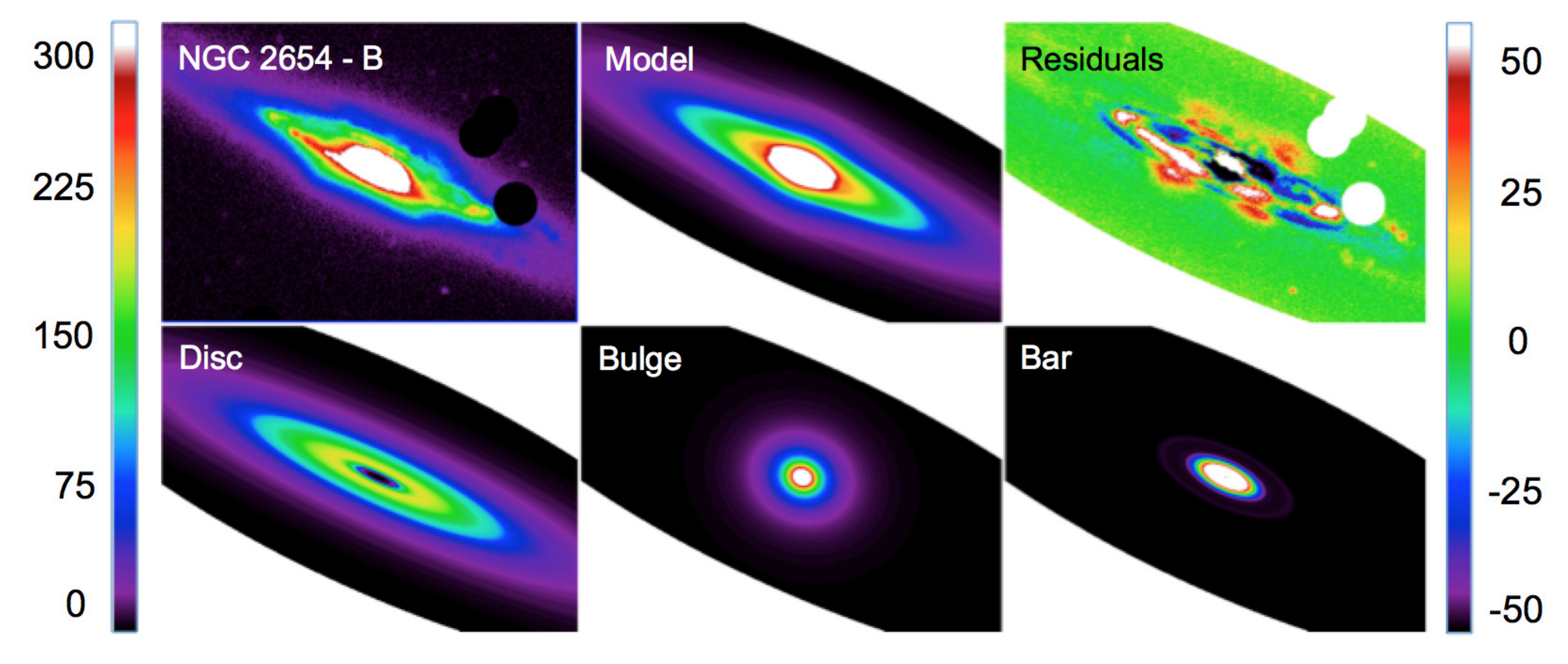}
  	\caption{Same as Fig.~\ref{n6674_b} but for NGC~2654. Each frame is 2.21$\arcmin$~$\times$~1.63$\arcmin$.}
  	\label{n2654_b}
\end{figure*}

There is good agreement across the four photometric bands on the values of all of the model parameters except for the bar length, which we discuss below. We find, on average, a P.A.$_{\mathrm{disc}}$\ of $\sim$66\degr, an inclination of $\sim$79\degr, a P.A.$_{\rm bar}$ of $\sim$62\degr, an $\epsilon_{\mathrm{bar}}$ of $\sim$0.57, a bulge effective radius of $\sim$12$\arcsec$,  an $\epsilon_{\mathrm{bulge}}$ of $\sim$0.23, and a bulge S\'{e}rsic index of $\sim$2. The disc contributes roughly 60\% of the total galaxy light, the bar roughly 15\%, and the bulge roughly 25\%.  The galaxy image, three-component model, (galaxy-model) residuals, and each of the individual model components are shown in Fig.~\ref{n2654_b} for the \textit{B}-band observations.

\begin{figure}
	\center
	\includegraphics[width=\columnwidth]{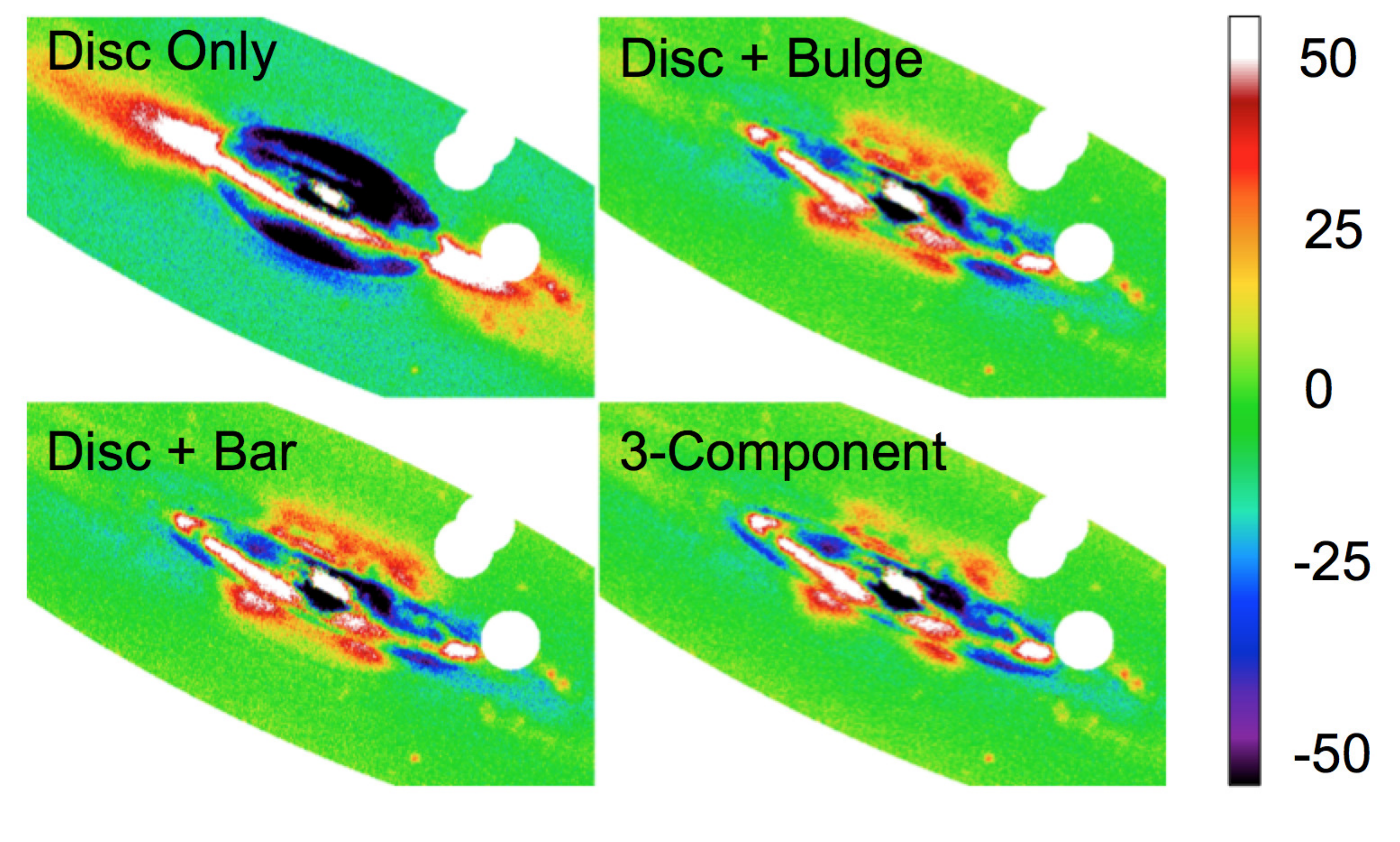}
	\caption{Same as Fig.~\ref{n6674_res} but for NGC~2654. Panels are 2.21$\arcmin$~$\times$~1.63$\arcmin$.}
	\label{n2654_res}
\end{figure}

Because NGC~2654 is quite inclined, the disc-only model was unable to converge on a realistic approximation of the galaxy. As the residuals in the top left panel of Fig.~\ref{n2654_res} show, the disc-only model attempted to fit the entire inner bulge with a rather face-on disc (\textit{i} $\sim$ 60\degr). 

Adding a second component to the galaxy model significantly improves upon the disc-only fit, as seen in the top right and bottom left panels of Fig.~\ref{n2654_res}. However, because the bar in NGC~2654 is hidden in the X-shaped bulge, both the disc+bulge and disc+bar models return nearly identical outputs and likewise, very similar residuals. 

There is a significant improvement in the residuals when all three components (disc, bulge, and bar) are included in the model. The areas directly above and below the galaxy centre, for example, are much better fit in this model than in any of the other three (see lower right panel of Fig.~\ref{n2654_res}). In addition, the three-component model provided the most consistency of all the parameters across all four photometric bands.

There are two additional features seen in the residuals that are worth mentioning: the ``X-shape'' and the ``ring.''  While the X-shape bulge is evident in our photometry, its presence becomes quite clear when looking at the residuals shown in Figs.~\ref{n2654_b} and \ref{n2654_res}. \texttt{DiskFit} attempts to fit this entire shape with a single bulge, resulting in a rather large, squashed ($\epsilon \sim$ 0.23) component. The other prominent feature in the residuals is the ring encircling the boxy-bulge. This has also been seen in Spitzer photometry from \citet{buta2015}. \texttt{DiskFit} is unable to model this component very well.

Our photometric modeling suggests that the bar in NGC~2654 runs nearly parallel to the major axis of the galaxy in the \textit{B}-band, but is offset roughly 3\degr northwards in the other three bands (though, within the errors, they, too, are consistent with being aligned with the disc major axis). We find considerable variation in the bar length between the four bands, with the \textit{V}-band being an extreme outlier. 

\begin{figure}
	\center 
    	\includegraphics[width=\columnwidth]{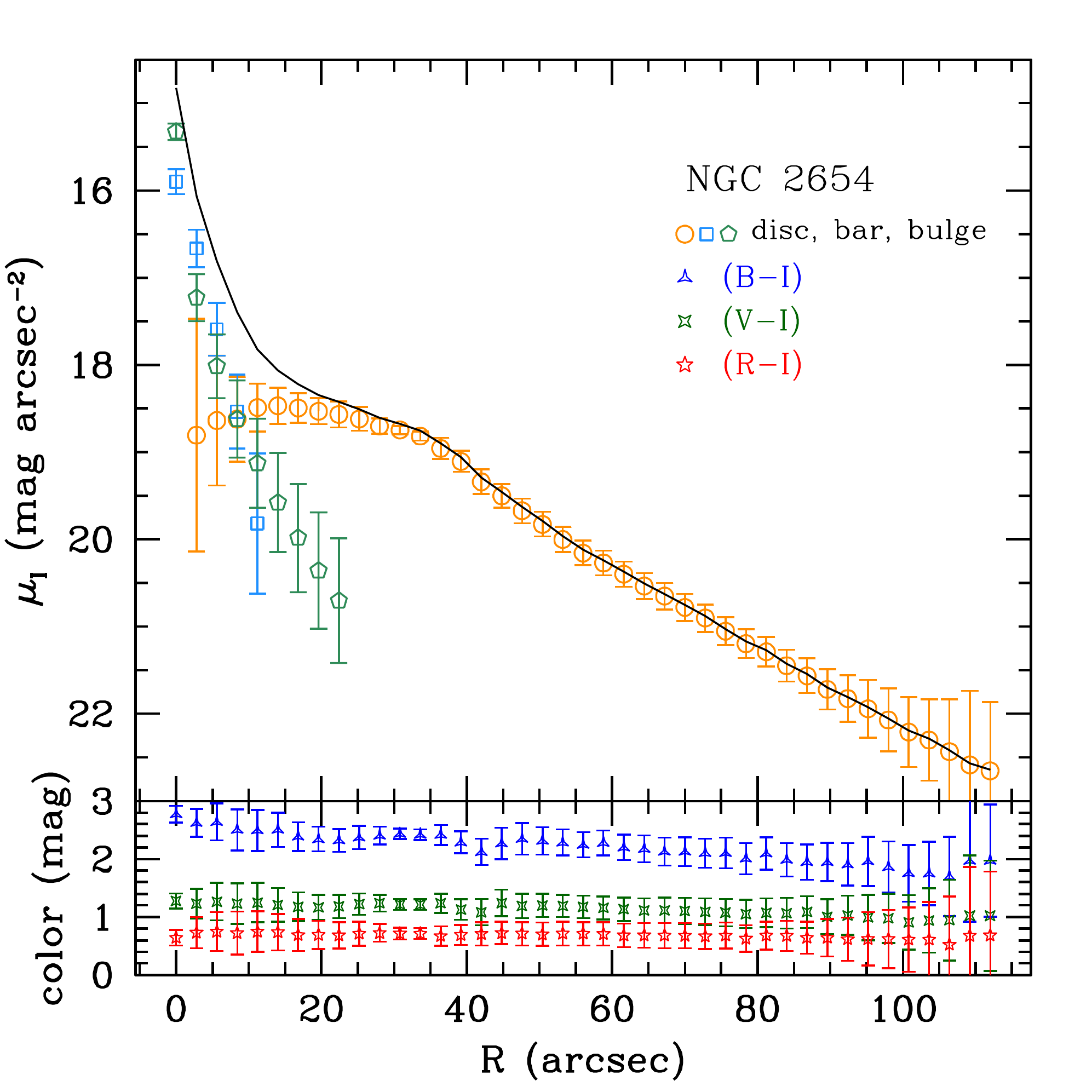}
  	\caption{Point types are the same as in Fig.~\ref{n6674_colour} but for NGC~2654. colour profiles same as in Figure \ref{n2841_colour}.}
  	\label{n2654_colour}
\end{figure}

The \texttt{DiskFit} \textit{I}-band surface brightness profile presented in Fig.~\ref{n2654_colour} shows that the inner 10$\arcsec$ of NGC~2654 are dominated by the light coming from the bulge and bar components. The profile exhibits a slight hump, or truncation, near 40$\arcsec$, and then proceeds to follow a near perfect exponential fall-off. This truncation is consistent with that of a Freeman Type II profile and is indicative of a late-type, barred spiral galaxy \citep{erwin2008, guitierrez2011}. The truncation we see at at 40$\arcsec$, however, was not found in a NIR study by \citet{florido2006}, who instead find only a pure exponential disc. 

The (\textit{B-I}), (\textit{V-I}), and (\textit{R-I}) colour profiles for NGC~2654 are also shown in Fig.~\ref{n2654_colour}.  We find a slight negative gradient in the (\textit{B-I}) colour, and relatively flat (\textit{V-I}) and (\textit{R-I}) colour profiles. The truncation at 40$\arcsec$ is also visible in the colour profiles, most notably in (\textit{B-I}).

\authorfix{Because NGC 2654 is quite inclined, it can be informative to use a secondary method independent of \texttt{DiskFit} to analyze the photometry. One example is the ``slit'' method of \citet{kuzio2009}. As we discuss in Section~\ref{sec:summary}, this is particularly useful for systems where \texttt{DiskFit} approaches or reaches its limits and fails. NGC 2654 is right at the limit of what \texttt{DiskFit} can accurately model, but still produces consistent models across four bands. It is therefore a good candidate for trying both techniques to test whether they provide consistent results.}

\subsubsection{Slit Photometry}
\label{sec:n2654slit}

\begin{figure*}
	\centering
	\includegraphics[scale=0.6]{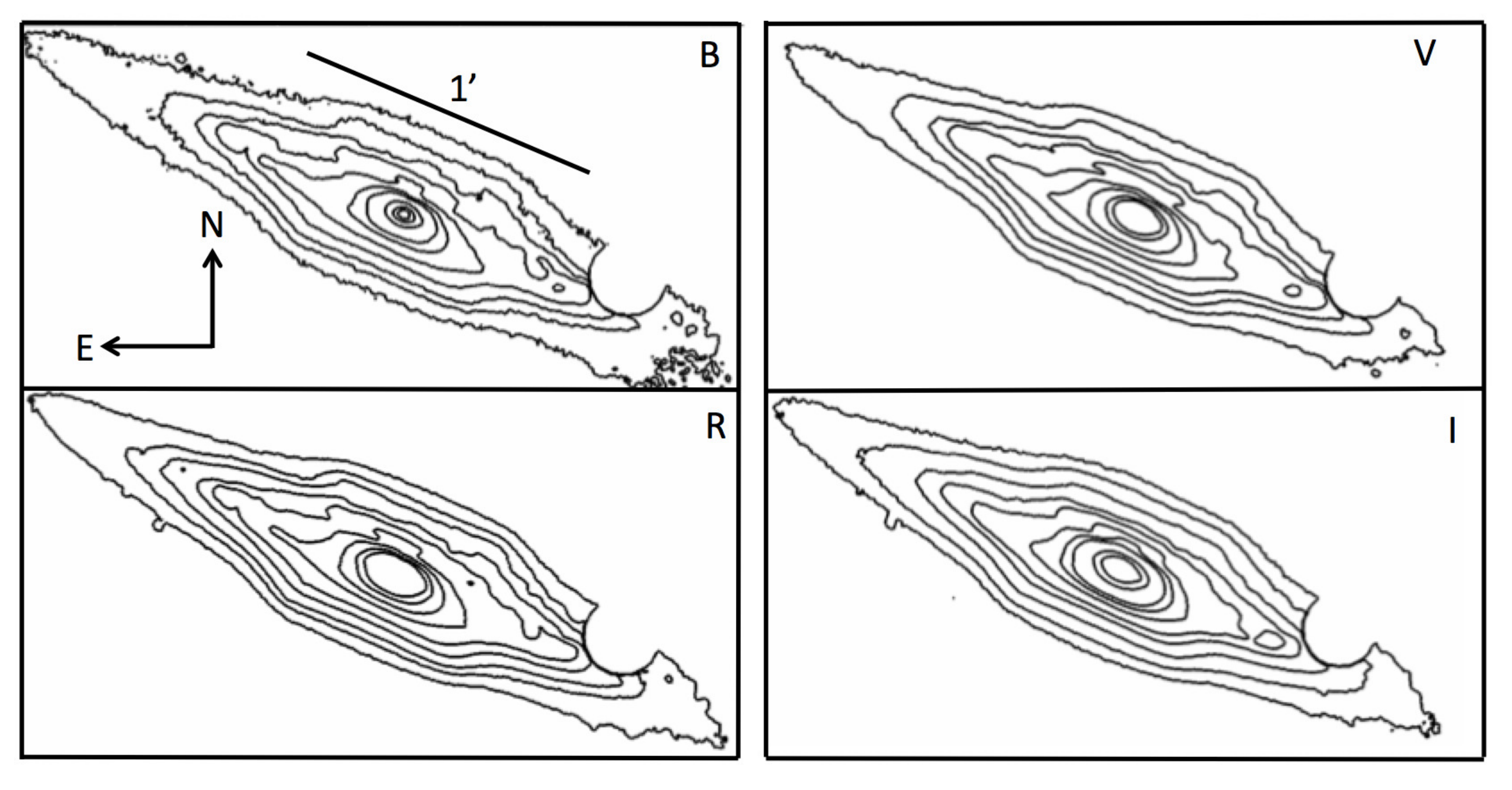}\\
	\caption{Contours of the centre of NGC 2654 in each of the four bands. The contour levels are at similar relative levels for each band. From these plots, the bulge of NGC 2654 is slightly peanut-shaped, most clearly seen as the pinch in the isophotes as they cross the minor axis. There is significant dust along the northern side of the galaxy.}
	\label{n2654_contour}
\end{figure*}

\authorfix{We begin our secondary photometric analysis of NGC 2654 by examining contour plots of the galaxy centre, shown in Figure \ref{n2654_contour}. Following the definitions of boxy/peanut shaped bulges in \citet{lutticke2000a} and \citet{kuzio2009}, we find that the bulge of NGC 2654 is subtly peanut-shaped. This can be most clearly seen in the pinched isophotes along the minor axis of the galaxy in the \textit{I}-band plot. We note, however, that the pinching in the isophotes is very slight and that the bulge is also quite boxy (isophotes remain parallel to the major axis as they cross the minor axis). A strongly peanut-shaped bulge indicates a bar is aligned side-on to our line of sight, whereas a boxy bulge is due to a bar at an intermediate angle between side-on and end-on \citep{kuzio2009}. Because we find the bulge in NGC 2654 to be both boxy and peanut-shaped, we infer that the bar is aligned slightly off of the disc major axis.}

\begin{figure*}
	\centering
	\includegraphics[scale=0.7]{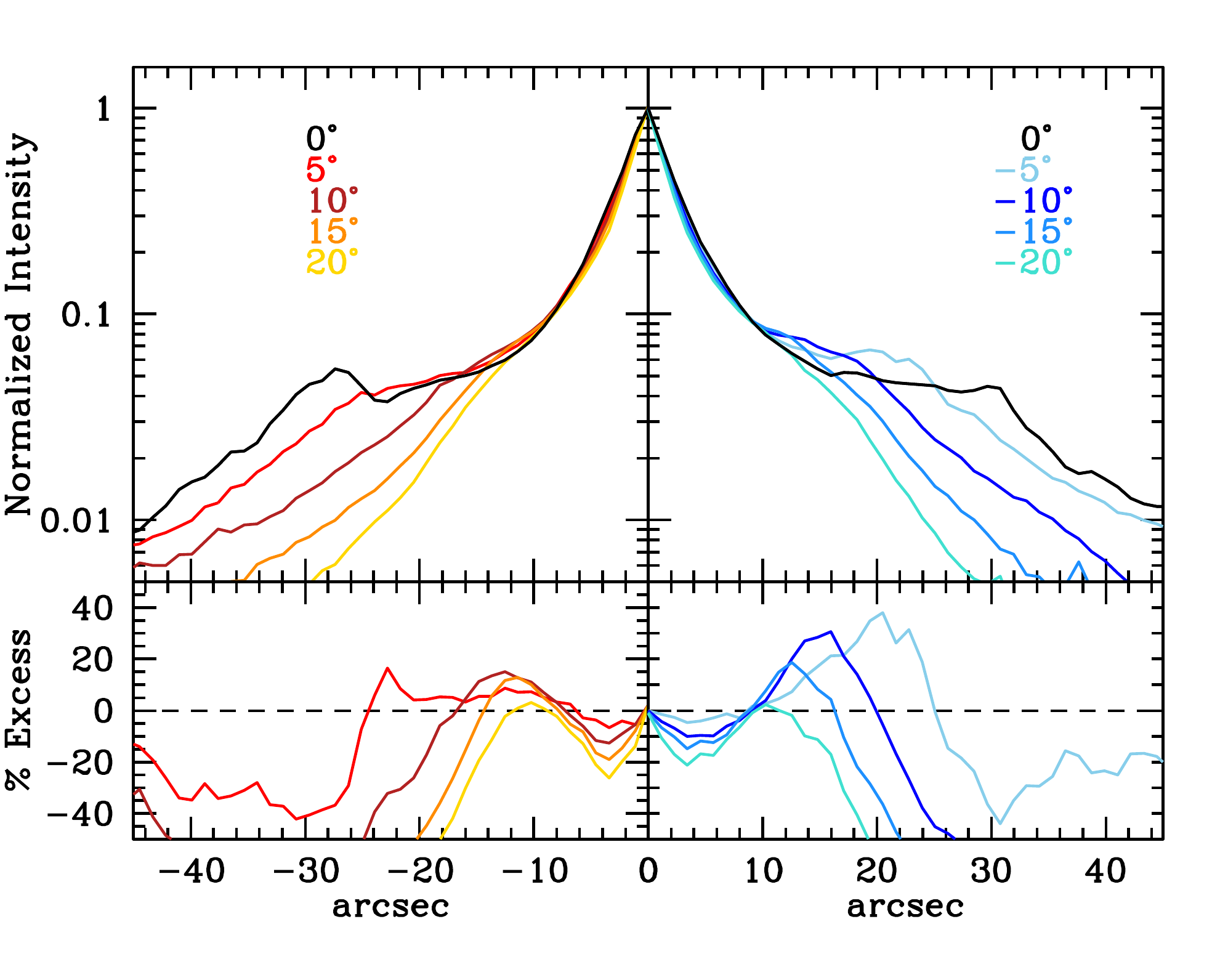} \\
	\includegraphics[scale=0.47]{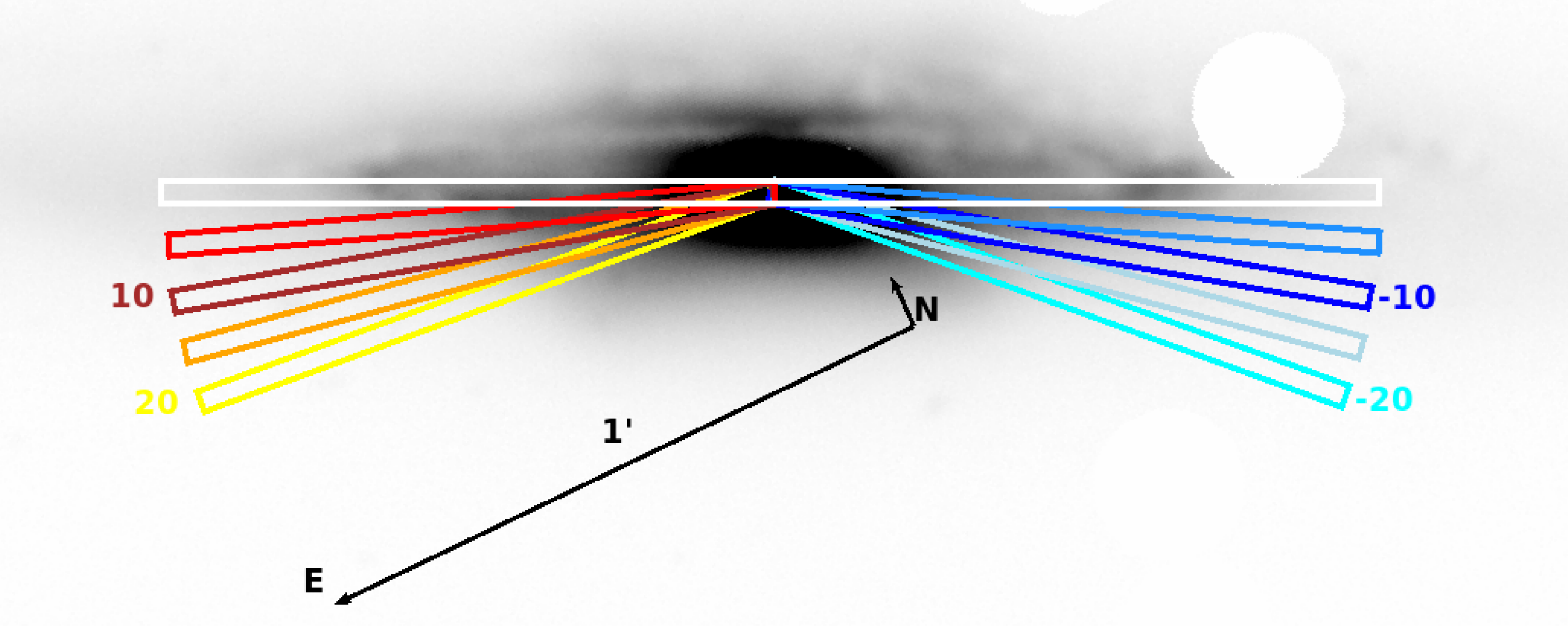}
	\caption{Angled slit photometry for NGC 2654. The top panels show the normalized intensity profiles along slits at angles to the major axis. Each slit is normalized to the value of the major axis in order to show any enhancements present in the various slits. The plots below the intensity profiles show the percent difference between the slit and major axis (i.e. $\frac{slit - major}{major}$). The extra light present in the -5\degr\ and -10\degr\ slits from 10$\arcsec$ to 25$\arcsec$ indicate that the bar is angled away from the disc major axis. The bottom panel shows the on-sky projection of the slits, coloured and labeled to match the top panel (major axis has been coloured white for easier viewing). Arrows show north and east. East arrow is one arcminute in length.}
	\label{n2654_angle_slit}
\end{figure*}

\authorfix{In order to more accurately determine the bar angle, we plot the intensity profiles along 5$\arcsec \times$100$\arcsec$ slits that are angled with respect to the major axis (see Figure \ref{n2654_angle_slit}). Given the dust on the north-western side of the galaxy (see Figures \ref{gal_pics} and \ref{n2654_contour}), we only plot the results for the eastern side. Slits with positive angles are on the NE side of the galaxy and slits with negative angles are on the SW side. Each profile has been normalized with respect to the major axis (black) in order to detect any enhancements that may be present in the slits. If a bar is present in NGC 2654 that is not aligned with respect to the disc major axis, it should leave a distinct signature in either, but not both, of the top panels of Figure \ref{n2654_angle_slit}.}

\authorfix{Specifically, we should see a light profile along a specific slit that exceeds that of the major axis on one side of the galaxy centre. In order to more clearly see any enhancement, we plot the \% Excess with respect to the major axis in the bottom portions of the top panels of Figure \ref{n2654_angle_slit}. If there is an excess in one of the angled slits due to the bar, the \% Excess will be positive. More specifically, the \% Excess is given by:
\begin{equation}
	\% {\rm Excess} (r) = \frac{{\rm Slit}(r) - {\rm Major}(r)}{{\rm Major}(r)}
\end{equation}}

\authorfix{We find a light excess of roughly 20\%-30\% in the -5\degr\ and -10\degr\ slits (top right panel of Fig.~\ref{n2654_angle_slit}). This indicates that the bar is slightly offset from the major axis. Based on a P.A.$_{\mathrm{disc}}$ of $\sim$65\degr, this means the bar is at a position angle between 55\degr\ and 60\degr. This offset from the major axis is slightly larger than that found by \texttt{DiskFit} (see Table~\ref{2654_table}).}

\authorfix{Looking at the top-panel of Fig.~\ref{n2654_angle_slit}, there also appears to be a light excess in the positive 5\degr\ and 10\degr\ slits, suggesting that the feature found by this method is symmetrical. However, the degree to which there is an excess in these slits is much less than what is found in the top-right panel. Therefore, it is likely that this method found the symmetrical bulge in both sides of the galaxy, contributing to an excess of $\sim$10\%, but also found the bar in the -5\degr\ and -10\degr\ slits.}

\subsubsection{Kinematics}
\label{sec:n2654kine}

\begin{figure*}
	\center
    	\hskip 10mm \includegraphics[width=0.46\textwidth]{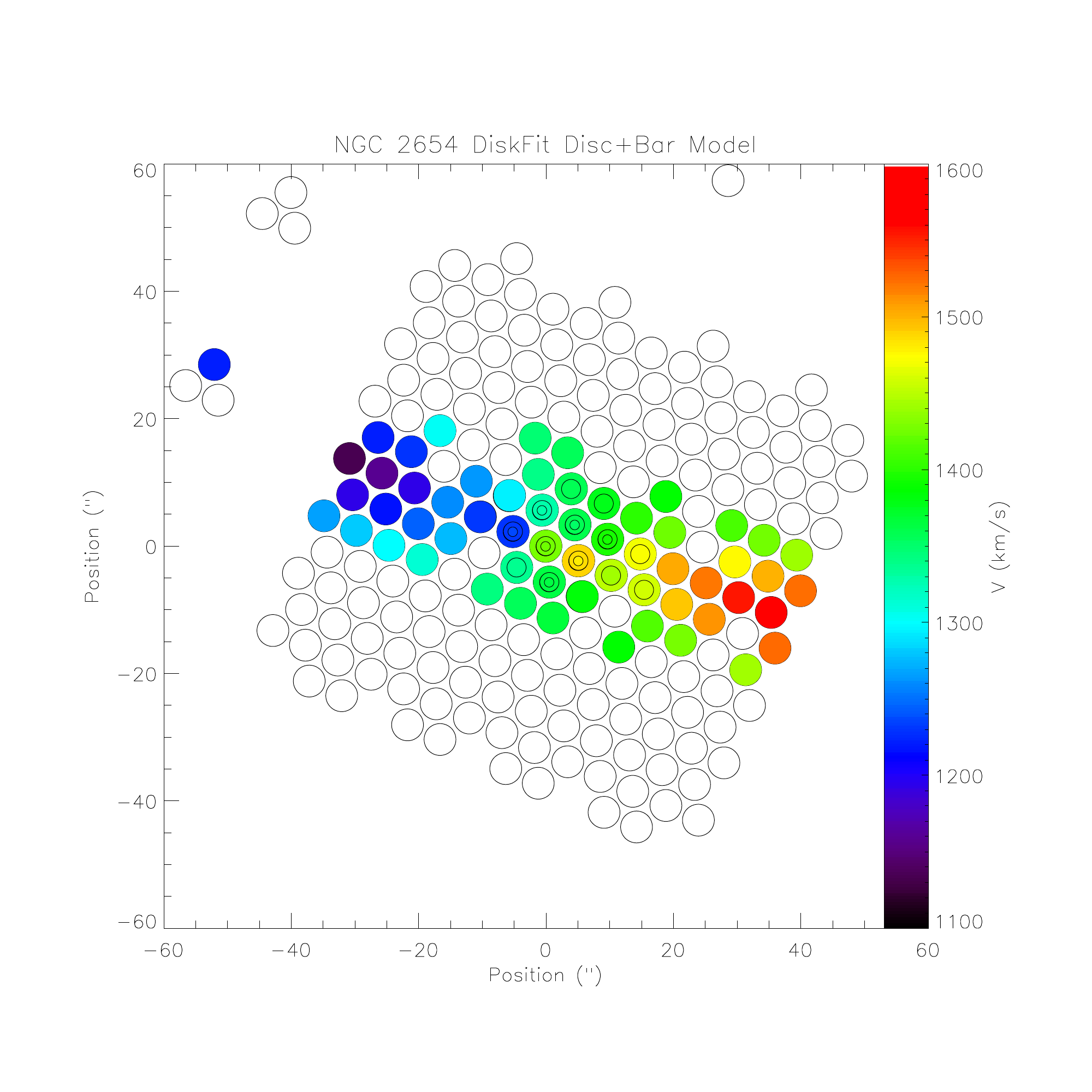} \hskip 2mm \includegraphics[width=0.46\textwidth]{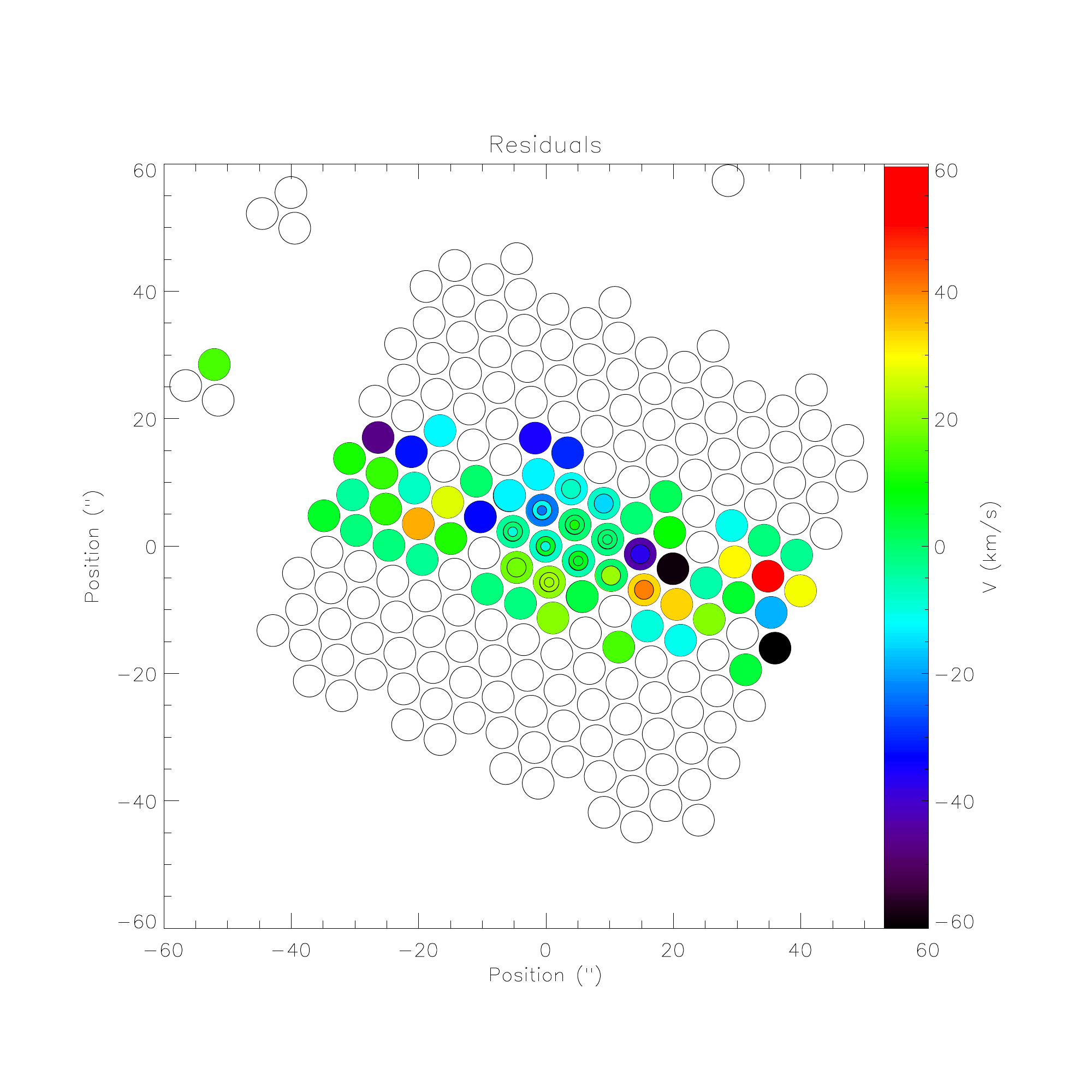}\\
	\hskip 2mm
    	\includegraphics[width=0.45\textwidth]{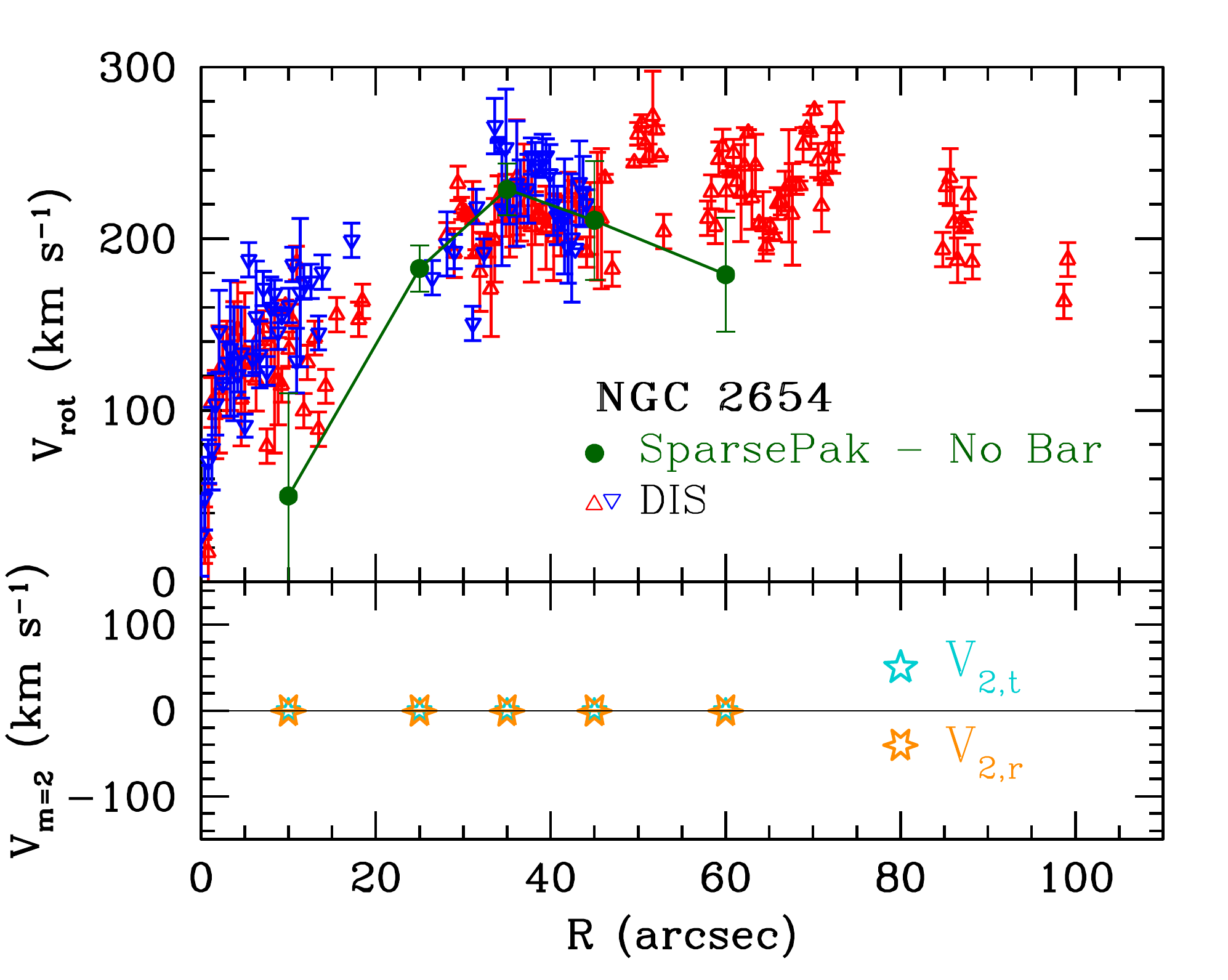} \hskip 2mm \includegraphics[width=0.45\textwidth]{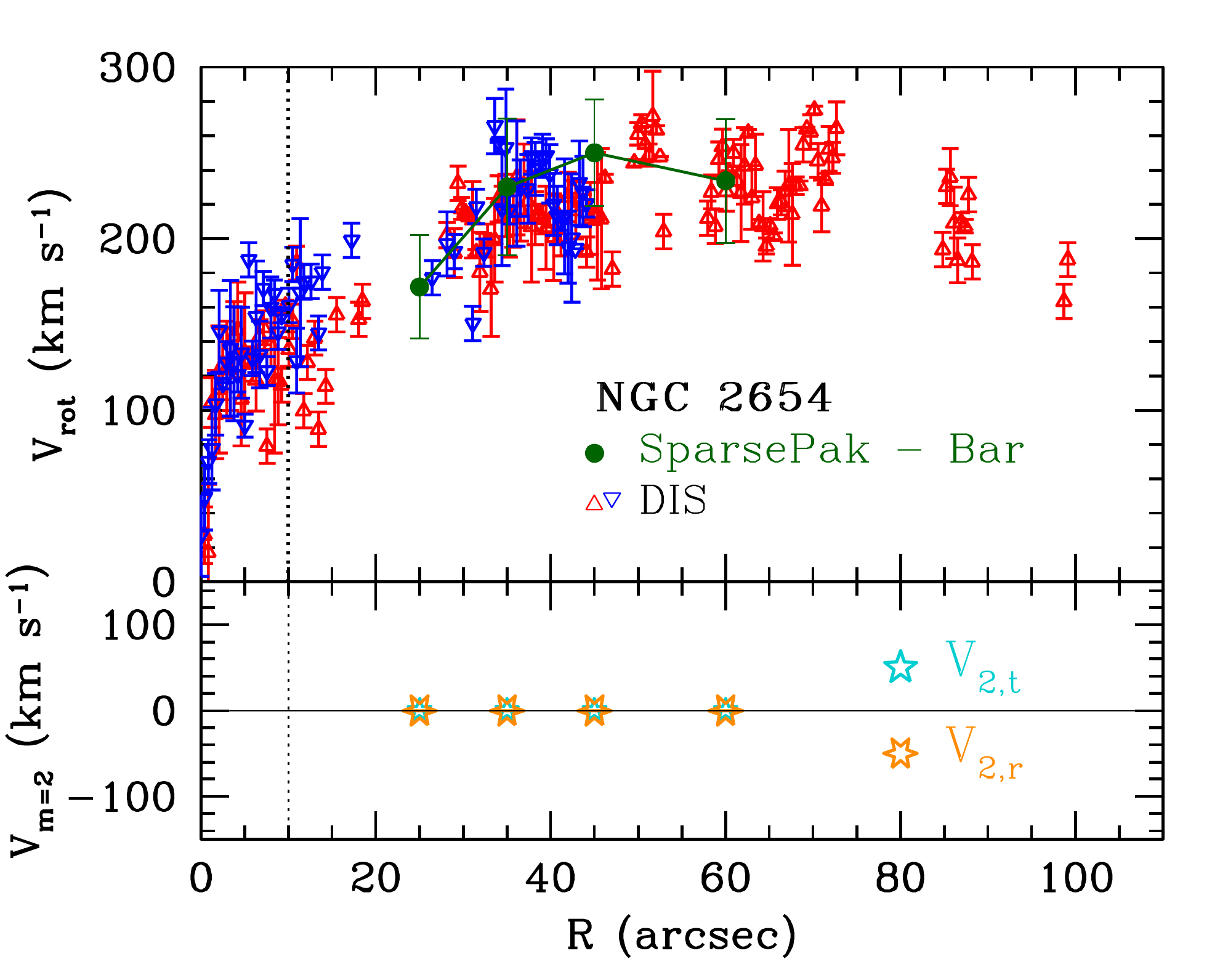}
  	\caption{Same as Fig.~\ref{n6674_kinematics} but for NGC~2654.}
  	\label{n2654_kinematics}
\end{figure*}

Our best-fitting \texttt{DiskFit} kinematic model of the SparsePak velocity field for NGC~2654 contains a disc and a bar; the model and residuals are shown in the top panels of Fig.~\ref{n2654_kinematics}. We find a disc position angle of 250\degr $\pm$ 3.26\degr, an inclination of 82.97\degr $\pm$ 1.69\degr, and a systemic velocity of 1356.38 $\pm$ 6.08 km s$^{-1}$. We find the bar to be roughly orthogonal to the major axis of the disc at a position angle of 156.84\degr$\pm$21.50\degr. Both the P.A.$_{\mathrm{disc}}$ and V$_{\mathrm{sys}}$ are roughly consistent with our DIS values.

The residuals for the disc+bar kinematic model are shown in the top right panel of Fig.~\ref{n2654_kinematics}. The model does a good job of matching the observed velocity field in the central region, as typical residuals are between $\pm$10 km s$^{-1}$. However, there are a number of poorly matched fibres, specifically north and directly west of the central region. The dark blue fibres (residuals of $\sim$ -40 km s$^{-1}$) to the north are located directly in the X-shape, and are likely members of the bar. The poorly matched fibres to the sides of the centre are likely caused by a lack of extended radial coverage of the galaxy. With the exception of the lone sky fibre located at -60$\arcsec$\ from the centre, our velocity field extends to only 40$\arcsec$. It is possible that this, in addition to the high inclination and poor spatial resolution from the large 5$\arcsec$ fibres, is causing the poor match to these fibres.

The SparsePak (filled blue circles) and DIS (open red triangles) rotation curves are shown in the bottom two panels of Fig.~\ref{n2654_kinematics}.  The SparsePak rotation curve in the left panel is from the disc-only \texttt{DiskFit} model and in the right panel is from the disc+bar model; the DIS rotation curve assumes only circular motion in both panels.

The DIS rotation curve has much higher spatial sampling than the SparsePak rotation curve(s) and extends almost twice as far as the IFU data. It shows a gentle rise in velocities up to $\sim$250 km s$^{-1}$ at 70$\arcsec$ before beginning to decline around 80$\arcsec$. This is consistent with the HI observations from \citet{noordermeer2007}. They find a much more extended rotation curve (out to nearly 400$\arcsec$) that also shows the decline around 80$\arcsec$, followed by a flattening at roughly 190 km s$^{-1}$. \authorfix{We do not find any significant asymmetry in the rotation curve for this galaxy.}

The SparsePak rotation curves look very different depending on whether or not a bar has been included in the \texttt{DiskFit} model.  Without a bar (bottom left panel of Fig.~\ref{n2654_kinematics}), the derived rotation curve is well below the observed DIS rotation curve at small radii, rapidly rises to overlap with the DIS rotation curve at $\sim$35$\arcsec$, and then abruptly drops again, falling below the long-slit rotation curve.  The derived SparsePak rotation curve from the disc+bar model, however, is in excellent agreement with the DIS rotation curve from 25$\arcsec$ out to 60$\arcsec$, even following the same gentle rise and fall of the velocities (see the bottom right panel of Fig.~\ref{n2654_kinematics}).

Though the length of the bar is indicated by the vertical dashed line at 10$\arcsec$ in the lower right panel of Fig.~\ref{n2654_kinematics}, we do not plot a rotation curve \textit{point} (or the values of the non-circular motions) at this radius.  While there is a significant improvement in the overall shape of the rotation curve when a bar is included, the error on the innermost ring is huge.  This is likely a combination of the very small number of data points in this first ring, only 9 compared with $\sim$20 in the other rings, and because the large SparsePak fibres are each encompassing a wide range of velocities.  Despite not being able to constrain the rotation and non-circular motions in the innermost regions of NGC~2654, we are confident that a bar must be present; without one, the rotation curve at intermediate and large radii is a poor match to long-slit and HI observations.

\subsubsection{Comparison of Photometric and Kinematic Models}
\label{sec:n2654compare}

At first glance, the P.A.$_{\mathrm{disc}}$\ from the photometry and kinematics appear to be inconsistent (Phot.~$\sim$~65\degr\ vs Kinem.~$\sim$~250\degr) but this is simply a result of convention. While it is customary for the kinematic position angle to be defined as the angle measured from north through east to the redshifted side of the galaxy, in photometry the position angle is simply measured from north through east to the first side of the galaxy that is encountered.  It is thus possible that kinematic and photometric position angles can be offset by 180\degr, as is the case with NGC~2654. 

The inclination found by the kinematic model ($\sim$83\degr) is slightly higher than that of the photometry ($\sim$79\degr). This is likely due to the fact that the kinematic data do not extend as far radially as the photometry (roughly 60$\arcsec$ compared to over 100$\arcsec$). 

While the parameters of the disc are well-matched between the photometry and kinematics, there is a striking disagreement over the bar position angle.  \authorfix{both} photometric results suggest a bar that is nearly parallel with the galaxy major axis, while the kinematics indicate that the bar is offset from the major axis by roughly 90\degr. Because NGC~2654 is quite inclined, this means that the photometric bar extends across our line-of-sight while the kinematic bar is being viewed closer to end-on.

As previously discussed in Section~\ref{sec:intro}, X-shaped or peanut-shaped bulges are indicative of bars seen edge-on rather than end-on. \authorfix{The `slit' method performed on NGC 2654 supports the results from \texttt{DiskFit} that the bar is nearly parallel to the disc major axis.} This would imply that our photometric results are self-consistent and likely more accurate than what is determined from the kinematics. This is further reinforced by the location of the bar ansae \authorfix{(or `ends')} seen in \citet{buta2015} which are found to be very close to the major axis of the galaxy. However, the bar ansae appear to be located at an angle slightly greater than the disc position angle ($\sim$68\degr), different from our photometric results. 

Given this discrepancy in the bar position angle, we re-ran the kinematic \texttt{DiskFit} models with the bar position angle held fixed at the photometric value to see how the rotation curve was affected. We find that this does not fix the inner rotational velocity, nor its very large error, in our SparsePak model rotation curve. In fact, there is no noticeable effect on the entire rotation curve between letting the bar position angle vary freely and holding it fixed. This implies that we cannot accurately constrain the bar position angle from the kinematic data.

Due to the combined difficulties resulting from a limited number of data and a rather high inclination, \texttt{DiskFit} is at the limit of what it can accurately model for the kinematics of NGC~2654. To further push our modeling and examine the effects of very high inclination, we last look at NGC~5746. 

\subsection{NGC~5746}
\label{sec:n5746}

The photometric analysis of NGC~5746 is shown in Figs.~\ref{n5746_contour} and \ref{n5746_angle_slit} and the results are presented in Table~\ref{5746_table}. The results of the best-fitting kinematic \texttt{DiskFit} model and DIS long-slit rotation curve are presented in Fig.~\ref{n5746_kinematics} and Table~\ref{5746_table}.    

\begin{table}
	\centering
	\caption{Best-fitting photometric and kinematic parameters for NGC 5746. See Section \ref{sec:n5746} for details.}
	\label{5746_table}
	\begin{tabular}{rccc}
		\hline
		Parameter		& Slit Photometry & SparsePak		&DIS\\
		\hline
	P.A.$_{\mathrm{disc}}$ (\degr) & ... & -9.13 $\pm$ 0.19 &  -10 \\
	\textit{i}$_{\mathrm{disc}}$ (\degr) & ... & 84.23 $\pm$ 0.25 & 82 \\
	$V_{\mathrm{sys}}$ (km s$^{-1}$) & ... & 1711.49 $\pm$ 3.04 & 1710 $\pm$ 15 \\
	\\
	P.A.$_{\mathrm{bar}}$ (\degr) & -5 & -9.85 $\pm$ 5.59 & ... \\
	$R_{\mathrm{bar}}$ (\arcsec) & 36 & 35 & ...\\
	\hline
	\end{tabular}
\end{table}

\subsubsection{Photometry}
\label{sec:n5746phot}

NGC~5746 is a very highly-inclined galaxy with a massive dust lane (see the lower left panel of Fig.~\ref{gal_pics}). This dust drastically separates the central bulge in two and greatly complicates any photometric decomposition.   

\texttt{DiskFit} struggles to find the galaxy centre and, because of the decrease in light on one side of the galaxy from the dust lane, attempts to fit a disc (to only one half of the galaxy) that is even more edge-on than the galaxy already is. Because of this, we were forced to handle the photometry for NGC~5746 differently than the other three galaxies in our sample \authorfix{and not use \texttt{DiskFit}. Instead, we use the same slit method we used on NGC 2654 in Section~\ref{sec:n2654slit}.}

Like NGC~2654, NGC~5746 has a boxy/peanut-shaped bulge that indirectly indicates that the galaxy contains a bar.  \authorfix{This technique will provide both} a qualitative and quantitative measure of the bar position angle and length. Unlike for the other galaxies in our sample, however, we will not be able to quantify the galaxy bulge or the amount of light coming from the various galaxy components, nor can we determine the disc inclination \authorfix{due to the inability of \texttt{DiskFit} to model the photometry of this galaxy.} 

\begin{figure*}
	\includegraphics[scale=0.65]{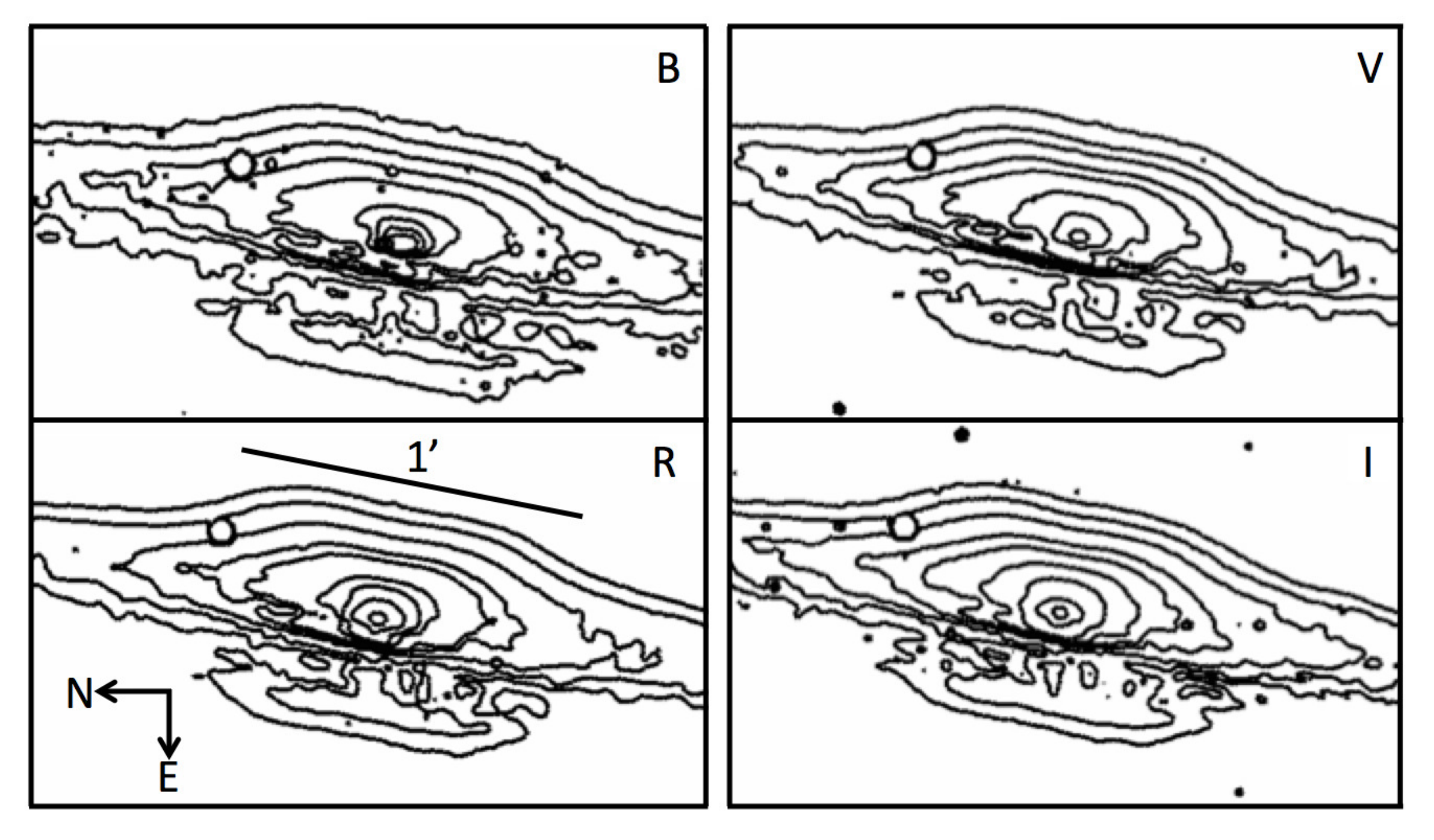}\\
	\caption{Contours of the centre of NGC 5746 in each of the four bands. The contour levels are at similar relative levels for each band. From these plots, the bulge of NGC 5746 is slightly peanut-shaped, most clearly seen as the pinch in the isophotes as they cross the minor axis. There is significant dust along the eastern side of the galaxy.}
	\label{n5746_contour}
\end{figure*}

We begin our photometric analysis of NGC 5746 by examining contour plots of the galaxy centre, shown in Figure \ref{n5746_contour}. \authorfix{As with NGC 2654, here we find the bulge in NGC 5746 to be slightly peanut-shaped. As previously discussed in Sec.~\ref{sec:n2654slit}, this indicates that the bar in NGC 5746 is slightly offset from the major axis of the galaxy.}

\begin{figure*}
	\centering
	\includegraphics[scale=0.7]{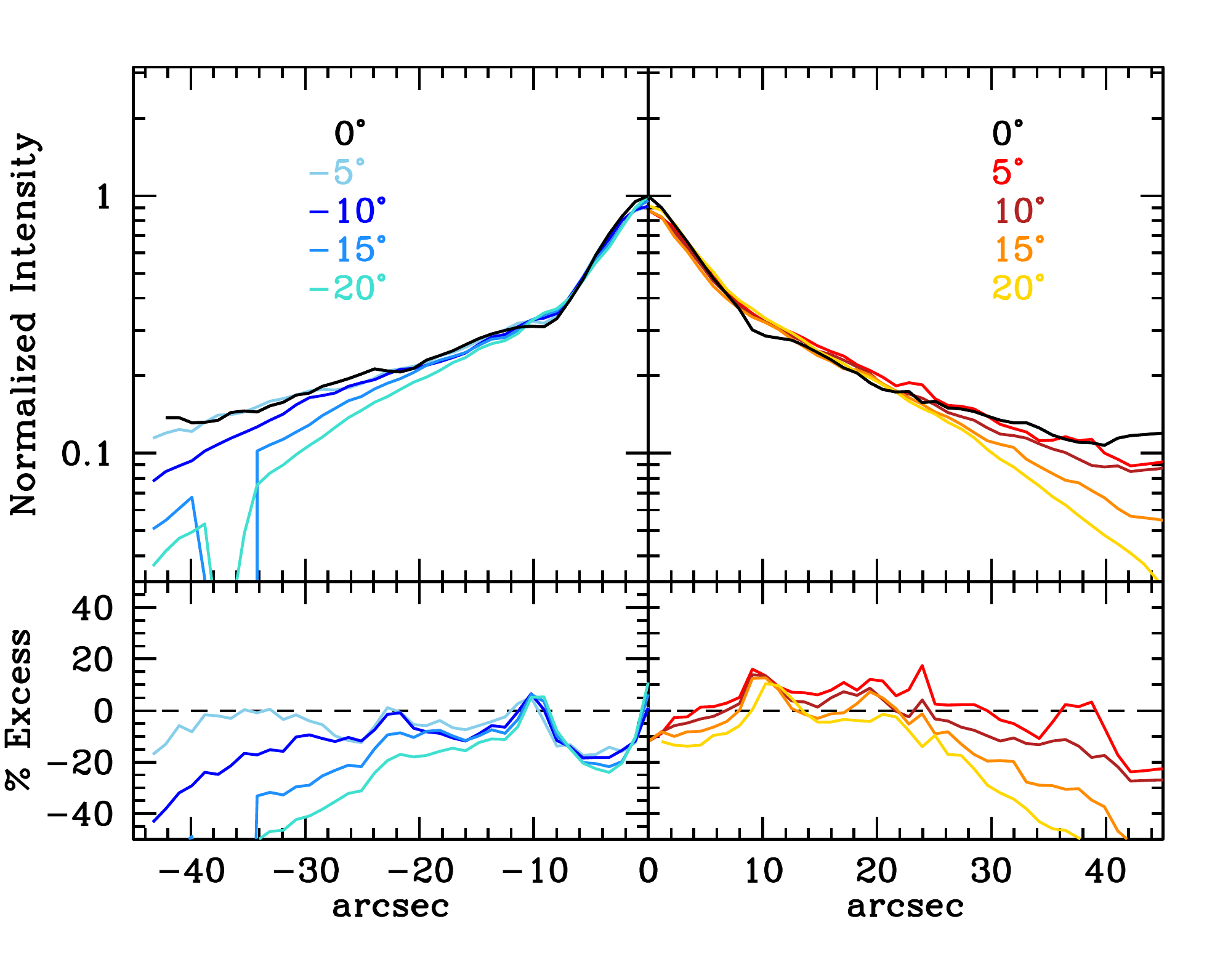} \\
	\includegraphics[scale=0.55]{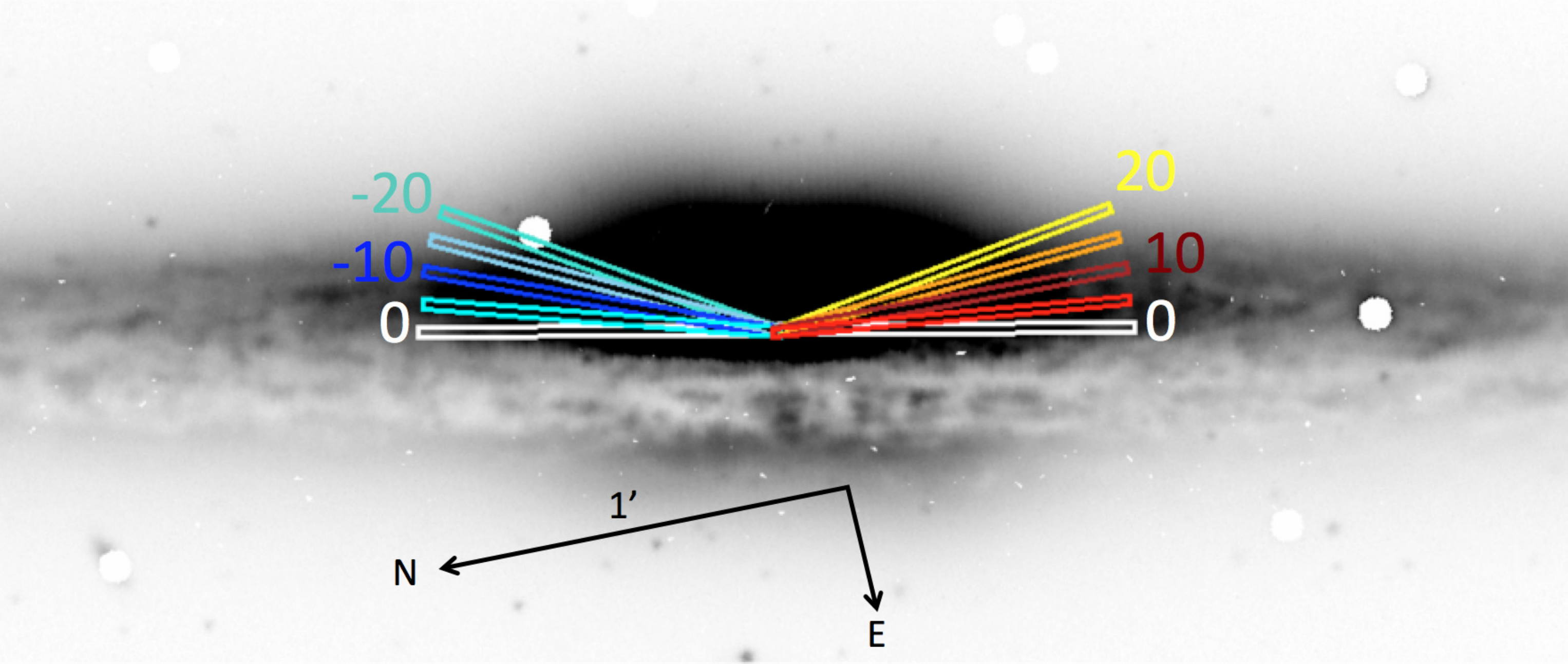}
	\caption{Angled slit photometry for NGC 5746. Same colours and format as Fig.~\ref{n2654_angle_slit}}
	\label{n5746_angle_slit}
\end{figure*}

\authorfix{Next, we plot the intensity profiles along 5$\arcsec \times$100$\arcsec$ slits that are angled with respect to the major axis (see Figure \ref{n5746_angle_slit}). Similar to Section~\ref{sec:n2654slit}, we only plot the results for the western side due the dust lane on the eastern side (see Figures \ref{gal_pics} and \ref{n5746_contour})}. Slits with positive angles are on the SW side of the galaxy and slits with negative angles are on the NW side.

\authorfix{Following the light \% Excess equation described in Section~\ref{sec:n2654slit}}, we find a light enhancement in the 5\degr\ (red) and 10\degr (brown) slits, from $\sim$7$\arcsec$ to $\sim$30$\arcsec$, most easily seen in the \% Excess plot. This enhancement is not visible in the top left panel, suggesting this is not due to a symmetric feature in the galaxy. This is consistent with a bar close to, but not quite, aligned with the galaxy major axis, offset by roughly 5\degr\ to 10\degr (northeast-southwest direction). From this plot, we can also approximate the length of the bar by identifying where the positive excess ends. Since the bar is so close to the major axis of NGC 5746, projection effects will be negligible. Therefore, we find the bar is roughly 30$\arcsec$ in length.

Based on our photometry analysis, we cannot make any quantitative claims about the bar in NGC 5746 to the degree that we could from \texttt{DiskFit} modeling. However, we can claim that there is a bar in NGC 5746 that is oriented in a north to south direction, which is consistent with a slightly peanut-shaped bulge. \authorfix{We are also confident in this result due to the success this method had with matching the \texttt{DiskFit} results for NGC 2654.}

\subsubsection{Kinematics}
\label{sec:n5746kine}

\begin{figure*}
	\center
    	\hskip 10mm \includegraphics[width=0.46\textwidth]{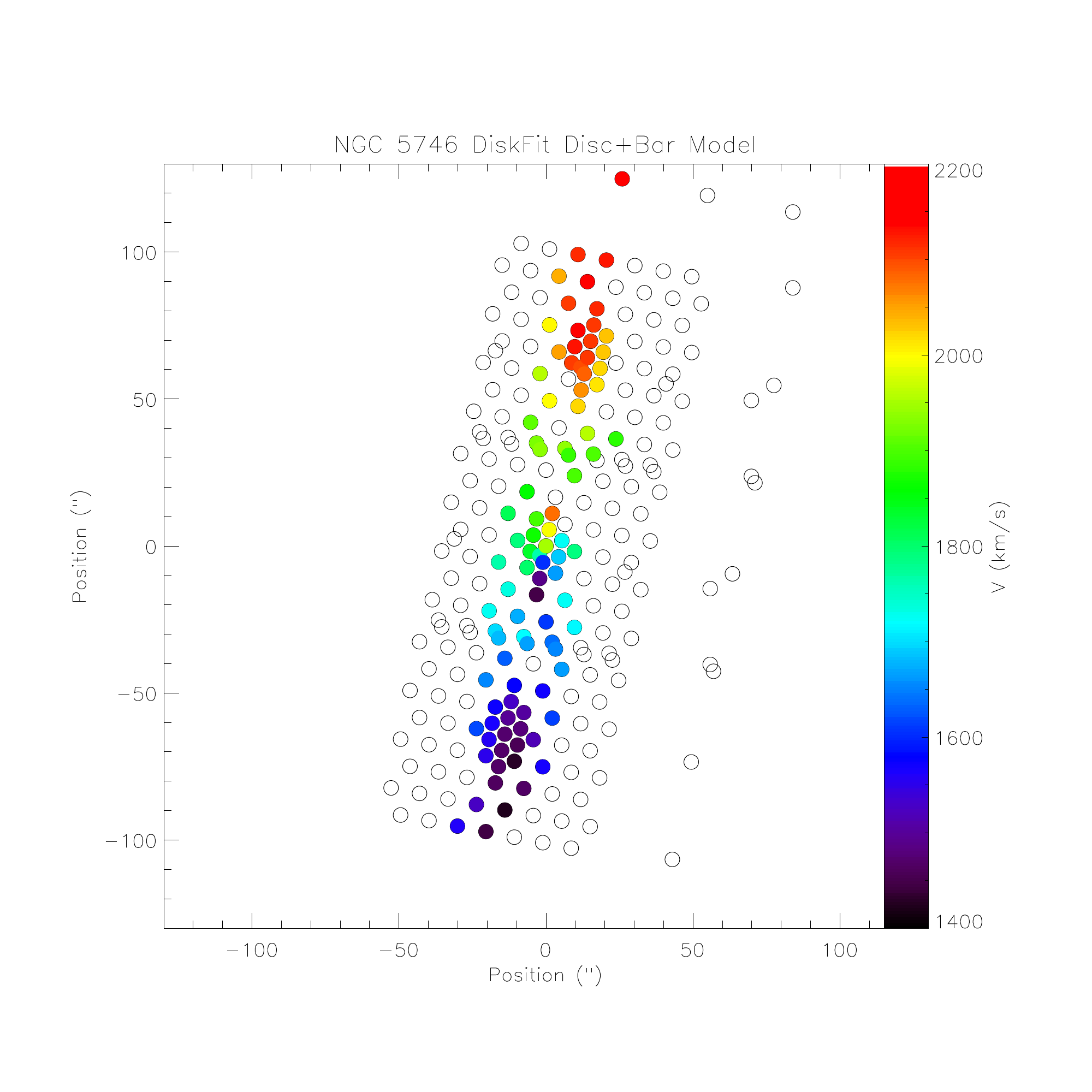} \hskip 2mm \includegraphics[width=0.46\textwidth]{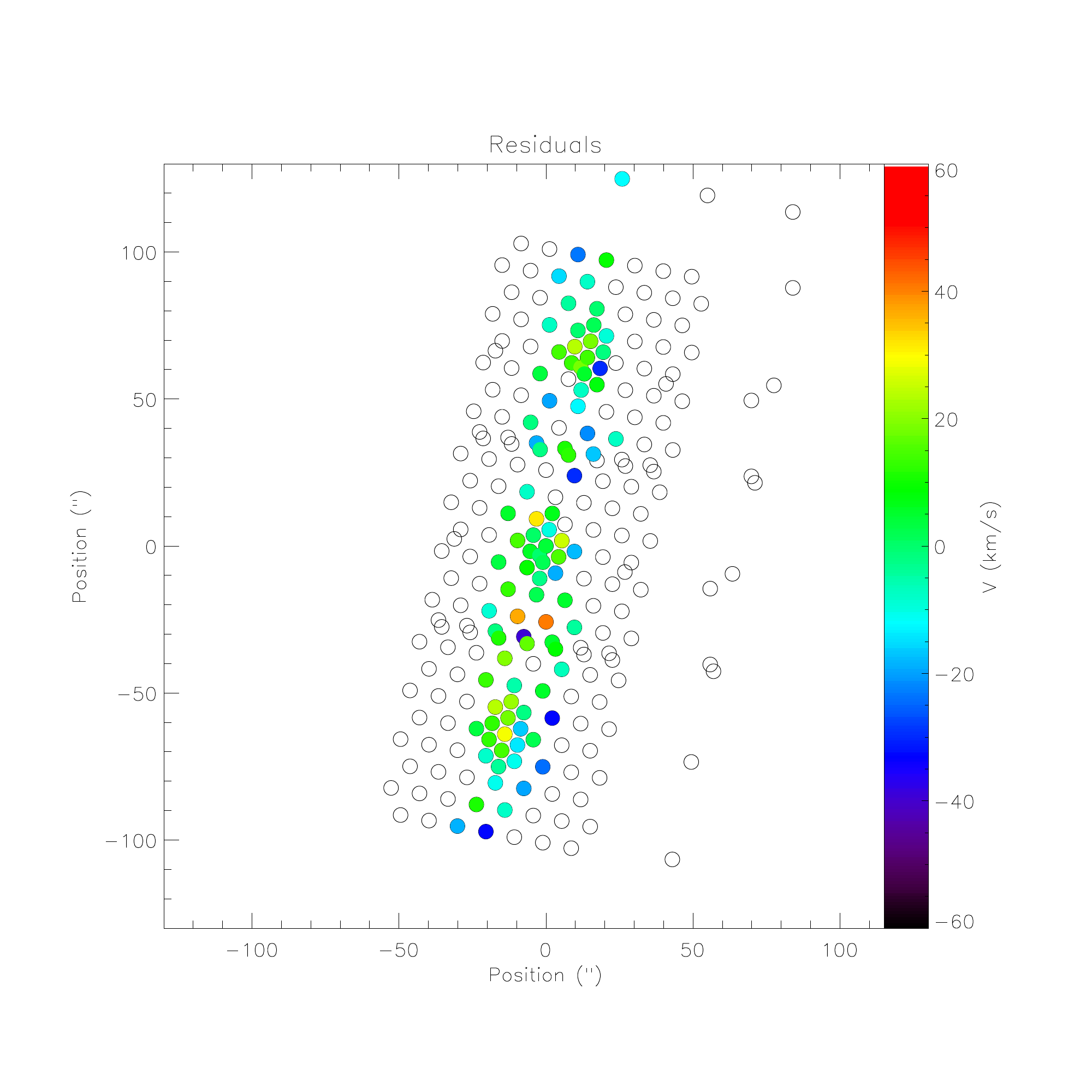}\\
	\hskip 2mm
    	\includegraphics[width=0.45\textwidth]{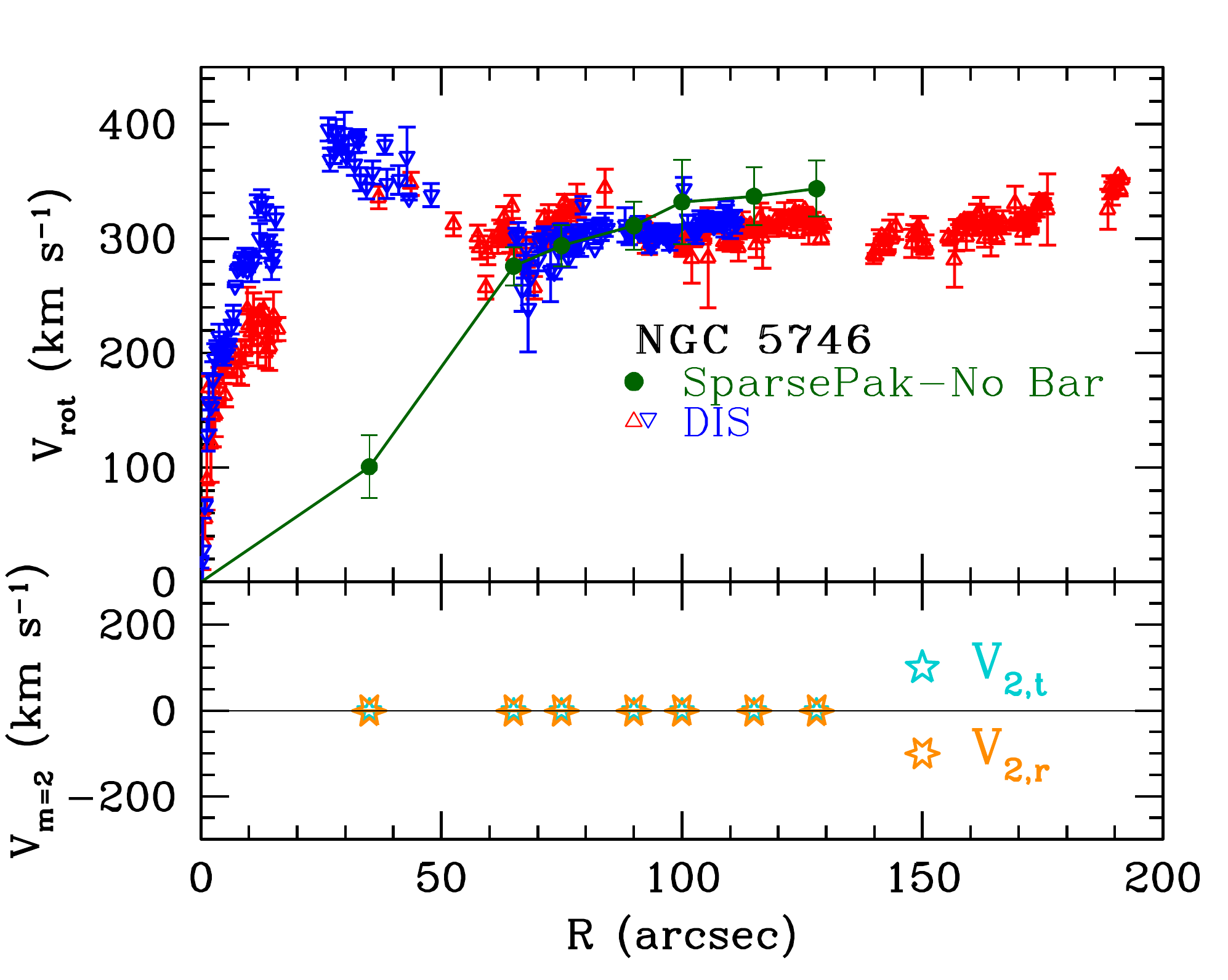} \hskip 2mm \includegraphics[width=0.45\textwidth]{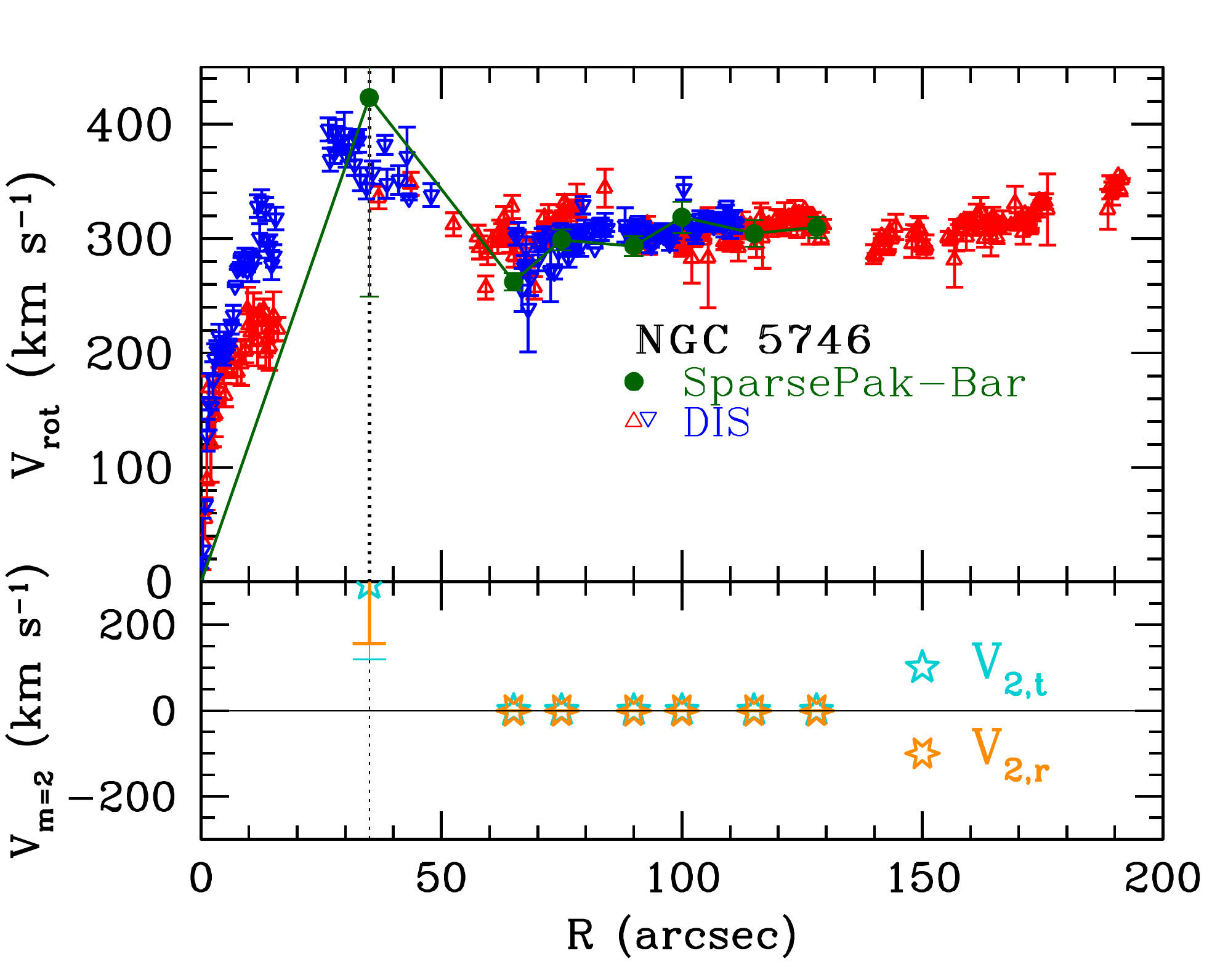}
  	\caption{Same as Fig.~\ref{n6674_kinematics} but for NGC~5746.}
  	\label{n5746_kinematics}
\end{figure*}

\texttt{DiskFit} has an easier time modeling the kinematic data for NGC~5746 than the photometry.  We determine that the best-fitting model of the SparsePak velocity field contains both a disc and a bar. We find a P.A.$_{\mathrm{disc}}$ of -9.13\degr $\pm$ 0.19, an inclination of 84.23\degr $\pm$ 0.25, a P.A.$_{\mathrm{bar}}$ of -9.85\degr $\pm$ 5.59, a bar radius of 35$\arcsec$, and a systemic velocity of 1711.49 $\pm$ 3.04 km s$^{-1}$.  The best-fitting kinematic model and residuals for NGC~5746 are shown in the top panels of Fig.~\ref{n5746_kinematics}. 

While there is a smattering of fibres with residuals larger than $\pm$30 km s$^{-1}$, they are not concentrated to any particular region of the velocity field. This suggests that there is not a particular region of the galaxy where the \texttt{DiskFit} disc+bar model outright fails.  This is not the case in the disc-only model; the residuals show that the model fails in the inner region of the galaxy, with typical residuals of $\sim$60 km s$^{-1}$ or higher. The outer regions of the galaxy fair better, although they are still worse than the disc+bar model: some areas have groups of fibres with residuals greater than 35 km s$^{-1}$.

In the bottom panels of Fig.~\ref{n5746_kinematics} we present the DIS long-slit (open red triangles) SparsePak (filled blue circles) rotation curves.  The SparsePak rotation curve in the left panel is for the \texttt{DiskFit} disc-only model and in the right panel is for the disc+bar model.

The DIS rotation curve is well-sampled and extends to nearly 200$\arcsec$.  It rises to $\sim$400 km s$^{-1}$ in the inner 30$\arcsec$, declines to $\sim$300 km s$^{-1}$ around 60$\arcsec$, and maintains this velocity for the extent of the data.  We find this rotation curve to be consistent with the H$\alpha$ rotation curve reported by  \citet{keel1996}. \authorfix{We find a somewhat significant asymmetry in the rotation curve for NGC 5746 in the inner 20$\arcsec$. In this region the approaching and receding sides of the rotation curve do not match and are offset from each other. However, the two sides agree quite well outside of this range.}

When measuring the emission lines in the long-slit spectra, we found that there is a very faint braiding, or figure-of-eight \citep{kuijken1995,kuzio2009}, structure in the H$\alpha$ and [NII] emission along the slit. This feature extends roughly 40$\arcsec$ from the galaxy centre and becomes roughly parallelogram in shape past 20$\arcsec$. This figure-of-eight pattern is known to be an observational signature of a bar hidden in peanut/boxy shaped bulges in edge-on galaxies. The characteristics of this structure are directly tied to the orientation of the bar relative to our line-of-sight. 

For example, if the bar is viewed end-on, the feature in the observed velocities will be very distinct, with two, separate components. One component will trace the majority of the gas (and stars) following the circular motions around the galaxy, and the other will follow orbits along the bar itself. These orbits will be seen as a straight, diagonal feature in the observed velocities, viewed as ``slower'' motions when compared to the circular motions (i.e.\ redshifted velocities on the blueshifted side and vice versa). When looking at a position-velocity diagram (PVD), these two components thus form a figure-of-eight pattern, intersecting in the galaxy centre, with the circular motions being more luminous than the bar motions.

For bar orientations not end-on, the two components in the observed velocities will not possess as clear of an intersection as previously mentioned. Instead of a figure-of-eight pattern, a parallelogram feature is seen. At bar orientations viewed side-on, there is difficulty in seeing a distinct enough feature in the observed velocities to separate out the bar motions.

We find a faint figure-of-eight pattern in our observed velocities, and infer a bar orientation that is at some intermediate angle between end-on and side-on, closer to end-on. The circular motions in the PVD are much more luminous than these other features, so we only measured these motions.

As has been the case for all 3 of our other galaxies, the shape of the SparsePak rotation curve for NGC~5746 changes dramatically depending on whether or not a bar is included in the \texttt{DiskFit} model. The disc-only rotation curve (bottom left panel of Fig.~\ref{n5746_kinematics}) does not match the DIS rotation curve well at all, failing spectacularly at small radii with velocities well below what is observed by DIS and exceeding the DIS velocities at radii larger than 100$\arcsec$. When a bar is included, however, the SparsePak rotation curve matches the DIS rotation curve extremely well (bottom right panel of Fig.~\ref{n5746_kinematics}). 

Given the very large errorbar on the innermost point of the disc+bar rotation curve, it is tempting to dismiss it.  However, one must consider that this ring is includes all  SparsePak data from 0-65$\arcsec$ (a total of $\sim$30 fibres). This radial range includes the entire bulge of the galaxy, seen as the very large hump in the rotation curve from 0-50$\arcsec$.  Within this region, NGC~5746 reaches a very large maximum rotational velocity before falling and flattening off. It is therefore not surprising that the error bar on the inner radii is very large. The associated non-circular motions at this radius (lower portion of the bottom right panel of Fig.~\ref{n5746_kinematics}) are similarly quite large.  

Given the improvements in the rotation curve when a bar is included in NGC~5746, we are confident that the kinematic data support the presence of a bar. Our \texttt{DiskFit} model implies the bar is aligned nearly perfectly with the galaxy's major axis. This is consistent with a bar viewed close to edge-on and is in agreement with the X-shape visible in the photometry. However, this result is somewhat in conflict with our findings from our long-slit observations, which suggested a bar more closely oriented along our line of sight, rather than edge-on.

\subsubsection{Comparison of Photometric and Kinematic Models}
\label{sec:n5746compare}

We find that the bar properties found by both our photometric analysis and kinematic modeling are roughly consistent. Our kinematic \texttt{DiskFit} modeling suggests the bar is parallel with the major axis and our photometric analysis suggests that the bar is roughly 5 - 10\degr\ away from the disc major axis. While we do see a figure-of-eight feature in our PVD which suggests the bar is close to end-on, there is a parallelogram shape present as well. This is likely indicating that the bar is at some intermediate angle, though this is still somewhat in conflict with our photometric analysis and \texttt{DiskFit} kinematic modeling.

Thus, our picture of NGC~5746 is that of a highly inclined galaxy ($\textit{i}$ = 84\degr) with a bar that runs at an angle in the north-south direction. This galaxy is a good example of the troubles that can arise when attempting to derive galaxy and bar parameters from either photometry or kinematics.

\section{Summary and Conclusions}
\label{sec:summary}

In this paper, we have presented new photometric and kinematic observations for four spiral galaxies: NGC~2654, NGC~2841, NGC~5746, and NGC~6674. We obtained \textit{B}, \textit{V}, \textit{R}, and \textit{I}  images using the ARCTIC and SPICAM imagers, H$\alpha$ velocity fields using the SparsePak IFU, and long-slit rotation curves using DIS. 

\authorfix{We have used the \texttt{DiskFit} code to model this new photometric and kinematic data and to explore the limits of the \texttt{DiskFit} technique. We have found that for moderately-inclined spiral galaxies, \texttt{DiskFit} modeling paints a consistent photometric and kinematic picture of a galaxy and that the \texttt{DiskFit} models are also consistent with previous results in the literature. We have found that \texttt{DiskFit} has difficulty photometrically modeling highly-inclined ($\sim$84\degr) galaxies containing large amounts of dust. In these cases, studying the shape of the galaxy isophotes using the ``slit technique'' is more informative. Finally, we find that having both types of data is vital for confidently determining bar and bulge parameters when the target galaxy is highly inclined and/or the features are obscured.} 

\authorfix{Our specific photometric and kinematic results for the galaxies are as follows:} we find that three of our galaxies (NGC~6674, NGC~2841, and NGC~2654) are best described by photometric \texttt{DiskFit} models that include a disc, bulge, and bar. The models for each galaxy are consistent across all four optical bands. For our most inclined galaxy, NGC~5746, we had to rely on an alternate technique to constrain the properties of the galaxy.  We are able to measure the bar length and position angle with this method, but we are unable to parameterize the galaxy bulge.

The kinematic \texttt{DiskFit} models indicate that all four of our galaxies  exhibit non-circular motions to some degree. We find that the rotation curves derived from the velocity field data are most consistent with the long-slit observations and previous HI data when a bar is included in the \texttt{DiskFit} models.

The \texttt{DiskFit} models of three of our galaxies NGC 2654, NGC 5746, and NGC 6674 are consistent with previous morphological classifications, all being known barred galaxies. Based on our \texttt{DiskFit} modeling of both the photometric and kinematic data, we reclassify NGC~2841 as a barred galaxy. 

\section*{Acknowledgements}

\authorfix{We thank the referee for their comments that helped us to clarify and improve this paper.} We have used DSS scans in this paper. Based on photographic data of the National Geographic Society -- Palomar Observatory Sky Survey (NGS-POSS) obtained using the Oschin Telescope on Palomar Mountain. The NGS-POSS was funded by a grant from the National Geographic Society to the California Institute of Technology. The plates were processed into the present compressed digital form with their permission. The Digitized Sky Survey was produced at the Space Telescope Science Institute under US Government grant NAG W-2166.  This research has made use of the SIMBAD database, operated at CDS, Strasbourg, France.

\bsp	
\label{lastpage}
\end{document}